\documentclass[journal,twoside,web]{ieeecolor}

\usepackage{jsen}
\usepackage[noadjust]{cite}
\usepackage{amsmath,amssymb,amsfonts}
\usepackage{graphicx}
\usepackage{textcomp}
\usepackage{wrapfig}
\usepackage{array}
\newcolumntype{M}[1]{>{\centering\arraybackslash}m{#1}}
\usepackage[caption=false,font=footnotesize]{subfig}
\usepackage[hyphens]{url}
\usepackage{stfloats}

\usepackage{booktabs}
\usepackage{nicematrix}
\NiceMatrixOptions{cell-space-top-limit = 1pt}
\usepackage[acronym,shortcuts]{glossaries}
\makeglossaries
\newacronym{cw}{CW}{Continuous Wave}
\newacronym{fmcw}{FMCW}{Frequency Modulated Continuous Wave}
\newacronym{bpsk}{BPSK}{Binary Phase-shift Keying}
\newacronym{fov}{FOV}{Field of View}
\newacronym{if}{IF}{Intermediate Frequency}
\newacronym{mimo}{MIMO}{Multiple Input Multiple Output}
\newacronym{los}{LOS}{Line of Sight}
\newacronym{pcb}{PCB}{Printed Circuit Board}
\newacronym[longplural={Angles Of Arrival}]{aoa}{AoA}{Angle of Arrival}
\newacronym{aod}{AoD}{Angle of Departure}
\newacronym{tof}{ToF}{Time of Flight}
\newacronym{rn}{RN}{Random Noise}
\newacronym{p}{P}{Pulse}
\newacronym{iq}{I/Q}{In/Quadrature}
\newacronym{sar}{SAR}{Synthetic Aperture Radar}
\newacronym{slar}{SLAR}{Side Looking Airborne Radar}
\newacronym{ic}{IC}{Integrated Circuit}
\newacronym{v2i}{V2I}{Vehicle to Infrastructure}
\newacronym{v2v}{V2V}{Vehicle to Vehicle}
\newacronym{dsrc}{DSRC}{Dedicated Short Range Communication}
\newacronym{obu}{OBU}{On-Board Unit}
\newacronym{rsu}{RSU}{Roadside Unit}
\newacronym{rss}{RSS}{Received Signal Strength}
\newacronym{cir}{CIR}{Channel Impulse Response}
\newacronym[longplural={Regions Of Interest}]{roi}{ROI}{Region Of Interest}
\newacronym{cfar}{CFAR}{Constant False Alarm Rate}
\newacronym{pca}{PCA}{Principal Component Analysis}
\newacronym{svm}{SVM}{Support Vector Machine}
\newacronym{t-sne}{t-SNE}{t-distributed Stochastic Neighbor Embedding}
\newacronym{adc}{ADC}{Analog-to-Digital Converter}
\newacronym{mle}{MLE}{Maximum Likelihood Estimation}
\newacronym{cnn}{CNN}{Convolutional Neural Network}
\newacronym{rnn}{RNN}{Recurrent Neural Network}
\newacronym{lstm}{LSTM}{Long Short Term Memory}
\newacronym{gru}{GRU}{Gated Recurrent Unit}
\newacronym{csvc}{C-SVC}{C-Support Vector Classification}
\newacronym{pdf}{PDF}{Probability Density Function}
\newacronym{icp}{ICP}{Iterative Closest Point}
\newacronym{ndt}{NDT}{Normal Distribution Transform}
\newacronym{roc}{ROC}{Receiver Operating Characteristic}
\newacronym{crlb}{CRLB}{Cramér-Rao Lower Bound}
\newacronym{rms}{RMS}{Root Mean Square}
\newacronym{rsa}{RSA}{RSS Series Analysis}
\newacronym{svd}{SVD}{Singular Value Decomposition}
\newacronym{svdd}{SVDD}{Support Vector Data Description}
\newacronym{mpm}{MPM}{Minimax Probability Machine}
\newacronym{som}{SOM}{Self-Organizing Map}
\newacronym{lvq}{LVQ}{Learning Vector Quantization}
\newacronym{knn}{k-NN}{k-Nearest Neighbors}
\newacronym{slam}{SLAM}{Simultaneous Localization And Mapping}
\newacronym{tdoa}{TDoA}{Time Difference of Arrival}
\newacronym{adam}{ADAM}{Adaptive Moment}
\newacronym{nms}{NMS}{Non-Maximum Suppression}
\newacronym{tcn}{TCN}{Temporal Convolutional Network}
\newacronym{bislr}{BiSLR}{Bioinformatic and Systematic Literature Review}
\newacronym{ddbr}{DDBR}{Dual-differential Background Removal.}
\newacronym{stft}{STFT}{Short-time Fourier transform}
\newacronym{fft}{FFT}{Fast Fourier Transform}
\newacronym{cpi}{CPI}{Coherent Processing Interval}
\newacronym{emd}{EMD}{Empirical Mode Decomposition}
\newacronym{imf}{IMF}{Intrinsic Mode Function}
\newacronym{mser}{MSER}{Maximally Stable Extremal Region}
\newacronym{csi}{CSI}{Channel State Information}
\newacronym{mvdr}{MVDR}{Minimum Variance Distortionless Response}
\newacronym{music}{MUSIC}{MUltiple SIgnal Classification}
\newacronym{samv}{SAMV}{Sparse Asymptotic Minimum Variance}
\newacronym{hmm}{HMM}{Hidden Markov Model}
\newacronym{mti}{MTI}{Moving Target Indication}
\newacronym{llse}{LLSE}{Linear Least Squares Estimation}
\newacronym{lrmf}{LRMF}{Low Rank Matrix Factorization}
\newacronym{fwt}{FWT}{Fast Wavelet Transform}
\newacronym{mrc}{MRC}{Maximal Ration Combining}
\newacronym{dtdoa}{DTDOA}{Double Time Difference Of Arrival}

\newcommand\Tstrut{\rule{0pt}{2.6ex}}   
\newcommand\TstrutTwo{\rule{0pt}{3.6ex}}   

\def\BibTeX{{\rm B\kern-.05em{\sc i\kern-.025em b}\kern-.08em
    T\kern-.1667em\lower.7ex\hbox{E}\kern-.125emX}}

\definecolor{abstractbg}{rgb}{0.89804,0.94510,0.83137}
\begin{document}

\title{Millimeter Wave Sensing: A Review of Application Pipelines and Building Blocks}

\author{Bram van Berlo, Amany Elkelany, Tanir Ozcelebi, and Nirvana Meratnia
\thanks{B.R.D. van Berlo, A.M.A. Elkelany, T. Ozcelebi, and N. Meratnia are with the Interconnected Resource-aware Intelligent Systems cluster, Department of Mathematics and Computer Science, Eindhoven University of Technology, 5600 MB Eindhoven, The Netherlands (e-mail: b.r.d.v.berlo@tue.nl; a.m.a.elkelany@tue.nl; t.ozcelebi@tue.nl; n.meratnia@tue.nl). Correspondence should be addressed to B.R.D. van Berlo.}}

\IEEEtitleabstractindextext{%
\fcolorbox{abstractbg}{abstractbg}{%
\begin{minipage}{\textwidth}%
\begin{wrapfigure}[17]{r}[-12pt]{4in}%
\includegraphics[width=4in]{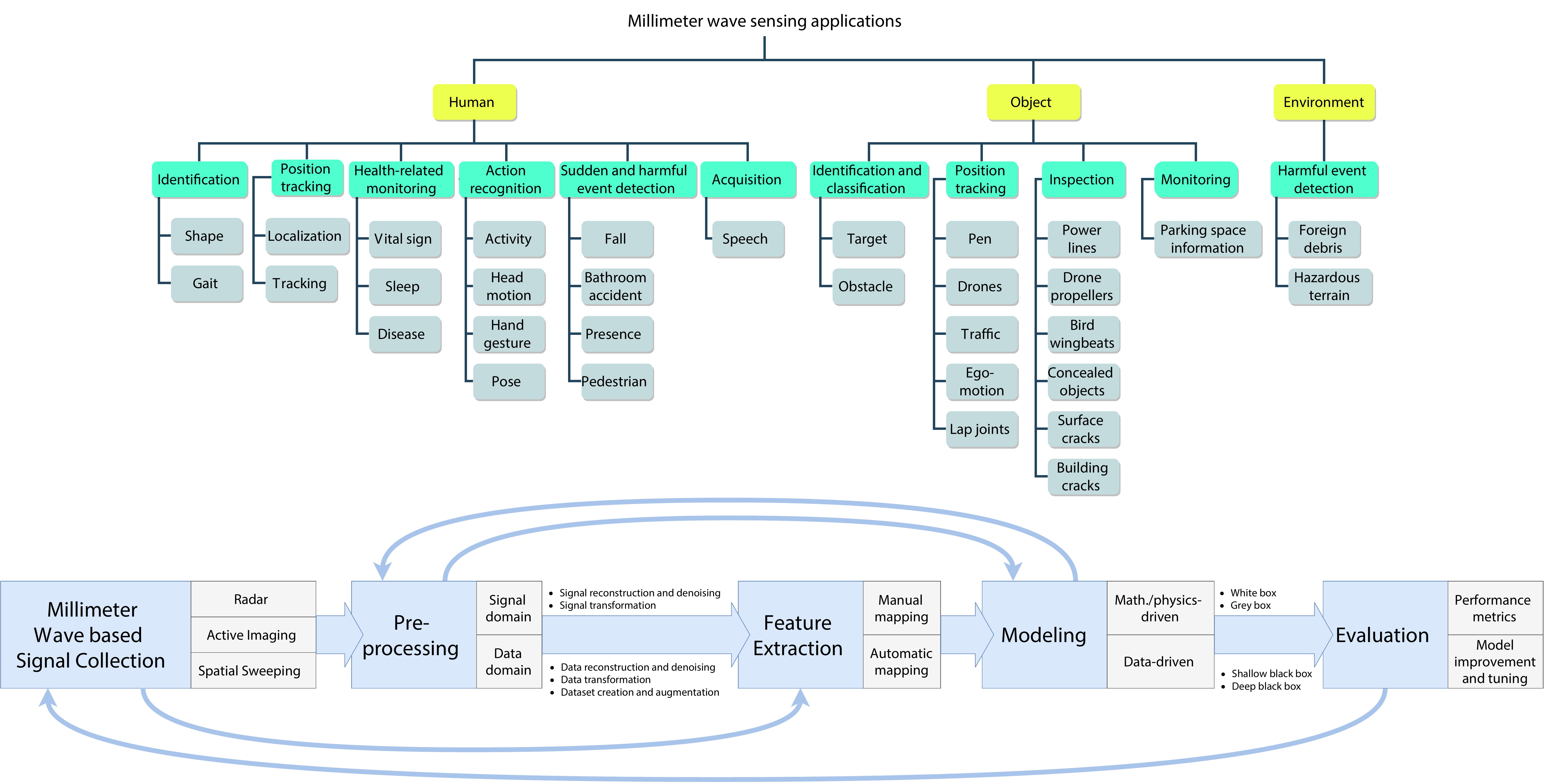}%
\end{wrapfigure}%
\begin{abstract}
The increasing bandwidth requirement of new wireless applications has lead to standardization of the millimeter wave spectrum for high-speed wireless communication. The millimeter wave spectrum is part of 5G and covers frequencies between 30 and 300 GHz corresponding to wavelengths ranging from 10 to 1 mm. Although millimeter wave is often considered as a communication medium, it has also proved to be an excellent `sensor', thanks to its narrow beams, operation across a wide bandwidth, and interaction with atmospheric constituents. In this paper, which is to the best of our knowledge the first review that completely covers millimeter wave sensing application pipelines, we provide a comprehensive overview and analysis of different basic application pipeline building blocks, including hardware, algorithms, analytical models, and model evaluation techniques. The review also provides a taxonomy that highlights different millimeter wave sensing application domains. By performing a thorough analysis, complying with the systematic literature review methodology and reviewing 165 papers, we not only extend previous investigations focused only on communication aspects of the millimeter wave technology and using millimeter wave technology for active imaging, but also highlight scientific and technological challenges and trends, and provide a future perspective for applications of millimeter wave as a sensing technology.
\end{abstract}

\end{minipage}}}

\maketitle

\section{Introduction}
\label{sec:introduction}
\IEEEPARstart{I}{ntroduction} of new wireless applications, their higher service quality requirements, and a significant increase of application users have put continuous demand on digital wireless communication bandwidth. To address this, Netflix and YouTube, for instance, have reduced video streaming quality in Europe to mitigate data traffic peaks in the 2.4 and 5 GHz WiFi bands. These peaks are caused by an increased amount of video and movie streams due to home confinement measures taken during the COVID-19 pandemic~\cite{gold_internet_breaking}. In the future, 8K video streaming will require a minimum data rate of 50 Mbps per TV. In case multiple simultaneous streams are launched on the same network, gigabit connections will approach their limit~\cite{hendrickson_8k_tv}. Maintaining the required network bandwidth and processing speed remains challenging for cloud gaming as a service~\cite{10.1007/978-3-642-40276-0_10}.

The increasing bandwidth requirement has lead to standardization of the millimeter wave spectrum for high-speed wireless communication in the IEEE 802.11ad, 802.11aj, and 802.11ay amendments~\cite{802_11_ad,802_11_aj,802_11_ay}. The millimeter wave is also part of the fifth-generation mobile communication technology (5G)~\cite{6894452,7010530}. Although use of the millimeter wave spectrum for communication is often directly associated with 5G, there are key differences. The millimeter wave spectrum is just one part of what 5G networks use to provide higher data rates. It covers frequencies between 30 and 300 GHz corresponding to wavelengths ranging from 10 to 1 mm~\cite{Hemadeh2018,8373698}, located between the centimeter wave and terahertz wave spectrums.

Thanks to covering a wide bandwidth and utilizing a short wavelength, the millimeter wave spectrum is able to provide higher data rates compared to other widely used wireless technologies such as WiFi. Compared to the spectrum in which 5 GHz WiFi operates, the bandwidth of the millimeter wave spectrum is 10x higher, i.e., 270 GHz versus 27 GHz. Under the assumption that environmental and legislative limitations do not exist, this means that the number of channels that can be created in the millimeter wave spectrum is also 10x more. Because millimeter waves operate at much higher frequencies compared to the WiFi bands, the wavelengths are much shorter. Consequently, the size of electronic components can be reduced. This, in turn, causes the beam emitted by electronic components to be much narrower. The narrow beams, in combination with greater signal attenuation compared to the WiFi bands, allow increased communication density~\cite{8373698}, i.e., number of messages communicated in parallel over separate links with the same carrier frequency in a limited area.

Although millimeter wave is often considered as a communication medium, it has also proved to be an excellent \textit{`sensor'} for humans, objects, and environmental sensing~\cite{8373698, Hemadeh2018}, thanks to its narrow beams, operation across a wide bandwidth, and interaction with atmospheric constituents. Narrow beams result in greater sensing resolution and directivity. A wide bandwidth and specific penetration, reflection, and attenuation reactions to different materials allow distinguishing different objects and humans~\cite{ti_mmwave_sensors,5876482}. 

However, using millimeter waves in sensing applications also has disadvantages. Signals at extremely high frequencies suffer from significant attenuation. Millimeter waves can therefore hardly be used for long-distance applications~\cite{Akyildiz2018}. The financial cost of deploying a millimeter wave sensing system, e.g., on a vehicle, is still high even though this cost is projected to decrease in the future~\cite{mmwave_system_deployment_cost}. Millimeter waves also suffer significant penetration loss through solid materials like concrete. As rain-drops are comparable in size to the wavelength of millimeter waves, heavy rain can cause considerable attenuation due to scattering~\cite{8373698, Hemadeh2018}.

The applications that use millimeter waves utilize a wide variety of hardware and algorithms for data collection, data pre-processing, feature extraction, feature analysis, analytical modeling, and modeling evaluation. This paper provides a comprehensive overview and analysis of millimeter wave sensing application pipelines and basic building blocks, including hardware, algorithms, analytical models, and model evaluation techniques, offers an insight into challenges and trends, and provides a future perspective for applications of millimeter wave as a sensing technology. Therefore, the review has struck a balance between the amount of details that are provided for each millimeter wave sensing application pipeline building block and does not elaborate on details of each application domain and algorithm independently. Due to the multidisciplinary nature of millimeter wave sensing application pipelines, the review offers technical details for each reader depending on familiarity with a certain field. For example, data collection and a part of data pre-processing are topics of electrical engineering, while analytical modeling and modeling evaluation are topics of data science and artificial intelligence. The main contributions of the paper are:

\begin{itemize}
    \item Extending previous investigations focused \textit{only on} communication aspects of the millimeter wave technology and using millimeter wave technology for active imaging. To the best of our knowledge, this is the first review that completely covers millimeter wave sensing application pipelines and pipeline building blocks.
    
    \item Providing a taxonomy that highlights different millimeter wave sensing application domains.
    
    \item Making a comprehensive review and analysis of hardware, algorithms, analytical models, and model evaluation techniques for each of the identified millimeter wave sensing application pipeline building blocks.
    
    \item Identifying commonalities, gaps, and shortcomings of current studies and solutions focusing on the use of millimeter wave as a sensing technology.
    
    \item Highlighting scientific and technological challenges and trends, and providing a future perspective for applications of millimeter wave as a sensing technology.
\end{itemize}

The review is organized as follows: Section~\ref{sec:survey_methodology} explains the review methodology. Section~\ref{sec:application_taxonomy} establishes an application taxonomy based on the research papers that report on millimeter wave sensing. Section~\ref{sec:application_pipeline} identifies the common building blocks that make up the application pipelines presented in the millimeter wave sensing papers. It also presents a thorough analysis of the wide variety of hardware, algorithms, analytical models, and model evaluation techniques across these papers in the respective building blocks. Section~\ref{sec:challenges_future_trends} identifies and explains the challenges and trends, and provides a future perspective for applications of millimeter wave as a sensing technology. Finally, Section~\ref{sec:conclusion} concludes the paper.

\section{Review Methodology}
\label{sec:survey_methodology}
\begin{figure}[t]
    \centering
    \includegraphics[width=6cm]{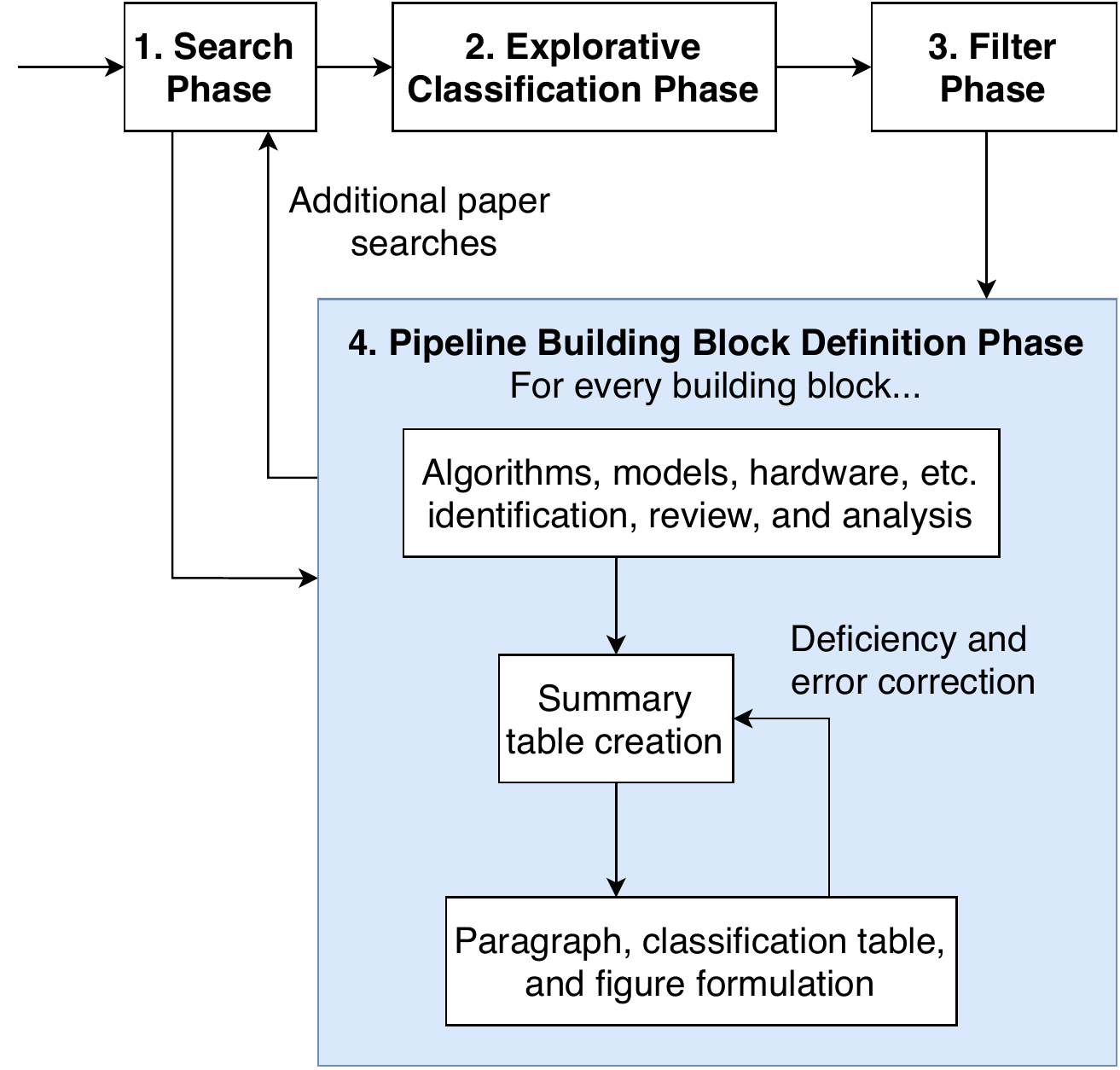}
    \caption{Review process block diagram}
    \label{fig:Search}
\end{figure}

We have conducted this review with a specific process involving four sequential phases. These phases are depicted in Figure~\ref{fig:Search}. The process is based on the iterative \gls{bislr} spiral model developed by Mariano et al.~\cite{mariano2017guide}. The differences compared to the iterative \gls{bislr} spiral model are explained below. Afterwards, the four sequential process phases are explained.

The iterative \gls{bislr} spiral model starts with a protocol definition phase. In this phase, main and specific research questions, research objectives, and inclusion and exclusion criteria are defined~\cite{mariano2017guide}. We omit the definition of main and specific research questions. The defined research objectives can be found in Section~\ref{sec:introduction}. Inclusion and exclusion criteria are defined in a later paper filter phase. During the reference collection phase, the iterative \gls{bislr} spiral model suggests to select scientific databases, develop and evaluate search keywords, and iteratively repeat the phase with different scientific databases and search keywords in case the research objectives cannot be reached~\cite{mariano2017guide}. We initially performed a rigurous reference collection, i.e., the search phase. Later during the pipeline building block definition phase, several small search phases were performed iteratively. In the data evaluation phase, the iterative \gls{bislr} spiral model performs title, abstract, diagonal, and full-text reading to determine if papers should be included in the list of papers used for data collection. The iterative \gls{bislr} spiral model finishes with data collection and narrative synthesis~\cite{mariano2017guide}. We used title, abstract, and diagonal reading for, and collected data during, the explorative paper classification phase and used full-text reading, collected data, and performed narrative synthesis during the application pipeline building block definition phase.

\begin{figure*}[ht]
    \centering
    \includegraphics[width=16cm]{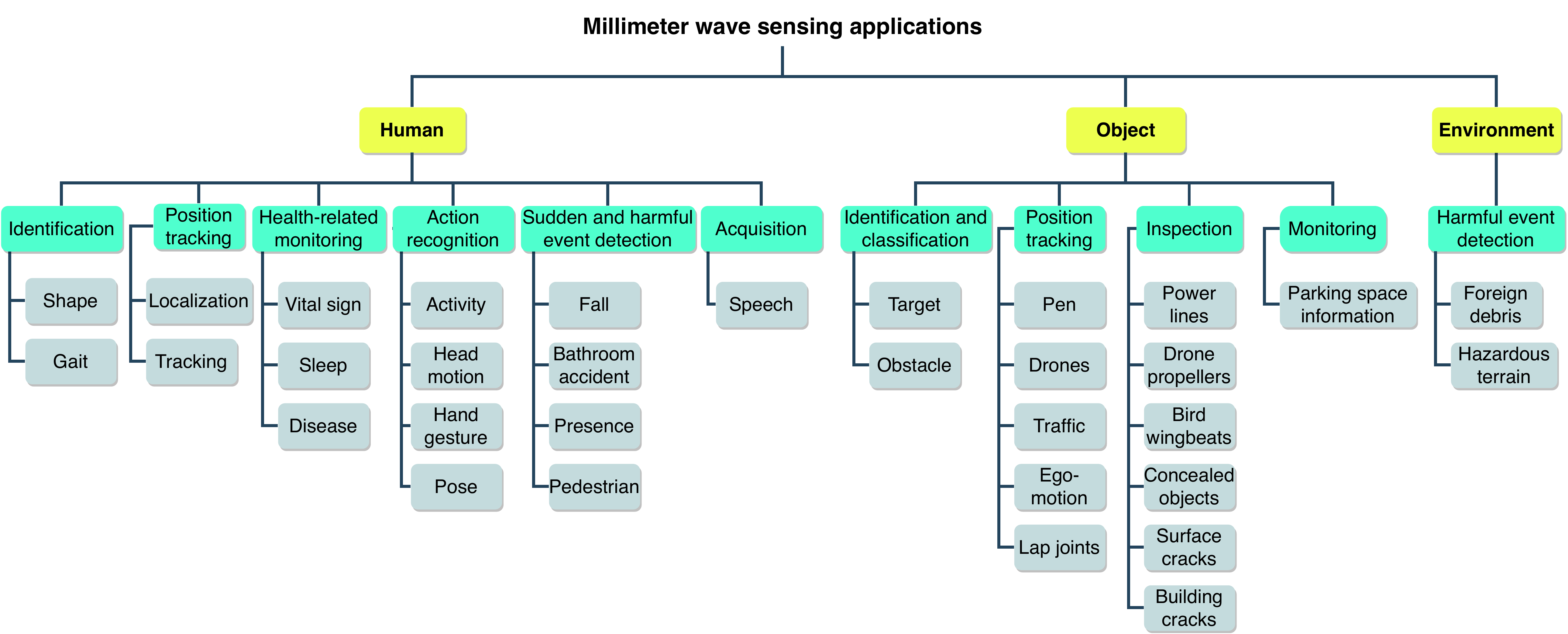}
    \caption{Taxonomy of millimeter wave sensing applications}
    \label{fig:application_taxonomy}
\end{figure*}

In the first phase, we searched for relevant papers published and indexed in digital libraries such as IEEE Xplore, ACM, Elsevier, and SPIE. Because millimeter wave technology is used across different sensing applications that use different terminologies, we used diverse search terminologies. Therefore, finding relevant papers required using synonyms of terms, which we categorized as exact, similar, narrower, and broader terms. Similar terms for `millimeter wave' include `mmwave', `mm-wave' and `mmw'. Narrower terms include `v-band', `w-band', `Ka-band', `Ka band', `v band' and `w band'. Broader terms include `microwave', `micro-Doppler', `Doppler' and `radar'. Several gesture recognition papers only refer to the specific technology used for tracking in the title. The narrower term for gesture recognition used in various papers is `Soli'. Term combinations used for searching were made with exact, similar, and narrower terms and consisted of a millimeter wave term and application type term. Based on the application domains addressed in the papers, we made an application taxonomy shown in Figure~\ref{fig:application_taxonomy}. Application type terms were derived from this taxonomy. Term combination examples include `millimeter wave tracking', `mm-wave gesture', and `w-band detection'. Broader millimeter wave terms such as `microwave' and `radar' resulted in too many hits that mostly included papers outside the scope of this review. Several millimeter wave sensing papers used in the review identify their content with broader millimeter wave terms. These papers were cited in the reference lists of millimeter wave papers found during paper search.

In the second phase we classified over 140 papers found during the first phase to identify relevant information. The classification was based on the most commonly identified elements, such as application domain addressed, objectives, target group, dataset size, modeling technique, methodology, deployment environment, evaluation parameters, and opportunities and challenges. The research papers featured varying publication dates from the year 1994 up until 2020. Based on the results obtained in the explorative phase, we made an application taxonomy (see Section~\ref{sec:application_taxonomy}) as well as an application pipeline consisting of several generic pipeline building blocks used in these papers.

The third phase was used to determine inclusion criteria for this review. After studying over 140 papers, we discovered some papers claiming to cover millimeter wave systems while they do not. In some of these papers, the carrier frequency utilized falls outside the 30 - 300 GHz band~\cite{36,64}. We also excluded papers written in other languages than English~\cite{83}. Papers focusing on radiometry (passive sensing and imaging)~\cite{10,30,40,45,47,51,55,56,58,60,84,111,114,145,154,156}, spectroscopy~\cite{89,90}, interferometry~\cite{109}, and utilization of waveguide technology~\cite{97,98,120} were also excluded. Radiometry and spectroscopy do not focus on actively emitting and measuring effects on millimeter waves for sensing. The waveguide technology has been used to design electric probes for interacting with membrane systems, corrosion, sintering process, etc.~\cite{97,98,120}. The technology has not been used explicitly for guiding millimeter waves to and from an antenna. Interferometry, rather than measuring effects on millimeter waves, measures wave interference using a multi-radar setup with special radar configurations~\cite{109} for sensing.

Radiometry measures electromagnetic radiation originating from humans or objects with receiver setups~\cite{79}. One group of radiometers creates passive millimeter wave images. These images are bi-dimensional radiation maps of a scene. Several studies focus on the creation of these systems~\cite{47,156}. In addition, several studies perform multiple object detection and tracking~\cite{30}, hidden object detection~\cite{10,40,45,55,60,114}, and military target detection~\cite{56} with analytical models that take passive millimeter wave images as input data. Another group of radiometers generates one dimensional radiation profiles. In~\cite{111}, a passive radiometric temperature profile is a one dimensional output voltage signal which can be generated in either power detection mode or correlation mode. In power detection mode, the output signal depends on antenna temperature which is proportional to the radiometric temperature of humans or objects within the radiometer's \gls{fov}. In correlation mode, output signals from two receivers in power detection mode are used in a correlator to produce an output signal which does not contain objects which mostly reflect and scatter radiation rather than radiate it. The detection capability of detection algorithms that can be used with radiometric temperature profiles has also been tested~\cite{111}. Millimeter wave hardware and a prediction algorithm that can generate human body emitted energy output traces and detect weapons and explosives~\cite{51,154} have also been created. Yujiri et al.~\cite{58} measure radiation temperature profiles of buried mines. Other uses of radiometry include analysis of sun brightness temperature and precipitating cloud extinction by means of analytical modeling and a sun-tracking radiometer~\cite{84}, and creation of an analytical model for predicting relative humidity profiles from clear-air radiances~\cite{145}.

Spectroscopy is a kind of radiometry in which the interaction between radiation and matter is measured~\cite{90}. Schmalz et al.~\cite{89} use gas spectroscopy to perform breath analysis. Schmalz et al. explain that ``a typical gas spectrometer consists of a radiation source, an absorption cell, a detector, and optical elements. The radiation is transmitted through an absorption cell, which is filled with a gas at a particular pressure and impinges on a detector, which generates an output voltage or current''~\cite{89}. Spectroscopy has also been used to analyze solar system objects by probing temperature and molecular abundance in planetary atmospheres~\cite{90}.

In the fourth phase, we defined the application pipeline building blocks. For every building block, we identified, reviewed, and analyzed the designed (or used) algorithms, models, hardware, and summarized findings in tables that map to the individual papers. Using the summary tables, paragraphs, comparative tables, and figures were formulated (see Section~\ref{sec:application_pipeline}) that explain the application pipeline building blocks. Occasionally, during paragraph, comparative table, and figure formulation, summary table deficiencies and errors in the summary table were discovered. These deficiencies and errors were corrected by revising the summary table and afterwards updating the associated paragraphs, comparative tables, and figures. This was an iterative process, during which extra papers were searched (phase 1) to cover the field as much as possible and to include all relevant papers. As a result, 25 additional papers were found (on top of the first 140 papers). These papers were inspected with the inclusion criteria (phase 2) and used during application pipeline building block definition. Therefore, 165 papers in total were analyzed during the literature review.

\section{Application Taxonomy}
\label{sec:application_taxonomy}
Millimeter wave sensing applications can be classified based on the entity (i.e. human, object, or the environment) being sensed, application goal, and application context. Figure~\ref{fig:application_taxonomy} represents the applications that we have identified based on these criteria. Application types and goals in sensing humans with millimeter waves include: identification, position tracking, action recognition, health-related monitoring, sudden and harmful event detection, and acquiring speech data. Application types and goals in sensing objects include: object identification and classification, position tracking, object inspection, and monitoring information derived from object characteristics. The environment is sensed in applications whose types and goals have been mainly focused on detection of harmful events related to space landing and airport runways.

\begin{table*}[hb]
    \centering
    \caption{Summary of data collection devices using radar technology. The modulation scheme and measured variable abbreviations are explained throughout Section~\ref{sec:data_collection_radar}.} \label{tab:data_collection_radar_summary}
    \tiny\sffamily
    \setlength{\tabcolsep}{2pt}
    \begin{NiceTabular}{@{}lcccccccccc|ccccc|ccccM{12mm}M{8.5mm}M{12mm}M{6mm}M{20mm}c@{}}[code-before =\rowcolors{3}{}{gray!12}]
         \noalign{\hrule height 0.8pt}
         & \Block{1-10}{\textbf{Carrier frequency (GHz)}} & & & & & & & & & & \Block{1-5}{\textbf{Modulation scheme}} & & & & & \Block{1-10}{\textbf{Measured variables}} & & & & & & & & & \Tstrut\\[1.2ex]
         & 30-39 & 40-49 & 50-59 & 60-69 & 70-79 & 80-89 & 90-99 & 120-129 & 160-169 & 220-229 & P & RN & BPSK & CW & FMCW & \(R\) & \(v_r\) & \(\theta\) & \(\varphi\) & \(\Delta\phi_t / f_d\) across time & Reflective intensity & Attenuation power & \acs{if} amplitude & Local min/max received signal amplitudes & \acs{rss} \\[1.7ex] \hline

           \cite{2} &   &   &   & X &   &   &   &   &   &   &   &   &   &   & X & X & X &   &   &   &   &   &   &   &   \\ 
           \cite{4} &   &   &   &   & X &   &   &   &   &   &   &   &   &   & X & X & X & X &   &   &   &   &   &   &   \\ 
           \cite{5} &   &   &   &   & X &   &   &   &   &   &   &   &   &   & X & X & X &   & X &   &   &   &   &   &   \\ 
           \cite{7} &   &   &   & X &   &   &   &   &   &   &   &   & X &   & X & X & X &   &   &   &   &   &   &   &   \\ 
           \cite{8} &   &   &   &   & X &   &   &   &   &   &   &   &   &   & X & X & X & X & X &   &   &   &   &   &   \\ 
           \cite{9} &   &   &   &   & X &   &   &   &   &   &   &   &   &   & X & X & X & X & X &   &   &   &   &   &   \\ 
          \cite{11} &   &   &   &   & X &   &   &   &   &   &   &   &   &   & X & X & X & X &   & X &   &   &   &   &   \\ 
          \cite{13} &   &   &   & X &   &   &   &   &   &   &   &   &   &   & X & X & X &   &   &   &   &   &   &   &   \\ 
          \cite{14} &   &   &   &   & X &   &   &   &   &   &   &   &   &   & X & X & X & X &   &   &   &   &   &   &   \\ 
          \cite{15} &   &   &   &   &   &   &   &   &   &   & X &   &   &   &   & X & X & X &   &   &   &   &   &   &   \\ 
          \cite{16} &   &   &   &   & X &   &   &   &   &   & X &   &   &   &   & X & X & X &   &   &   &   &   &   &   \\ 
          \cite{17} &   &   &   &   & X &   &   &   &   &   &   &   &   &   & X & X & X & X &   & X &   &   &   &   &   \\ 
          \cite{18} &   &   &   &   &   &   &   &   &   &   &   &   &   & X &   &   &   & X &   & X &   &   &   &   &   \\ 
          \cite{21} &   &   &   & X &   &   &   &   &   &   &   &   &   &   & X & X & X & X &   &   &   &   &   &   &   \\ 
          \cite{23} &   &   &   & X &   &   &   &   &   &   &   &   &   &   & X & X &   & X &   &   &   &   &   &   &   \\ 
          \cite{24} &   &   &   &   & X &   &   &   &   &   &   &   &   &   & X & X &   &   &   & X &   &   &   &   &   \\ 
          \cite{25} &   &   &   &   & X &   &   &   &   &   &   &   &   &   & X & X &   & X &   &   &   &   &   &   &   \\ 
          \cite{28} &   &   &   &   &   &   & X &   &   &   &   &   &   &   &   &   &   &   &   & X &   &   &   &   &   \\ 
          \cite{29} &   &   &   &   &   &   &   &   &   & X &   &   &   & X &   &   &   &   &   & X &   &   &   &   &   \\ 
          \cite{31} &   &   &   &   & X &   &   &   &   &   & X &   &   &   &   & X & X & X &   &   &   &   &   &   &   \\ 
          \cite{34} &   &   &   &   & X &   &   &   &   &   &   &   &   &   & X & X &   & X &   & X &   &   &   &   &   \\ 
          \cite{35} &   &   &   & X &   &   &   &   &   &   &   &   &   &   & X & X & X &   &   &   &   &   &   &   &   \\ 
          \cite{37} & X &   &   &   &   &   &   &   &   &   &   &   &   & X &   &   &   &   &   & X &   &   &   &   &   \\ 
          \cite{42} &   &   &   &   &   &   & X &   &   &   &   &   &   & X &   &   &   &   &   & X &   &   &   &   &   \\ 
          \cite{43} &   &   &   &   &   &   &   &   &   &   &   &   &   &   &   & X & X &   &   &   &   &   &   &   &   \\ 
          \cite{44} &   &   &   &   &   &   &   &   &   &   &   &   &   &   &   & X & X &   & X &   &   &   &   &   &   \\ 
          \cite{50} &   &   &   &   &   &   & X &   &   &   &   &   &   & X &   &   &   &   &   &   & X &   &   &   &   \\ 
          \cite{54} &   &   &   &   & X &   &   &   &   &   &   &   &   &   & X &   &   &   &   &   &   &   &   &   &   \\ 
          \cite{57} &   &   &   &   & X &   &   &   &   &   &   &   &   &   & X & X &   &   &   &   &   &   &   &   &   \\ 
          \cite{62} &   &   &   &   & X &   &   &   &   &   &   &   &   &   & X & X & X & X &   &   &   &   &   &   &   \\ 
          \cite{63} &   &   &   &   & X &   &   &   &   &   &   &   &   &   & X & X &   &   & X &   &   &   &   &   &   \\ 
          \cite{65} &   &   &   &   & X &   &   &   &   &   &   &   &   &   & X & X &   &   &   &   &   &   &   &   &   \\ 
          \cite{69} &   &   &   &   & X &   & X &   &   &   &   &   &   &   & X & X &   &   &   &   &   &   &   &   &   \\ 
          \cite{70} &   &   &   &   & X &   &   &   &   &   &   &   &   &   & X &   &   &   &   & X &   &   &   &   &   \\ 
          \cite{71} &   &   &   &   & X &   &   &   &   &   &   &   &   &   & X & X & X & X & X &   &   &   &   &   &   \\ 
          \cite{72} &   &   &   &   & X &   &   &   &   &   &   &   &   &   & X & X & X & X &   &   &   &   &   &   &   \\ 
          \cite{74} &   &   &   &   & X &   &   &   &   &   &   &   &   &   & X &   & X &   &   &   &   &   &   &   &   \\ 
          \cite{75} &   &   &   &   &   &   &   &   &   &   &   &   &   & X &   &   &   &   &   & X &   &   &   &   &   \\ 
          \cite{76} &   &   &   & X &   &   &   &   &   &   &   &   &   &   &   &   &   &   &   &   &   &   &   &   &   \\ 
          \cite{77} &   &   &   & X &   &   &   &   &   &   &   &   &   &   & X & X & X &   &   & X &   &   &   &   &   \\ 
          \cite{78} &   &   & X &   &   &   &   &   &   &   &   &   &   &   &   &   &   &   &   & X &   &   &   &   &   \\ 
          \cite{81} &   &   &   &   & X &   &   &   &   &   &   &   &   &   & X & X &   & X &   &   &   &   &   &   &   \\ 
          \cite{82} &   &   &   &   & X &   &   &   &   &   &   &   &   &   &   & X &   &   &   &   &   &   &   &   &   \\ 
          \cite{85} &   &   &   &   & X &   &   &   &   &   &   &   &   &   & X & X &   & X & X &   &   &   &   &   &   \\ 
          \cite{86} &   &   &   &   & X &   &   &   &   &   &   &   &   &   & X & X &   & X &   &   &   &   &   &   &   \\ 
          \cite{87} &   &   &   &   &   &   &   &   & X &   & X &   &   &   &   &   & X &   &   &   &   &   &   &   &   \\ 
          \cite{88} &   &   &   &   &   &   & X &   &   &   & X &   &   &   &   & X &   &   &   &   &   &   &   &   &   \\ 
          \cite{91} &   &   &   &   & X &   &   &   &   &   &   &   &   &   & X & X & X & X & X &   &   &   &   &   &   \\ 
          \cite{92} &   &   &   &   & X &   &   &   &   &   &   &   &   &   & X & X & X & X &   &   &   &   &   &   & X \\ 
          \cite{93} &   &   &   & X &   &   &   &   &   &   &   &   &   &   &   & X & X &   &   &   &   &   &   &   &   \\ \noalign{\hrule height 0.8pt}
    \end{NiceTabular}
\end{table*}

\begin{table*}[ht]
    \ContinuedFloat
    \centering
    \caption{Continued}
    \tiny\sffamily 
    \setlength{\tabcolsep}{2pt}
    \begin{NiceTabular}{@{}lcccccccccc|ccccc|ccccM{12mm}M{8.5mm}M{12mm}M{6mm}M{20mm}c@{}}[code-before =\rowcolors{3}{}{gray!12}]
         \noalign{\hrule height 0.8pt}
         & \Block{1-10}{\textbf{Carrier frequency (GHz)}} & & & & & & & & & & \Block{1-5}{\textbf{Modulation scheme}} & & & & & \Block{1-10}{\textbf{Measured variables}} & & & & & & & & & \Tstrut\\[1.2ex]
         & 30-39 & 40-49 & 50-59 & 60-69 & 70-79 & 80-89 & 90-99 & 120-129 & 160-169 & 220-229 & P & RN & BPSK & CW & FMCW & \(R\) & \(v_r\) & \(\theta\) & \(\varphi\) & \(\Delta\phi_t / f_d\) across time & Reflective intensity & Attenuation power & \acs{if} amplitude & Local min/max received signal amplitudes & \acs{rss} \\[1.7ex] \hline

          \cite{94} &   &   &   &   & X &   &   &   &   &   &   &   &   &   & X & X &   &   &   & X &   &   &   &   &   \\ 
          \cite{96} &   &   &   &   & X &   &   &   &   &   &   &   &   &   & X & X & X & X &   &   &   &   &   &   &   \\ 
          \cite{99} & X &   &   & X &   &   &   &   &   &   &   &   &   &   & X & X &   & X &   &   &   &   &   &   &   \\ 
         \cite{100} &   &   &   &   & X &   &   &   &   &   &   &   &   &   & X & X & X & X &   &   &   &   &   &   &   \\ 
         \cite{102} &   &   &   & X &   &   &   &   &   &   &   &   &   &   & X & X & X & X &   &   &   &   &   &   &   \\ 
         \cite{103} &   &   &   &   & X &   &   &   &   &   &   &   &   &   & X & X & X & X & X &   &   &   &   &   &   \\ 
         \cite{104} &   &   &   &   & X &   &   &   &   &   &   &   &   &   & X & X &   & X &   &   &   &   &   &   &   \\ 
         \cite{105} &   &   &   &   & X &   &   &   &   &   &   &   &   &   & X & X &   & X &   &   &   &   &   &   &   \\ 
         \cite{106} &   &   &   &   &   &   & X &   &   &   &   &   &   &   & X & X &   &   &   &   &   &   &   &   &   \\ 
         \cite{110} &   &   &   &   &   &   & X &   &   &   & X &   &   &   &   & X &   & X &   &   &   &   &   &   &   \\ 
         \cite{112} &   &   &   &   &   &   &   &   &   &   &   & X &   &   &   & X &   &   &   & X &   &   &   &   &   \\ 
         \cite{113} &   &   &   &   & X &   &   &   &   &   &   &   &   &   & X & X & X & X & X &   &   &   &   &   &   \\ 
         \cite{115} & X &   &   &   &   &   &   &   &   &   & X &   &   &   &   &   &   &   &   &   &   &   &   & X &   \\ 
         \cite{121} &   &   &   &   &   &   & X &   &   &   &   &   &   & X &   &   &   &   &   & X &   &   &   &   &   \\ 
         \cite{122} &   &   &   &   & X &   &   &   &   &   &   &   &   &   & X & X &   & X &   & X &   &   &   &   &   \\ 
         \cite{123} &   &   &   &   &   & X &   &   &   &   &   &   &   &   & X & X &   &   &   & X &   &   &   &   &   \\ 
         \cite{124} &   &   &   &   & X &   &   &   &   &   &   &   &   &   & X & X &   &   &   & X &   &   &   &   &   \\ 
         \cite{125} &   &   &   & X &   &   &   &   &   &   &   &   &   & X &   &   &   &   &   & X &   &   &   &   &   \\ 
         \cite{126} & X &   &   &   &   &   &   &   &   &   &   &   &   & X &   &   &   &   &   & X &   &   &   &   &   \\ 
         \cite{127} &   & X &   &   &   &   &   &   &   &   &   &   &   & X &   &   &   &   &   & X &   &   &   &   &   \\ 
         \cite{128} & X &   &   &   &   &   &   &   &   &   &   &   &   &   &   &   &   &   &   & X &   &   &   &   &   \\ 
         \cite{129} &   &   &   &   &   &   & X &   &   &   &   &   &   & X &   &   &   &   &   & X &   &   &   &   &   \\ 
         \cite{132} &   &   &   &   & X &   &   &   &   &   &   &   &   &   & X & X & X & X &   &   &   &   &   &   &   \\ 
         \cite{133} &   &   &   &   & X &   &   &   &   &   &   &   &   &   & X & X & X & X &   &   &   &   &   &   &   \\ 
         \cite{134} &   &   &   &   &   &   & X &   &   &   &   &   &   & X &   &   & X &   &   &   &   &   &   &   &   \\ 
         \cite{136} &   &   &   &   &   &   & X &   &   &   &   &   &   &   & X & X &   &   &   &   &   &   &   &   &   \\ 
         \cite{137} & X &   &   &   &   &   &   &   &   &   &   &   &   &   &   & X &   & X &   &   &   &   &   &   &   \\ 
         \cite{138} &   &   &   &   &   &   & X &   &   &   &   &   &   &   & X & X & X &   &   &   &   &   &   &   &   \\ 
         \cite{139} &   &   &   &   & X &   &   &   &   &   &   &   &   &   & X & X & X & X & X &   &   &   &   &   &   \\ 
         \cite{140} &   &   &   &   & X &   &   &   &   &   &   &   &   &   & X & X & X & X &   &   &   &   &   &   &   \\ 
         \cite{144} &   &   &   &   &   &   &   & X &   &   &   &   &   &   & X & X & X &   &   &   & X &   &   &   &   \\ 
         \cite{146} &   &   &   &   & X &   &   &   &   &   &   &   &   &   & X & X & X &   &   &   &   &   &   &   &   \\ 
         \cite{147} &   &   &   &   & X &   &   &   &   &   &   &   &   &   & X & X & X & X & X &   & X &   &   &   &   \\ 
         \cite{148} &   &   &   & X &   &   &   &   &   &   &   &   &   &   &   & X & X &   &   &   &   &   &   &   &   \\ 
         \cite{149} &   &   &   & X &   &   &   &   &   &   &   &   &   &   & X & X & X & X &   &   &   &   &   &   &   \\ 
         \cite{150} &   &   &   & X &   &   &   &   &   &   &   &   &   &   & X & X & X &   &   &   &   &   &   &   &   \\ 
         \cite{151} &   &   &   & X &   &   &   &   &   &   &   &   &   &   & X & X &   & X & X &   &   &   &   &   &   \\ 
         \cite{152} &   &   &   &   & X &   &   &   &   &   &   &   &   &   & X & X &   &   &   & X &   &   &   &   &   \\ 
         \cite{153} &   &   &   &   & X &   &   &   &   &   &   &   &   &   & X & X & X & X & X &   &   &   &   &   &   \\ 
         \cite{157} &   &   &   &   & X &   &   &   &   &   &   &   &   &   & X & X & X & X &   &   &   &   &   &   &   \\ 
         \cite{158} &   &   &   &   &   &   &   &   &   &   &   &   &   &   & X & X &   & X &   &   &   &   &   &   &   \\ 
         \cite{159} &   &   &   &   & X &   &   &   &   &   &   &   &   &   &   &   & X & X &   &   &   &   &   &   &   \\ 
         \cite{160} &   &   &   &   & X &   &   &   &   &   &   &   &   &   &   & X & X & X &   &   &   &   &   &   &   \\ 
         \cite{161} &   &   &   &   & X &   &   &   &   &   & X &   &   &   &   & X & X & X &   &   &   & X &   &   &   \\ 
         \cite{162} &   &   &   &   & X &   &   &   &   &   &   &   &   &   & X & X &   & X &   &   &   &   & X &   &   \\ 
         \cite{163} &   &   &   & X &   &   &   &   &   &   &   &   &   &   & X &   &   &   &   &   &   &   & X &   &   \\ 
         \cite{164} &   &   &   & X &   &   &   &   &   &   &   &   &   &   & X & X & X & X & X &   &   &   &   &   &   \\ 
         \cite{165} &   &   &   & X &   &   &   &   &   &   &   &   &   &   & X &   &   &   &   &   &   &   & X &   &   \\
         \cite{166} &   &   &   &   & X &   &   &   &   &   &   &   & X &   &   & X &   &   &   &   &   &   &   &   &   \\ \noalign{\hrule height 0.8pt}
    \end{NiceTabular}
\end{table*}

\section{Application Pipeline}
\label{sec:application_pipeline}
This section presents the five common building blocks found in application pipelines of the reviewed papers: data collection, pre-processing, feature extraction, analytical modeling, and modeling evaluation. For each building block, we review and analyze relevant papers and provide a comparative table highlighting their focus. We use these tables to identify commonalities and gaps of employed techniques and methodologies and to highlight challenges.

\subsection{Collection Systems}
\label{sec:data_collection}
The first building block in millimeter wave sensing applications is data collection, in which millimeter wave related measurements are collected using a variety of different measurement systems. In this section, we first briefly explain different data collection approaches and their fundamentals, including suitable antenna types and designs, and then describe the variables that are measured. Hardware calibration~\cite{3,42,92,99,102,118} and physical noise reduction~\cite{125} are considered to be outside the scope of this paper and will therefore not be addressed.

\subsubsection{Suitable Antenna Types and Designs}
Among the papers reporting on measurement systems, a variety of different suitable antenna types and designs have been identified that are used in these measurement systems. To prevent confusion in later sections, we decouple the concept of transmitters and receivers from transmit and receive antenna's in the elaboration on suitable antenna types and designs. This means that any transmitter or receiver has a certain amount (at least one) of transmit or receive antenna's depending on the antenna type and design. Suitable antenna types include, but are not limited to, horn antenna~\cite{12,26,27,32,37,42,50,66,69,73,95,107,112,117,118,125,130,134,135,136,137,138,142,143}, yagi antenna~\cite{95}, lens antenna~\cite{28,121}, reflectarray antenna~\cite{85,86}, cassegrain antenna~\cite{88,129}, (on-chip integrated) patch antenna~\cite{13,18,33,59,78,117,150,151,165}, and parabolic antenna~\cite{126,128}. Suitable antenna designs include, but are not limited to, phased array antenna design~\cite{1,6,20,26,27,66,80,101}, frequency scanned antenna design~\cite{87}, fan and pencil beam antenna design~\cite{138}, gaussian beam antenna design~\cite{106}, and omni-directional antenna design~\cite{32}.

By far most measurement systems identified in the papers either use on-chip integrated patch, i.e. microstrip or printed, antennas or series fed patch antenna arrays. Even though this is not evident when looking at the suitable antenna types and designs that are actually reported, many papers list commercial measurement systems, without listing the used antenna type and design, that use on-chip integrated patch antennas or series fed patch antenna arrays. A patch antenna is advantageous compared to other antenna types because it is low cost, can easily be integrated in printed circuit boards of various sizes, and is easy to mass produce~\cite{8373698}. A series fed patch antenna array is an array of multiple patch antenna's connected via a single feed line in series and is used to create a beam pattern with certain characteristics that cannot be created with a single patch antenna~\cite{radar_basics,Chen2019}.

The phased array antenna design is an antenna array design in which densely packed small gain antenna's are phase shifted by a separate analog or digital phase shift module to create an overall high gain beam that can be steered without mechanically rotating the antenna's~\cite{radar_basics}. This antenna array design is very important in millimeter wave communication systems since, due to experiencing more penentration loss than centimeter wave communication systems and solely relying on \gls{los} propagation, transmitter and receiver beams constantly have to be re-aligned~\cite{8373698}. This antenna array design can also be used in ubiquitous sensing systems that rely on rotation for measuring certain variables to steer the sensing system to a sensing position of interest without experiencing noise caused by mechanical rotation. More information regarding the other antenna types and designs can be found in~\cite{8373698,radar_basics}.

\subsubsection{Radar}
\label{sec:data_collection_radar}
Radars emit electromagnetic radiation signals, which are either reflected or scattered from targets with a smooth or rough surface, respectively. The differences between the emitted and received signals are of interest to the sensing application~\cite{radar_basics}. To understand the basic principles of the radar approaches, we consider in this section situations in which (i) there is a single object or human in the radar's \gls{fov} and (ii) no noise artifacts exist. We will deal with issues related to multiple targets, noise artifacts, increasing sensing resolution, and isolating phase shift components in Section~\ref{sec:data_preprocessing}. Table~\ref{tab:data_collection_radar_summary} presents an overview of papers that used radar for data collection in millimeter wave sensing applications. These papers are compared based on their carrier frequency, modulation scheme, and measured variables, each of which is explained below:

\begin{itemize}
    \item The carrier frequency indicates what part of the millimeter wave frequency band has been explored and to what extent. It is also strongly correlated to the attainable measurement distance in two ways. Firstly, the higher the carrier frequency, the faster the attenuation of the transmitted signal. Secondly, there are exceptions to the first correlation in the form of several attenuation peaks. For example, applications operating at 60-69 GHz do so under signal absorption of the oxygen in the atmosphere. This allows, for example, mobile phone hand gesture recognition in compact populated areas. Human position tracking applications typically operate at around 70-79 GHz. This frequency interval, compared to 60-69 GHz, allows measurements at greater distances~\cite{1456317}.
    \item The modulation scheme dictates which variables the radar can measure and how. Certain modulation schemes have been used to test joint communication and sensing~\cite{166}.
    \item Measured variables strongly depend on application type and its objective. Identification, position tracking, action recognition, etc. rely on range, velocity and angle information~\cite{8,9}. Health monitoring and speech acquisition rely on phase and Doppler shift information in time~\cite{34,125,129}. Object identification and classification employ \gls{if} signal variations across different \gls{if} channels~\cite{163,165}.
\end{itemize}

\begin{figure}[h]
    \centering
    \includegraphics[width=\linewidth]{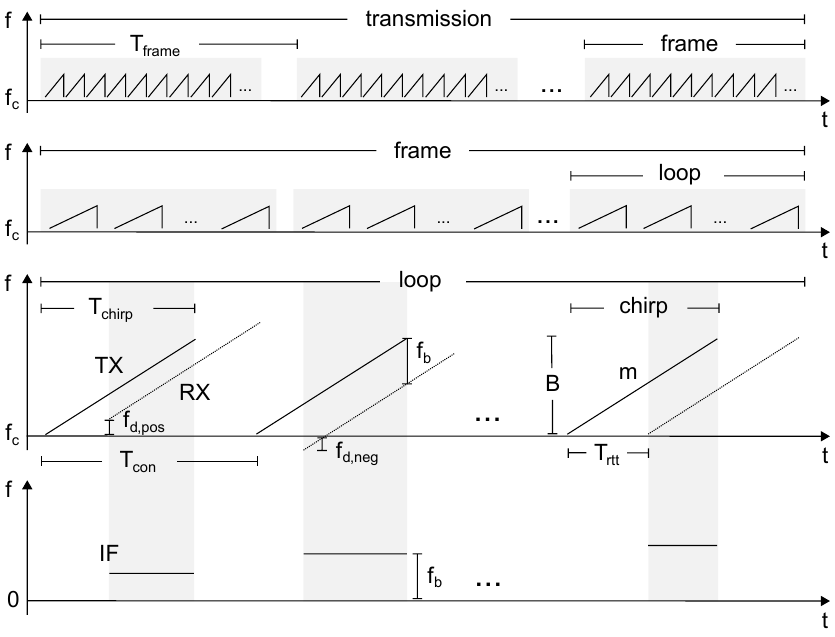}
    \caption{FMCW radar frequency sawtooth modulation across time. Adopted from~\cite{constapel2019practical}.}
    \label{fig:fmcw_sawtooth_modulation}
\end{figure}

The most widely used radar type in millimeter wave sensing applications is the \gls{fmcw} radar, which was commercialized recently with utilization of the \gls{ic} technology~\cite{radar2017,7,tef810x}. Almost all papers found that use this radar modulate the signal frequency according to a sawtooth pattern, which converts the signal into a continuous stream of chirps~\cite{constapel2019practical}. A minority of papers modulate the signal frequency according to a triangular pattern~\cite{85,106}. An explanation of this frequency modulation pattern can be found in~\cite{radar_basics}. Sawtooth frequency modulation is depicted in Figure~\ref{fig:fmcw_sawtooth_modulation}.

\begin{figure}[h]
    \centering
    \includegraphics[width=\linewidth]{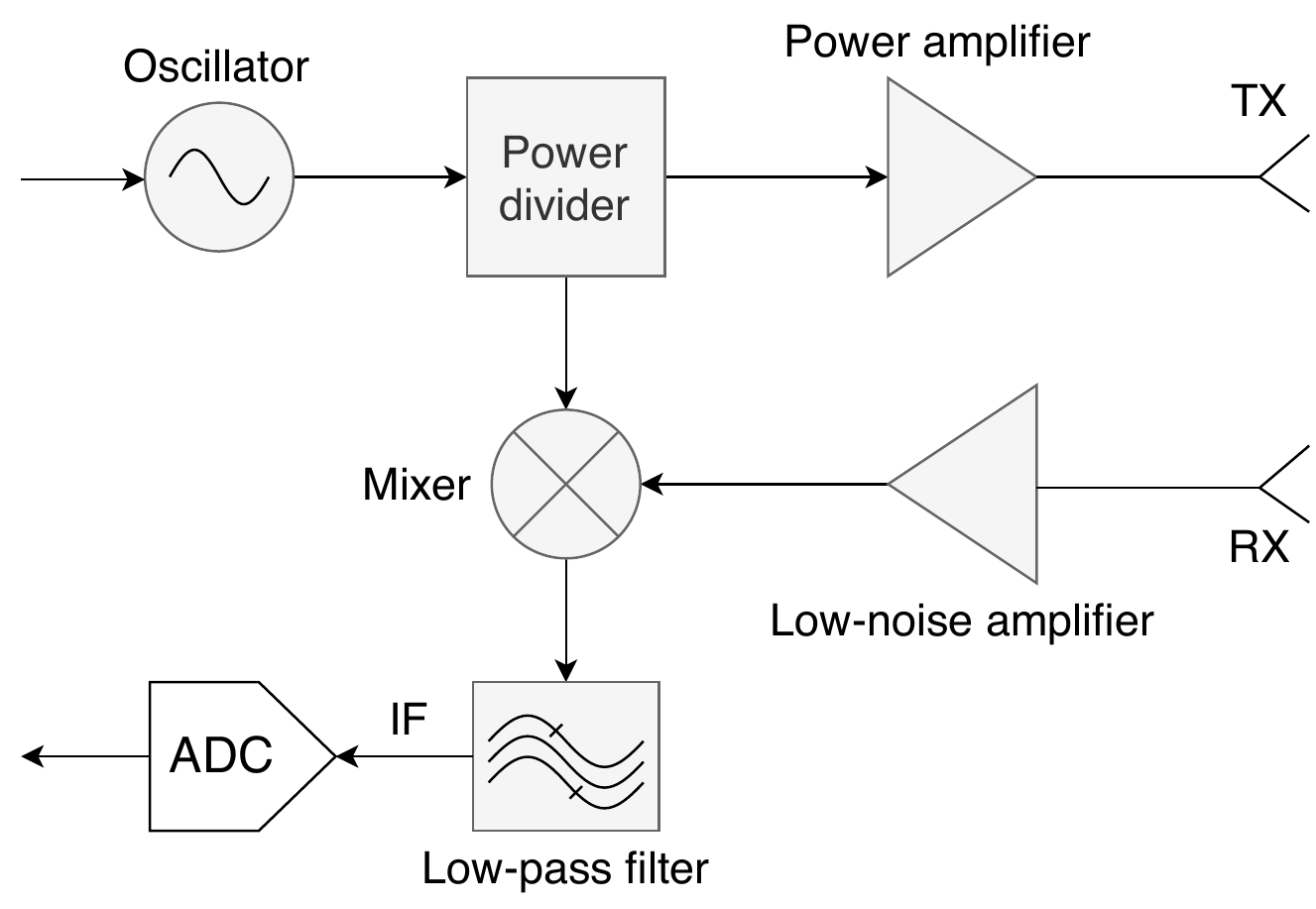}
    \caption{Block diagram of fundamental components of CW radar. Adopted from~\cite{essay70986}.}
    \label{fig:fmcw_block_diagram}
\end{figure}

A chirp is defined as a part of a trigonometric function across a limited time window with length \(T_{chirp}\). In this time window, the signal frequency is increased linearly across a bandwidth \(B\) with slope \(m\) using a voltage-controlled oscillator. The transmitted chirp at transmitter TX is then reflected or scattered from a target and received at receiver RX after a round trip time \(T_{rtt}\). This causes the received chirp to show a frequency deviation \(f_b\) compared to the transmitted chirp at a specific time instant. An \gls{if} signal (a.k.a. beat or baseband signal) operating at constant frequency deviation \(f_b\) can be obtained using a down-conversion mixer and low-pass filter in sequence as shown in Figure~\ref{fig:fmcw_block_diagram}. This operation returns a correct output in time window \(T_{chirp} - T_{rtt}\), where the transmitted and received chirps overlap~\cite{radar2017,essay70986}. Given a trigonometric function \(x = A \cos (2 \pi ft + \phi)\) for the transmitted chirp \(x_{tx}\) and the received chirp \(x_{rx}\), the \gls{if} signal \(x_{if} = [(A_{rx} \cdot A_{tx})/2] \cos [2\pi (f_{tx} - f_{rx})t + (\phi_{tx} - \phi_{rx})]\)~\cite{mixer_intro}. Using frequency deviation \(f_b = f_{tx} - f_{rx}\), the range \(R\) between the radar and the target is computed using Equation~\ref{eqn:range_calculation}~\cite{essay70986}. The symbol \(c\) denotes a constant representing the speed of light. Due to its unmodulated constant signal frequency, a \gls{cw} radar is unable to measure range.

\begin{equation}
\label{eqn:range_calculation}
    R = \frac{f_b \cdot c}{2 \cdot \frac{B}{T_{chirp}}}
\end{equation}

Radial velocity \(v_r\) of the target can be obtained by using consecutive chirps separated by time window \(T_{con}\) emitted across a loop. In case the target is in motion, the \gls{if} signal resulting from the chirps will experience a significant phase difference \(\Delta\phi_t\) relative to the previous loop. The resulting frequency difference is indiscernible for small motion changes. The radial velocity with one chirp per loop is computed using Equation~\ref{eqn:velocity_calculation} (left). The symbol \(\lambda_c\) refers to the radar carrier wavelength. Another approach that can be used to obtain radial velocity is to measure a frequency difference between the transmitted and the received chirps, which is commonly known as the Doppler frequency shift \(f_d\). The radial velocity is then computed using Equation~\ref{eqn:velocity_calculation} (right). The symbol \(f_c\) refers to the radar carrier frequency. In case of \gls{fmcw} modulation, the carrier frequency and the wavelength refer to the starting frequency and the wavelength of a transmitted chirp~\cite{radar2017,essay70986}. Other modulation techniques can measure the phase difference, the Doppler shift, and the radial velocity by using the same techniques and associated equations. For example, \gls{p} modulation can measure a Doppler shift between a pair of transmitted and received pulses~\cite{ee_times_pulse_doppler}. The authors in~\cite{37} recognize that when using a bistatic radar configuration, in which the transmitter and the receiver are placed perpendicular to each other, Doppler shifts due to yaw, pitch, and roll movements of the head are more clearly distinguishable. For radars using an \gls{iq}-phase mixer, the phase difference \(\Delta\phi_{t, \gls{iq}} = \tan^{-1} (x_q / x_i)\). Symbols \(x_q, x_i\) denote the quadrature-phase and in-phase \gls{if} signals coming from the \gls{iq} mixer~\cite{28,42,129}.

\begin{align}
\label{eqn:velocity_calculation}
    v_r = \frac{\lambda_c\Delta\phi_t}{4\pi T_{con}} && v_r = \frac{f_d \cdot c}{2 \cdot f_c}
\end{align}

\begin{figure}[!h]
    \centering
    \includegraphics[width=\linewidth]{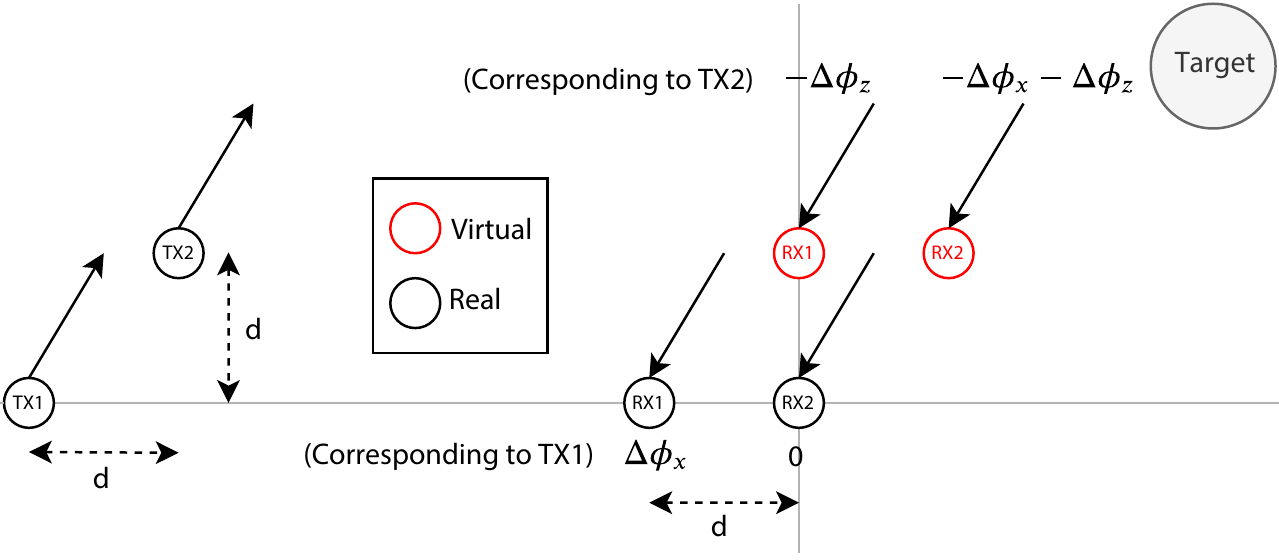}
    \caption{Uniform linear \gls{mimo} radar layout including two transmitters and receivers. Partly adopted from~\cite{ti_mimo_radar}.}
    \label{fig:mimo_radar_layout}
\end{figure}

The \gls{aoa} of the reflected or scattered signal in both the elevation and the azimuth dimensions (bearing angle) is calculated based on \gls{mimo} radar principles~\cite{ti_mimo_radar,153,4201876}. Rather than having one transmitter and one receiver, multiple transmitters and receivers are used. Almost all radars explained in the papers use a uniform linear layout. An example can be found in Figure~\ref{fig:mimo_radar_layout}. A minimum redundancy layout~\cite{25} and a uniform circular~\cite{137} layout are also explored. Real and virtual receivers, spaced according to a matrix structure on a horizontal surface, observe signal reflections coming from a target with a distance \(d\) apart from one another along the front and/or right direction. The signal reflection observed at a given receiver has to travel a certain distance further or shorter along a vertical and/or horizontal direction in relation to one specific receiver in the array to arrive at the receiver. This distance in the vertical and/or horizontal direction is unique to every receiver. The difference in distance between two neighboring receivers in the vertical or horizontal direction directly corresponds to a phase shift component denoted by \(\Delta\phi_{\in \{x, z\}} = [2\pi d \sin(\theta)] / \lambda_c\). A phase shift component overview is presented in Figure~\ref{fig:mimo_radar_layout}. The azimuth or elevation angle is calculated by using Equation~\ref{eqn:angle_calculation}~\cite{radar2017} with the phase shift component in the horizontal or vertical direction. Some approaches use multiple radars. For example,~\cite{85,113} use azimuth angles from two different radars, in which one was rotated 90 degrees counter-clockwise compared to the other. The rotation makes the azimuth angle correspond to the elevation angle. The data collection systems in~\cite{85,86,158} rotate the radar to obtain angle measurements.

\begin{align}
    \label{eqn:angle_calculation}
    \theta = \sin^{-1}\left(\frac{\lambda_c\Delta\phi_x}{2\pi d}\right) &&  \varphi = \sin^{-1}\left(\frac{\lambda_c\Delta\phi_z}{2\pi d}\right) 
\end{align}

A \gls{rn} radar transmits noise signals. \gls{bpsk} radars are normally used for communicating a bit stream. The bit value determines the phase offset \(\phi\) (0 or \(\pi\)) of the transmitted signal \(x_{tx}\). Both radar types use cross-correlation between the transmitted and received signals. This cross-correlation relates to a time delayed version (delayed by \(2R/c\)) of the transmitted signal to obtain an auto-correlation function. When plotted, this function shows a peak that corresponds to a target's range~\cite{112,166}. The noise radar in~\cite{112} measures a Doppler frequency shift based on a tone embedded into the noise signal.

\begin{table*}[ht]
    \centering
    \caption{Summary of data collection devices using active imaging approach. The modulation scheme, measured variable and scanning method abbreviations are explained throughout Section~\ref{sec:data_collection_imaging}.}
    \label{tab:data_collection_imaging_summary}
    \tiny\sffamily
    \setlength{\tabcolsep}{2pt}
    \begin{NiceTabular}{@{}lccccccc|M{8mm}M{6mm}M{12mm}M{5mm}|ccc|M{12mm}M{8.5mm}M{6mm}M{8mm}M{5mm}cM{6mm}M{10mm}M{10mm}@{}}[code-before =\rowcolors{3}{}{gray!12}]
    \noalign{\hrule height 0.8pt}
     & \Block{2-7}{\textbf{Carrier frequency (GHz)}} & & & & & & & \Block{2-4}{\textbf{Scanning method}} & & & & \Block{1-3}{\textbf{Modulation}} & & & \Block{2-9}{\textbf{Measured variables}} & & & & & & & & \TstrutTwo\\[-1.2ex]
     &   &   &   &   &   &   &   &   &   &   &   & \Block{1-3}{\textbf{scheme}}  &   &   &   &   &   &   &   &   &   &   &   \\[1.2ex]
     & 30-39 & 50-59 & 60-69 & 70-79 & 80-89 & 90-99 & 260-269 & Mechan- ical line array & Mechan- ical plane & Mechanical, single radar & Free space & P & AM & FMCW & Scattering signal power distribution & Scattering coefficient & Scatter- ing signal & Visibility function & \gls{if} signal & \(v\) & mean elevation & roughness property & Reflection intensity \\[2.7ex] \hline
     
     \cite{3}   & X &   &   &   &   &   &   & X &   &   &   &   &   & X &   & X &   &   &   &   &   &   &   \\ 
     \cite{19}  &   &   &   &   &   &   &   &   &   &   & X & X &   &   &   &   &   & X &   &   &   &   &   \\ 
     \cite{33}  &   &   &   &   &   & X &   &   & X &   &   &   &   &   &   &   &   & X &   &   &   &   &   \\ 
     \cite{38}  &   &   &   &   &   & X &   &   &   & X &   &   &   & X &   &   &   &   & X &   &   &   &   \\ 
     \cite{67}  &   &   &   & X &   &   &   &   &   &   & X & X &   &   &   &   &   &   & X &   &   &   &   \\ 
     \cite{68}  &   &   &   &   &   & X &   &   &   &   & X & X &   &   &   &   &   &   & X &   &   &   &   \\ 
     \cite{80}  & X &   &   &   &   & X &   &   &   &   & X & X &   &   &   &   &   &   &   & X & X & X &   \\ 
     \cite{95}  &   &   & X &   &   &   &   & X &   &   &   &   & X &   & X &   &   &   &   &   &   &   &   \\ 
     \cite{117} &   &   &   & X &   &   &   & X &   &   &   &   & X &   &   &   &   &   &   &   &   &   & X \\ 
     \cite{118} &   & X &   &   &   &   &   &   &   & X &   &   &   & X &   & X &   &   &   &   &   &   &   \\ 
     \cite{119} &   &   &   &   & X &   &   &   &   & X &   &   &   & X &   &   &   &   & X &   &   &   &   \\ 
     \cite{130} &   &   &   & X &   &   &   & X &   &   &   &   &   & X &   &   &   &   & X &   &   &   &   \\ 
     \cite{142} &   &   &   &   &   &   & X &   &   & X &   & X &   &   &   &   & X &   &   &   &   &   &   \\
     \cite{155} &   &   &   & X &   &   &   &   &   & X &   &   &   & X &   &   &   &   & X &   &   &   &   \\ \noalign{\hrule height 0.8pt}
    \end{NiceTabular}
\end{table*}

\acrlong{p} radars emit a powerful high gain signal pulse for a short time period. This period is called the pulse width. Afterwards, the radar waits for a given rest time to receive reflections before emitting another pulse. The range can be determined based on the round trip time between the emitted pulse and received reflection of this pulse in the rest time: \((c \cdot T_{rtt})/2\). In addition to \gls{mimo} radar principles, the azimuth and elevation angles can also be retrieved from the radar's own directivity compared to a baseline direction denoting north or the ground respectively~\cite{radar_basics}. The authors in~\cite{87} measure mean velocity based on the correlation between two pulses under an unambiguous pulse interval for a sloped terrain. Pulse radars normally work well in long distance applications~\cite{radar_basics}. One exception to this was found in~\cite{115}, in which breast cancer detection was performed successfully. The authors in~\cite{7} observe that pulse radar does not provide enough resolution for tasks requiring high resolution measurements such as gesture recognition.

\subsubsection{Active Imaging}
\label{sec:data_collection_imaging}
Active imaging is a specific type of radar-based approach that obtains measurement results in the form of map-like images~\cite{radar_basics}. The radar scans a given area. For every position, the radar emits a signal, measures the reflections returned, and based on a measured variable assigns, for example, a color or gray-scale value to the pixel corresponding to that position. Most active imaging research between 2002-2014 was focused on data collection devices for monitoring cracks in civil infrastructures, detecting concealed objects on the human body, and measuring hazardous landing terrain~\cite{80,87}. Review papers on active imaging used for detecting explosives and monitoring civil infrastructures include~\cite{79,116}. Active imaging recently received attention again in the context of map-like image deep learning and applying existing techniques using a cost effective method in commercial applications~\cite{33,67,68,39} for applications such as parking space monitoring and analyzing objects in enclosed packaging for food quality control and non-invasive fault detection. Table~\ref{tab:data_collection_imaging_summary} is an extension of previous reviews~\cite{79,116}, providing an overview of active imaging approaches used for data collection in millimeter wave sensing applications. In addition to the carrier frequency, the modulation scheme, and the measured variables, we consider the scanning method here as well.

The scanning method defines how the measurement system moves from one position to another to obtain, for example, a color or gray-scale value for every pixel in the map-like image. One such method is mechanical scanning where radar devices are mechanically moved in fixed directions using a rail system. The literature covers three types of mechanical scanning. In line array scanning, a 1-dimensional array of antenna's on a straight line scan in a single direction back and forth (i.e., up/down or right/left)~\cite{3,95,117,130}. In plane scanning, antennas arranged into a 2-dimensional plane scan the target from different distances by moving perpendicular to the plane~\cite{33}. Some papers employ scanning a target by moving a small radar system in two directions on a virtual plane~\cite{38,118,119,142,155}. In the so called free space scanning method, the forward motion of a radar connected to a moving device, such as a drone or planetary lander, is used to construct images. Sensing application pipelines using this type of scanning either use a large antenna array to cover big areas~\cite{80} or a limited number of ingrained low-cost radar device antennas~\cite{67,68}.

The free space scanning methods found in the literature are commonly referred to as \gls{sar}. A few millimeter wave application papers present application pipelines containing data collection devices that do not belong to active imaging by means of free space scanning and mention use of \gls{sar}~\cite{130} or \gls{sar} pre-processing methods~\cite{119,142,155}. We consider these data collection devices not part of \gls{sar} since \gls{sar} originates from \gls{slar} (a free space scanning method). \gls{sar} provides a solution to the impractically long antenna or use of extremely short wavelengths, resulting in severe atmospheric attenuation, required for sufficient azimuth resolution in \gls{slar} images. \gls{sar} essentially \textit{synthesizes} a very long antenna to obtain high resolution images~\cite{07110101}. More information regarding \gls{sar} can be found in~\cite{sar_intro,07110101}. Raw \gls{sar} images are severely out of focus since they do not  represent spatial information correctly yet and therefore need additional pre-processing which is elaborated on in Section~\ref{sec:signal_transform}.

\begin{table*}[ht]
    \centering
    \caption{Summary of data collection devices using spatial sweeping approach. The modulation scheme, measured variable and sweeping method abbreviations are explained throughout Section~\ref{sec:data_collection_sweeping}.} 
    \label{tab:data_collection_sweeping_summary}
    \tiny\sffamily
    \setlength{\tabcolsep}{5pt}
    \begin{NiceTabular}{@{}lccc|ccccc|cccccccc@{}}[code-before =\rowcolors{3}{}{gray!12}]
    \noalign{\hrule height 0.8pt}
     & \Block{1-3}{\textbf{Carrier frequency (GHz)}} & & & \Block{1-5}{\textbf{Sweeping method}} & & & & & \Block{1-8}{\textbf{Measured variables}} & & & & & & & \Tstrut\\[1.2ex]
     & 30-39 & 60-69 & 70-79 & Fixed & Rotational & Radio Tomography & V2I & Base Station Exchange & CIR & \acs{rss} & Phase & Reflection loss  & \(f_d\) across time & ToF & AoD & AoA \\[0.4ex] \hline
     
     \cite{1}   &   & X &   & X &   &   &   &   & X &   &   &   &   &   &   &   \\ 
     \cite{6}   &   & X &   & X &   &   &   &   &   & X & X &   &   &   &   &   \\ 
     \cite{12}  &   & X &   & X & X &   &   &   &   & X &   &   &   &   &   &   \\ 
     \cite{20}  &   & X &   &   & X &   &   &   &   & X &   &   &   &   &   &   \\ 
     \cite{22}  &   &   & X &   &   &   &   & X &   &   &   &   &   & X & X &   \\ 
     \cite{26}  &   & X &   & X & X &   &   &   &   & X &   & X &   &   &   &   \\ 
     \cite{27}  &   & X &   &   &   &   &   & X &   &   &   &   &   &   &   & X \\ 
     \cite{32}  &   & X &   &   &   &   & X &   &   &   &   &   & X &   &   &   \\ 
     \cite{66}  &   & X &   & X & X &   &   &   &   & X &   & X &   &   &   &   \\ 
     \cite{73}  &   & X &   &   & X &   &   &   &   & X &   &   &   &   &   &   \\ 
     \cite{101} &   & X &   &   &   &   &   & X &   &   &   &   &   & X &   &   \\ 
     \cite{107} &   & X &   &   &   & X &   &   &   & X &   &   &   &   &   &   \\ 
     \cite{131} & X &   &   &   &   &   & X &   &   &   &   &   &   & X & X & X \\ 
     \cite{135} &   & X &   &   &   &   &   & X &   & X &   &   &   &   &   &   \\ 
     \cite{141} & X &   &   &   &   &   &   & X &   & X &   &   &   &   &   & X \\
     \cite{143} &   & X &   &   & X &   &   &   & X &   &   &   &   &   &   &   \\ \noalign{\hrule height 0.8pt}
    \end{NiceTabular}
\end{table*}

The \gls{if} signal is introduced in Section~\ref{sec:data_collection_radar}. Radars using \gls{p} modulation obtain an \gls{if} signal through a concept called pulse compression~\cite{68,sar_intro}, during which the pulse frequency is modulated in a similar pattern as \gls{fmcw} modulation. Pulse compression allows transmission powers comparable to a pulse with a long pulse width while simultaneously retaining the range resolution which is only attainable with a short pulse width in the context of \gls{sar}. Low transmission power results in low receiver signal detectability and measurement precision~\cite{mahafza-1998}. The \gls{if} signal is either used directly in pre-processing~\cite{68} or its amplitude~\cite{67,68,119}, magnitude~\cite{38,155}, and/or phase values~\cite{38,67,68,119} are used to color the image or for further pre-processing. The authors in~\cite{130} measure a lap-joint position with an \gls{if} signal component. The component accounts for amplitude variation based on radiation characteristics and the target's shape and a phase variation based on the target's position at the center frequency.

Other less frequently measured variables include a visibility function and a scattering coefficient. Visibility is defined as a correlation value between two non-directive receivers. Both mechanical~\cite{33} and free space~\cite{19} approaches have been used for measuring these variables. The scattering coefficient can be considered as an energy ratio between transmitted and received energy~\cite{118}. When a signal scatters due to a rough surface, more energy will be observed by the receiver compared to when most energy is reflected away from the receiver due to a smooth reflection surface~\cite{sar_intro}.

\subsubsection{Spatial Sweeping}
\label{sec:data_collection_sweeping}
Spatial sweeping is an approach where signal response metrics measured in the context of the millimeter wave radio communication between sender and receiver systems are analyzed. Changes in the measured metrics are caused by the interaction between the target of interest to be sensed and the communication signals. Spatial sweeping refers to the observation that almost all collection systems in Table~\ref{tab:data_collection_sweeping_summary} have transmitter and/or receiver components that spatially move and rotate during communication. Movement and rotation naturally increase the system's sensing \gls{fov}. Table~\ref{tab:data_collection_sweeping_summary} gives an overview of spatial sweeping approaches used for data collection in millimeter wave sensing application. In addition to the carrier frequency and measured variables, we consider here the used sweeping method as well.

The sweeping method defines how senders and receivers move and rotate during communication with each other while at the same time being used for measurement procedures. Table~\ref{tab:data_collection_sweeping_summary} shows that several sweeping methods are quite generic while others are more specific. This is because certain systems are tested for specific real-life contexts while other systems are tested for determining feasibility of, for example, rotational and fixed communication for sensing. A variety of sweeping methods exist, such as rotational sweeping, fixed sweeping, and radio tomography. Most rotational sweeping methods rely on a transmitter rotating itself to transmit a signal at beam angles within a given area. During this time, receivers make a variable measurement at one angle of arrival if the angle is not in line with the current beam angle. This process is repeated sequentially for a number of angles of arrival to form a measurement matrix~\cite{12,26,66,73}. The authors in~\cite{20} use a robot that moves and rotates. The authors in~\cite{143} use a fixed position where both the transmitter and receiver are located. Measurements are taken at a given location and associated angles~\cite{20,143}. In fixed sweeping, position and direction of both transmitters and receivers are fixed. The authors in~\cite{1} put a transmitter and receiver in direct line-of-sight, while authors in~\cite{6} use a quasi-omni transmit antenna to cover a measurement area. Certain rotational sweeping methods fall back to fixed sweeping by rotating a transmitter and receivers towards a fixed angle to perform additional measurements~\cite{12,26,66} while the angle remains fixed. Radio tomography uses a grid network of transceivers to cover a measurement area~\cite{107}. \gls{v2i} communication relies on fast \gls{dsrc} between a stationary \gls{rsu} and \gls{obu} inside a moving vehicle~\cite{v2i_explained}. Base station exchange covers a measurement area via a set of moving receivers~\cite{22,27,135,141}.

\subsubsection{Gaps and Challenges}
From Tables~\ref{tab:data_collection_radar_summary},~\ref{tab:data_collection_imaging_summary} and~\ref{tab:data_collection_sweeping_summary}, a number of interesting observations can be made. Firstly, most data collection systems operate in the 60-79 GHz frequency band. This leaves ample room for performing research and data collection on the 30-59 GHz and 80-299 GHz bands. Secondly, most radar systems stick to variables that are delivered out of the box by commercialized systems. These include range, radial velocity and \glspl{aoa}. It is yet to be seen and investigated whether additional variables such as \gls{if} signal amplitude variation across time and \gls{rss} provide increased sensing accuracy to a wide variety of applications.

\begin{figure*}[ht]
    \centering
    \includegraphics[width=\textwidth]{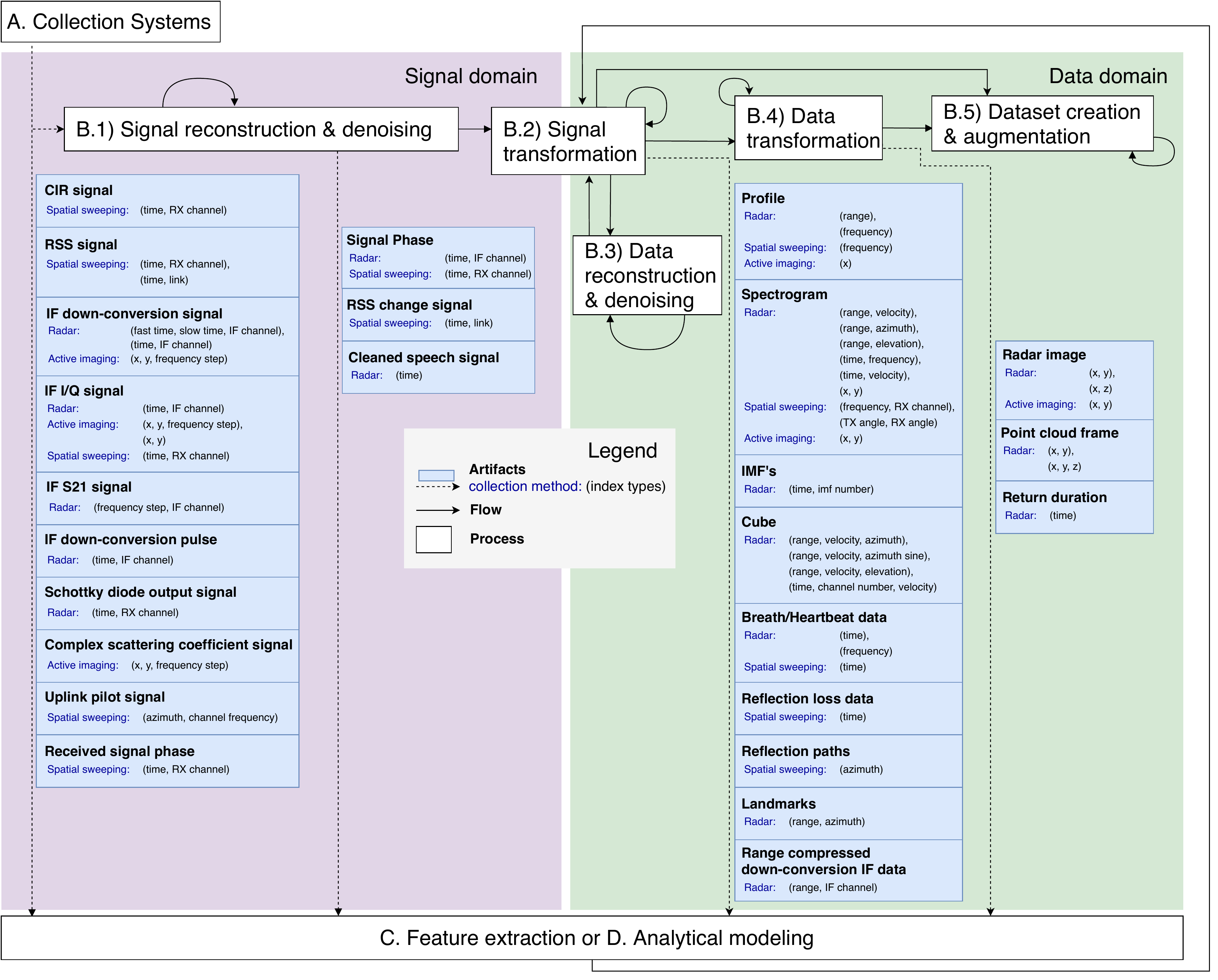}
    \caption{Abstract pre-processing pipeline diagram. It does not show spatial sweeping across time (see Section~\ref{sec:data_collection_sweeping}). It also omits artifact parts extracted with pre-processing and additional artifact features that are normally not extracted together with the artifact. Artifact part examples include profile, spectrogram, cube bins and radar image parts~\cite{61}. Feature examples include velocity~\cite{147,164} and intensity~\cite{147} with point cloud frames and vehicle location~\cite{67,68}, empty parking site~\cite{67,68}, obstacles~\cite{68}, length-width ratio~\cite{67}, barycenter location~\cite{67}, and intensity~\cite{113} with radar images.}
    \label{fig:pre-processing-pipeline-diagram}
\end{figure*}

\subsection{Pre-processing}
\label{sec:data_preprocessing}
Using the collection systems mentioned in Section~\ref{sec:data_collection}, discrete signal vectors in the time domain are sampled via an \gls{adc}. For the majority of sensing applications, these raw signal vectors cannot be used to infer information of interest to the sensing application. Therefore, raw signal vectors are passed through a pre-processing pipeline to extract data types that can be used to infer information. An abstract pipeline overview can be found in Figure~\ref{fig:pre-processing-pipeline-diagram}. This section reports on the pre-processing methods that are mentioned in the papers. Simulation with a ray tracing tool~\cite{22,27,101,141}, single-radar simulation~\cite{166}, multi-radar simulation~\cite{110}, imaging simulation~\cite{33}, and radar data modeling~\cite{19,49,53} are considered to be outside the scope of this review. The pipeline overview does not accurately represent the exact flow that every methodology follows and position in the pipeline where artifacts are extracted since parts are skipped, not reported, etc. It covers the predominant processes, flows, and places where artifacts are extracted. The pre-processing space has been divided into two domains: signal and data. The signal domain encompasses all pre-processing methods that are applied directly on the \gls{adc} sampled signal vectors. Once signal vectors are passed through a signal transformation method for the first time, the artifacts resulting from these methods are referred to as data and are subsequently processed in the data domain. Several pre-processing methods reported in this section are based on analytical modeling. Because these models are used to execute pre-processing tasks, i.e., not used to extract higher level information relevant to the millimeter wave application, they are considered to be pre-processing methods. Several application pipelines directly retrieve artifacts from the collection system hardware or use a pre-processing pipeline retrieved from a paper. The papers associated to these pipelines do not elaborate on the pre-processing methods that are executed prior to retrieving the artifacts. These pipelines and associated papers are summarized in Table~\ref{tab:abstracted-pre-processing-pipelines}. Several papers elaborate on additional pre-processing methods that are executed after retrieving the artifacts. These methods are explained in the respective pre-processing subsections.

\begin{table}[h]
    \centering
    \caption{Application pipelines that directly retrieve artifacts from the collection system hardware or use a pre-processing pipeline building block retrieved from a paper.}
    \label{tab:abstracted-pre-processing-pipelines}
    \tiny\sffamily
    \setlength{\tabcolsep}{8pt}
    \begin{NiceTabular}{@{}llll@{}}[code-before =\rowcolors{3}{}{gray!12}[respect-blocks]]
        \noalign{\hrule height 0.8pt}
        & \Block{2-1}{\textbf{Artifact(s)}} & \Block{2-1}{\textbf{(index types)}} & \textbf{Collection} \Tstrut\\
        &                                      &                                        & \textbf{method/Paper} \\[1.2ex] \hline
        \cite{2} & Spectrogram & (range, velocity) & Radar \\ 
        \cite{9} & Point cloud frame & (x, y, z) & Radar \\ 
        \cite{11}  & Spectrogram & (range, velocity) & Radar \\ 
        \Block{2-1}{\cite{13}}  & \gls{if} \gls{iq} signal, & (time, \gls{if} channel), & \Block{2-1}{Radar} \\
                                & Spectrogram & (range, velocity) & \\ 
        \cite{15}  & \(R, \theta, \text{and } v_r\) & Undefined & Radar \\ 
        \cite{16}  & Long. and lat. \(R \text{ and } v\) & Undefined & Radar \\ 
        \Block{2-1}{\cite{23}} & Profile & (range) & \Block{2-1}{Radar}\\
         & \(\theta\) & Undefined & \\ 
        \cite{31}  & Long. and lat. \(R \text{ and } v\) & Undefined & Radar \\ 
        \cite{32}  & Doppler shift data & (time, TX channel) & \cite{8756930} \\ 
        \cite{35}  & Spectrogram & (range, velocity) & Radar \\ 
        \cite{39}  & Radar image & (x, z) & Active imaging \\ 
        \cite{41}  & Radar image & (x, z) & Active imaging \\ 
        \cite{44}  & \(R, \varphi, \text{and } v_r\) & Undefined & Radar \\ 
        \cite{46}  & Radar image & (x, z) & Active imaging \\ 
        \cite{52}  & Radar image & (x, z) & Active imaging \\ 
        \cite{54}  & Heartbeat data & (time) & Radar \\ 
        \cite{61}  & Radar image & (x, z) & Active imaging \\ 
        \cite{63}  & Spectrogram & (x, z) & \cite{amano_nikkei_electronics} \\ 
        \cite{67}  & Radar image & (x, y) & Active imaging \\ 
        \cite{68}  & Radar image & (x, y) & Active imaging \\ 
        \Block{2-1}{\cite{72}} & Point cloud frame & (x, y) & \Block{2-1}{Radar} \\
                              & \(R, v_r, \theta, \text{and power}\) & Undefined & \\ 
        \cite{77}  & Spectrogram & (range, velocity) & \cite{8378688} \\ 
        \cite{81}  & Spectrogram & (range, azimuth) & Radar \\ 
        \cite{82}  & Profile & (range) & Radar \\ 
        \Block{2-1}{\cite{87}}  & Range compressed & \Block{2-1}{(range, \gls{if} channel)} & \Block{2-1}{Radar} \\ 
                                   & down-conversion \gls{if} data &                              &                       \\
        \cite{91}  & Point cloud frame & (x, y, z) & Radar \\ 
        \cite{93}  & Spectrogram & (range, velocity) & Radar \\ 
        \Block{2-1}{\cite{100}} & Point cloud frame & (x, y, z) & \Block{2-1}{Radar} \\
                                & \(v_r\) & Undefined & \\ 
        \cite{103} & Point cloud frame & (x, y, z) & Radar \\ 
        \cite{105} & \(R \text{ and } \theta\) & Undefined & Radar \\ 
        \cite{108} & Reflection intensity & Undefined & Radar \\ 
        \Block{2-1}{\cite{112}} & Profile & (range) & \Block{2-1}{\cite{gallagher_2013_noise_radar}} \\
                                & Spectrogram & (time, frequency) & \\ 
        \cite{115} & Radar trace & Undefined & Radar \\ 
        \cite{119} & Spectrogram & (x, y) & Active imaging \\ 
        \cite{121} & Heartbeat data & (time) & \cite{Mikhelson2013} \\ 
        \cite{128} & Spectrogram & (time, frequency) & \cite{DING2009259} \\ 
        \cite{131} & Positioning data & Undefined & Spatial sweeping \\ 
        \cite{133} & \(R \text{ and } \theta\) & Undefined & Radar \\ 
        \cite{135} & \gls{rss} signal & (time, RX channel) & Spatial sweeping \\ 
        \Block{2-1}{\cite{136}} & Profile & (range) & \Block{2-1}{Mono-pulse pipeline} \\
                                & Waterfall chart & (time, range) & \\
        \Block{2-1}{\cite{137}} & Dynamic (time-varying) & \Block{2-1}{Undefined} & \Block{2-1}{Radar} \\ 
                                   & signal components      &                           &                       \\
        \Block{2-1}{\cite{139}} & Point cloud frame & (x, y, z) & \Block{2-1}{\cite{nuscenes2019}} \\ 
                                & \(v_r\) & Undefined & \\
        \Block{2-1}{\cite{147}} & Point cloud frame & (x, y, z) & \Block{2-1}{\cite{radar2017}}\\ 
                                & \(v_r \text{ and intensity}\) & Undefined & \\ 
        \cite{148} & Spectrogram & (range, velocity) & Radar \\
        \cite{157} & Spectrogram & (x, y) & Radar \\ 
        \cite{159} & \(v_r \text{ and } \theta\) & Undefined & Radar \\ 
        \cite{160} & \(R, v_r, \text{and } \theta\) & Undefined & Radar \\ 
        \cite{161} & \(R, v_r, \theta, \text{and power amplitude}\) & Undefined & Radar \\
        \cite{162} & \(R, \theta, \text{and power amplitude}\) & Undefined & Radar \\ \noalign{\hrule height 0.8pt}
    \end{NiceTabular}
\end{table}

\subsubsection{Signal Reconstruction and Denoising}
\label{sec:signal_reconstruct_denoise}
Signal reconstruction and denoising resolve signal corruption and spectral leakage. Signal corruption refers to observing a sampled signal that is drastically different from its theoretical definition. Changes are caused by superimposed noise from unwanted stationary and systematic reflections from a target and nearby objects~\cite{6,26,28,57,59,73,107,115,126,128,129,137}, high frequency static noise (noise that is concentrated around, and remains in, a high frequency range)~\cite{26,73,134}, phase wrapping around a certain value that causes signal jumps~\cite{28,42,57}, DC offset~\cite{26,73,134}, hardware noise~\cite{29,126,128,129}, harmonic noise~\cite{126,129}, and channel noise~\cite{129}. Spectral leakage refers to non-zero values that show up in a signal's frequency profile at frequencies other than the frequency components actually present in the signal after signal transformation. Signal reconstruction and denoising resolve signal corruption and spectral leakage differently.

\begin{table*}[ht]
    \centering
    \caption{Summary of signal reconstruction \& denoising pre-processing methods deployed in millimeter wave sensing pipelines. The denoising abbreviations are explained throughout Section~\ref{sec:signal_reconstruct_denoise}.}
    \label{tab:signal_reconstruct_denoise}
    \tiny\sffamily
    \setlength{\tabcolsep}{4pt}
    \begin{NiceTabular}{@{}lM{10mm}M{10mm}|M{10mm}M{15mm}M{10mm}M{10mm}M{10mm}M{10mm}cM{10mm}M{15mm}M{17mm}@{}}[code-before =\rowcolors{3}{}{gray!12}]
        \noalign{\hrule height 0.8pt}
        & \Block{1-2}{\textbf{Reconstruction}} & & \Block{1-10}{\textbf{Denoising}} & & & & & & & & & \TstrutTwo\\[1.2ex]
        & Phase regeneration & Phase unwrapping & Mean centering & Moving average filter & Highpass filter & Lowpass filter & Line fitting & Bandpass filter & DDBR & Windowing & Wavelet Packet Noise Reduction & Manual background subtraction \\[1.7ex] \hline

        \cite{6}   & X &   &   &   &   &   &   &   & X &   &   &   \\ 
        \cite{21}  &   &   &   &   &   &   &   &   &   & X &   &   \\ 
        \cite{26}  &   &   &   & X &   &   &   & X &   &   &   &   \\ 
        \cite{28}  &   & X & X &   & X & X &   & X &   &   &   &   \\ 
        \cite{29}  &   & X &   &   &   &   &   & X &   &   &   &   \\ 
        \cite{34}  &   &   &   &   &   &   &   &   &   & X &   &   \\ 
        \cite{37}  &   &   &   &   &   &   &   &   &   & X &   &   \\ 
        \cite{42}  &   & X &   &   &   &   &   &   &   &   &   &   \\ 
        \cite{57}  &   &   &   &   &   & X & X &   &   &   &   &   \\ 
        \cite{59}  &   &   &   &   & X &   &   &   &   &   &   &   \\ 
        \cite{66}  &   &   &   & X &   &   &   & X &   &   &   &   \\ 
        \cite{73}  &   &   &   &   &   &   &   & X &   &   &   & X \\ 
        \cite{85}  &   &   &   &   &   &   &   &   &   & X &   &   \\ 
        \cite{86}  &   &   &   &   &   &   &   &   &   & X &   &   \\ 
        \cite{102} &   &   &   &   &   &   &   &   &   & X &   &   \\ 
        \cite{104} &   &   &   &   &   &   &   &   &   & X &   &   \\ 
        \cite{107} &   &   &   &   &   &   &   &   &   &   &   & X \\ 
        \cite{115} &   &   &   &   &   &   &   &   &   &   &   & X \\ 
        \cite{118} &   &   &   &   &   &   &   &   &   & X &   &   \\ 
        \cite{126} &   &   &   &   &   &   &   &   &   &   & X &   \\ 
        \cite{128} &   &   &   &   &   &   &   &   &   &   &   & X \\ 
        \cite{132} &   &   &   &   &   &   &   &   &   & X &   &   \\ 
        \cite{134} &   &   &   &   &   &   &   & X &   & X &   &   \\ 
        \cite{137} &   &   &   &   &   &   &   &   &   &   &   & X \\ 
        \cite{138} &   &   &   &   &   &   &   &   &   & X &   &   \\ 
        \cite{140} &   &   &   &   &   &   &   &   &   & X &   &   \\ 
        \cite{150} &   &   &   &   &   &   &   &   &   & X &   &   \\
        \cite{153} &   &   &   &   &   &   &   &   &   & X &   &   \\ \noalign{\hrule height 0.8pt}
    \end{NiceTabular}
\end{table*}

Signal reconstruction artificially creates or alters a signal with help from a construction algorithm. Phase regeneration exploits the fact that the received signal phase exhibits a periodic pattern, even though it is not linear with respect to the target's moving distance. Phase regeneration creates an artificial phase shift signal by counting phase shifts of the measured signal, creating a predefined phase shift on every count, and linking two neighboring counts with a linear increasing or decreasing trend~\cite{6}. Phase unwrapping is not explained in~\cite{28,29,42}.

Signal denoising encompasses all methods that remove signal components from the sampled signal. A signal component, in the context of Fourier series, refers to a wave with a less complex waveform in a set of waves with a less complex waveform that reconstitute the sampled signal if summed together. Mean centering~\cite{28} is done by subtracting the geometric mean vector of a complex output trace from each sampled signal point. A moving average filter averages a number of the sampled signal points to produce a new signal point~\cite{Smith_dsp_guide_ch15}. The high, low, and bandpass filters attenuate signal components from the sampled signal that have a certain frequency. The highpass filter attenuates everything below a certain cutoff frequency, the lowpass filter attenuates everything above a certain cutoff frequency, and the bandpass filter attentuates everything outside of a predefined frequency band. Line fitting estimates a demodulated \gls{if} signal phase by fitting a straight line to the phase of a lowpass filtered demodulated \gls{if} signal version and obtaining the y-intercept of the demodulated \gls{if} signal with the straight line~\cite{57}. \gls{ddbr} takes two differentials. The first differential is computed with signal points at time \(t-1 \text{ and } t\) and the second differential with signal points at time \(t \text{ and } t+1\). \gls{ddbr} then adds the differentials to obtain a background canceled signal point. A standard signal window function used in combination with a signal transformation operation such as \gls{stft} can be understood as a brick-wall filter~\cite{37}. Cosine-sum and adjustable windows such as Hann~\cite{85}, Hamming~\cite{118,150}, Dolph-Chebyshev~\cite{86}, and Kaiser-Bessel~\cite{150} windows reduce spectral leakage in a signal transformation output~\cite{ni_instrument_fundamentals}. Wavelet packet noise reduction~\cite{126} removes signal noise by utilizing a signal enhancement scheme on a fast wavelet transformed signal representation and afterwards transforming the signal back to the time domain. The main idea behind manual background subtraction is to first obtain a sampled signal from an environment without any target and afterwards subtracting the sampled signal from a sampled signal obtained when a target is present in the environment~\cite{73,115}. Manual background subtraction in~\cite{107,128} averages multiple signals obtained from an environment without a target. Exponential averaging can also be used to obtain a sampled signal from an environment without a target~\cite{137}.

\subsubsection{Signal Transformation}
\label{sec:signal_transform}
Signal transformation is the most important pre-processing operation of almost every millimeter wave sensing application pipeline. Without signal transformation, many of the artifacts depicted in Figure~\ref{fig:pre-processing-pipeline-diagram} would not exist. Signal transformation refers to transforming a time domain signal, through use of mathematical operations, into a data structure that represents certain aspects of the signal in either the time or frequency domain.

\begin{table*}[ht]
    \centering
    \caption{Summary of signal transformation pre-processing methods deployed in millimeter wave sensing pipelines. The abbreviations are explained throughout Section~\ref{sec:signal_transform}.}
    \label{tab:signal_transform}
    \tiny\sffamily
    \setlength{\tabcolsep}{2pt}
    \begin{NiceTabular}{@{}lM{10mm}M{10mm}ccM{10mm}M{10mm}M{10mm}M{10mm}|cccccM{10mm}M{10mm}M{10mm}cM{10mm}@{}}[code-before =\rowcolors{3}{}{gray!12}]
        \noalign{\hrule height 0.8pt}
        & \Block{1-8}{\textbf{Time domain}} & & & & & & & & \Block{1-10}{\textbf{Frequency domain}} & & & & & & & & & \TstrutTwo\\[1.2ex]
        & Reflection path extraction & Impulse response convolution & EMD & Synchronization & Surface normal calculation & Peak detection & Reflection loss extraction & Range compression & FFT & STFT & Beamformer & MLE & MSER & Visual saliency detection & Landmark extraction & Pulse integration & POSP & Frequency compression \\[2.7ex] \hline

        \cite{1}   &   &   &   &   &   &   &   &   & X &   &   &   &   &   &   &   &   &   \\ 
        \cite{3}   &   &   &   &   &   &   &   &   & X &   &   &   &   &   &   &   & X &   \\ 
        \cite{4}   &   &   &   &   &   &   &   &   & X &   &   &   &   &   &   &   &   &   \\ 
        \cite{5}   &   &   &   &   &   &   &   &   & X &   &   &   &   &   &   &   &   &   \\ 
        \cite{7}   &   &   &   &   &   &   &   &   & X &   &   &   &   &   &   &   &   &   \\ 
        \cite{8}   &   &   &   &   &   &   &   &   & X &   &   &   &   &   &   &   &   &   \\ 
        \cite{14}  &   &   &   &   &   &   &   &   & X &   & X &   &   &   &   &   &   &   \\ 
        \cite{17}  &   &   &   &   &   &   &   &   & X &   &   &   &   &   &   &   &   &   \\ 
        \cite{20}  & X &   &   &   &   &   &   &   &   &   &   &   &   &   &   &   &   &   \\ 
        \cite{21}  &   &   &   &   &   &   &   &   & X &   & X &   &   &   &   &   &   &   \\ 
        \cite{24}  &   &   &   &   &   &   &   &   & X &   &   &   &   &   &   &   &   &   \\ 
        \cite{25}  &   &   &   &   &   &   &   &   & X &   & X &   &   &   &   &   &   &   \\ 
        \cite{26}  &   &   &   &   &   & X & X &   & X &   &   &   &   &   &   &   &   &   \\ 
        \cite{29}  &   &   &   &   &   &   &   &   &   & X &   &   &   &   &   &   &   &   \\ 
        \cite{34}  &   &   &   &   &   &   &   &   & X &   & X &   &   &   &   &   &   &   \\ 
        \cite{37}  &   &   &   &   &   &   &   &   &   & X &   &   &   &   &   &   &   &   \\ 
        \cite{42}  &   &   &   &   &   &   &   &   &   & X &   &   &   &   &   &   &   &   \\ 
        \cite{43}  &   &   &   &   &   &   &   &   & X &   &   &   &   &   &   &   &   &   \\ 
        \cite{50}  &   &   &   &   &   & X &   &   &   &   &   &   &   &   &   &   &   &   \\ 
        \cite{57}  &   &   &   &   &   &   &   &   & X &   &   &   &   &   &   &   &   &   \\ 
        \cite{62}  &   &   &   &   &   &   &   &   & X &   &   &   &   &   &   &   &   &   \\ 
        \cite{65}  &   & X &   &   &   &   &   &   &   &   &   &   &   &   &   &   &   &   \\ 
        \cite{66}  &   &   &   &   &   & X & X &   & X &   &   &   &   &   &   &   &   &   \\ 
        \cite{67}  &   &   &   &   &   &   &   &   &   &   &   &   &   & X &   &   &   &   \\ 
        \cite{68}  &   &   &   &   &   &   &   &   &   &   &   &   & X & X &   &   &   &   \\ 
        \cite{69}  &   &   &   &   &   &   &   &   & X &   &   &   &   &   &   &   &   &   \\ 
        \cite{70}  &   &   &   &   &   &   &   &   & X & X &   &   &   &   &   &   &   &   \\ 
        \cite{71}  &   &   &   &   &   &   &   &   & X &   &   &   &   &   &   &   &   &   \\ 
        \cite{74}  &   &   &   &   &   &   &   &   & X & X &   &   &   &   &   &   &   &   \\ 
        \cite{75}  &   &   & X &   &   &   &   &   &   &   &   &   &   &   &   &   &   &   \\ 
        \cite{76}  &   &   &   &   &   &   &   &   & X &   &   &   &   &   &   &   &   &   \\ 
        \cite{77}  &   &   &   &   &   &   &   &   & X &   &   &   &   &   &   &   &   &   \\ 
        \cite{78}  &   &   &   &   &   &   &   &   & X &   &   &   &   &   &   &   &   &   \\ 
        \cite{80}  &   &   &   &   &   &   &   & X &   &   &   &   &   &   &   &   &   &   \\ 
        \cite{85}  &   &   &   &   &   &   &   &   & X &   &   &   &   &   &   &   &   &   \\ 
        \cite{86}  &   &   &   &   &   &   &   &   & X &   &   &   &   &   &   &   &   &   \\ 
        \cite{88}  &   &   &   &   &   &   &   &   & X &   &   &   &   &   &   & X &   & X \\ 
        \cite{92}  &   &   &   &   &   &   &   &   & X &   &   &   &   &   &   &   &   &   \\ 
        \cite{94}  &   &   &   &   &   &   &   &   & X & X &   &   &   &   &   &   &   &   \\ 
        \cite{96}  &   &   &   &   &   &   &   &   & X & X &   &   &   &   &   &   &   &   \\ 
        \cite{99}  &   &   &   &   &   &   &   &   & X &   &   &   &   &   &   &   &   &   \\ 
        \cite{102} &   &   &   &   &   &   &   &   & X &   & X &   &   &   &   &   &   &   \\ 
        \cite{104} &   &   &   &   &   &   &   &   & X &   & X &   &   &   &   &   &   &   \\ 
        \cite{106} &   &   &   &   &   &   &   &   & X &   &   &   &   &   &   &   &   &   \\ 
        \cite{112} &   &   &   &   &   &   &   &   &   & X &   &   &   &   &   &   &   &   \\ 
        \cite{113} &   &   &   &   &   &   &   &   & X &   &   &   &   &   &   &   &   &   \\ 
        \cite{117} &   &   &   & X &   &   &   &   &   &   &   &   &   &   &   &   &   &   \\ 
        \cite{118} &   &   &   &   &   &   &   &   & X &   &   &   &   &   &   &   &   &   \\ 
        \cite{119} &   &   &   &   & X &   &   &   &   &   &   &   &   &   &   &   &   &   \\ 
        \cite{122} &   &   & X &   &   &   &   &   & X &   &   &   &   &   &   & X &   &   \\ 
        \cite{123} &   &   &   &   &   &   &   &   & X &   &   &   &   &   &   &   &   &   \\ 
        \cite{124} &   &   &   &   &   &   &   &   & X &   &   &   &   &   &   &   &   &   \\ 
        \cite{125} &   &   &   &   &   &   &   &   & X &   &   &   &   &   &   &   &   &   \\ 
        \cite{129} &   &   &   &   &   &   &   &   &   & X &   &   &   &   &   &   &   &   \\ 
        \cite{130} &   &   &   &   &   &   &   &   & X &   &   &   &   &   &   &   &   &   \\ 
        \cite{132} &   &   &   &   &   &   &   &   & X &   & X &   &   &   &   &   &   &   \\ 
        \cite{134} &   &   &   &   &   &   &   &   & X & X &   &   &   &   &   &   &   &   \\ 
        \cite{137} &   &   &   &   &   &   &   &   &   &   &   & X &   &   &   &   &   &   \\ 
        \cite{138} &   &   &   &   &   &   &   &   & X & X &   &   &   &   &   &   &   &   \\ 
        \cite{140} &   &   &   &   &   &   &   &   & X &   &   &   &   &   &   &   &   &   \\ 
        \cite{142} &   &   &   &   &   &   &   &   & X &   &   &   &   &   &   &   &   &   \\ 
        \cite{144} &   &   &   &   &   &   &   &   & X &   &   &   &   &   &   &   &   &   \\ 
        \cite{146} &   &   &   &   &   &   &   &   & X &   &   &   &   &   &   &   &   &   \\ 
        \cite{149} &   &   &   &   &   &   &   &   & X &   &   &   &   &   &   &   &   &   \\ 
        \cite{150} &   &   &   &   &   &   &   &   & X &   &   &   &   &   &   &   &   &   \\ 
        \cite{151} &   &   &   &   &   &   &   &   & X &   &   &   &   &   &   & X &   &   \\ 
        \cite{152} &   &   &   &   &   &   &   &   & X & X &   &   &   &   &   & X &   &   \\ 
        \cite{153} &   &   &   &   &   &   &   &   & X &   &   &   &   &   &   & X &   &   \\ 
        \cite{155} &   &   &   &   &   &   &   &   & X &   &   &   &   &   &   & X &   &   \\ 
        \cite{157} &   &   &   &   &   &   &   &   &   &   & X &   &   &   &   &   &   &   \\ 
        \cite{158} &   &   &   &   &   &   &   &   &   &   &   &   &   &   & X &   &   &   \\
        \cite{164} &   &   &   &   &   &   &   &   & X &   &   &   &   &   &   &   &   &   \\ \noalign{\hrule height 0.8pt}
    \end{NiceTabular}
\end{table*}

The most widely used signal transformation operation is the \gls{fft}. In Section~\ref{sec:data_collection_radar}, it was explained that an \gls{if} signal operating at a certain beat frequency can be obtained when a single object or person is standing in a \gls{fmcw} radar's \gls{fov}. When multiple objects or persons are standing in the radar's \gls{fov}, the sampled \gls{if} signal will exhibit a more complex waveform. The \gls{if} signal is a constitution of multiple \gls{if} components because multiple reflections are obtained by the receiver channels. These components will have different beat frequencies if objects or persons are standing at different distances. When running the sampled \gls{if} signal through a \gls{fft} operation across the fast time dimension, the resulting data structure will be a one dimensional set of complex numbers, i.e., a profile. The explanation ignores the complex conjugates in the set of complex numbers~\cite{radar2017}. In-phase and quadrature-phase \gls{if} signals can be combined into a single complex signal~\cite{78} and used in the \gls{fft} operation. The magnitude of several complex numbers will show a peak compared to the other complex numbers. The associated frequency index of these complex numbers directly corresponds to a range value as indicated in Equation~\ref{eqn:range_calculation}. Range resolution, i.e., minimum required distance separation between persons or objects, has a direct relation with the bandwidth size across which chirp frequency is modulated~\cite{radar2017}.

Radial velocity, according to Section~\ref{sec:data_collection_radar}, can be determined with multiple chirps emitted across a loop. The profiles originating from the chirps are stacked on top of each other to form a two dimensional data structure made up of complex numbers. The newly introduced dimension is called slow time (a.k.a. \gls{cpi}). Across slow time, the complex number's phase, due to the motion of persons or objects, rotates at a constant rate. The phase of every complex number is a sum of phases corresponding to different objects or persons~\cite{radar2017}. After \gls{fft} operations have been performed across slow time at every frequency index, a spectrogram (a.k.a. heatmap) indexed by frequency and phase difference is obtained. The phase difference corresponds to a radial velocity value according to Equation~\ref{eqn:velocity_calculation}. Magnitude peaks in the spectrogram correspond to objects or persons with a certain range travelling at a given radial velocity. Radial velocity resolution, i.e., minimum required radial velocity separation between persons or objects, has a direct relation with the number of chirps per loop~\cite{radar2017}.

\gls{aoa}, as explained in Section~\ref{sec:data_collection_radar}, is determined through \gls{mimo} radar principles. The explanation assumes that spectrograms originating from the real and virtual receiver channels in Figure~\ref{fig:mimo_radar_layout} are available. Transmit techniques required to obtain spectrograms from real and virtual receivers channels are elaborated on in Section~\ref{sec:data_reconstruction_denoising}. Azimuth and elevation angles are computed in separate sets of \gls{fft} operations. To determine the azimuth angle, the spectrograms originating from real receiver channels RX1 and RX2 are stacked to form a three dimensional data structure. The newly introduced dimension is called \gls{if} channel. Across \gls{if} channel, like with determining radial velocity, the complex number's phase rotates. This is due to extra distance which has to be traversed by a signal reflection. After performing \gls{fft} operations for every element in the spectrogram across \gls{if} channel, a data cube is obtained. Magnitude peaks in the cube correspond to objects or persons with a certain range, radial velocity, and phase shift in the horizontal direction. The phase shift corresponds to azimuth angle according to Equation~\ref{eqn:angle_calculation} (left)~\cite{radar2017}. When looking at Figure~\ref{fig:mimo_radar_layout}, one may assume that the same \gls{fft} operations are valid for elevation angle on spectrograms from real receiver channel RX2 and virtual receiver channel RX1. In bigger uniform linear layouts, more signal transformation steps are required to retrieve the elevation angle. A phase shift isolation technique for a bigger uniform linear layout is presented in~\cite{mmwave_sdk_for_explanation}. The main idea is that after several \gls{fft} operation steps the elevation phase difference can be isolated by phasor (another way to represent a complex number) multiplication. \gls{aoa} resolution, i.e., minimum required angle separation between persons or objects, has a direct relation with the number of transmitter and receiver channels used in the \gls{mimo} radar layout~\cite{radar2017}.

Other uses of the \gls{fft} operation include retrieving \gls{csi} frequency domain profile~\cite{1}, conversion of complex scattering coefficient signal to frequency domain~\cite{3}, aid in applying bandpass filters to isolate signal components~\cite{26,66,123}, conversion of phase data to frequency domain~\cite{34,124}, conversion of \gls{rss} variance to frequency domain~\cite{66}, inverse \gls{fft} on \gls{if} S21 or complex scattering coefficient signal to retrieve range profile~\cite{69,118}, \gls{fft} on \gls{if} signal coming from \gls{cw} radar to determine breath/heartbeat data~\cite{78,125}, dimension reduction~\cite{130}, filtered frequency index dominant frequency determination~\cite{134}, \gls{fft} and inverse \gls{fft} for image reconstruction~\cite{142,155}, and retrieving range profile from uplink pilot signal~\cite{142}. The main difference between \gls{fft} and \gls{stft} is that \gls{stft} separates \gls{fft} operations in chunks across time. \gls{stft} has been used to retrieve phase information across time~\cite{29,42}, for creation of time and frequency spectrograms~\cite{37,129,134}, and to create range and micro-velocity or micro-Doppler spectrograms from a stack of range profiles~\cite{70,74,94,96,152} or directly from a signal containing Doppler information~\cite{112,138}. Micro-velocity spectrograms are used to analyze fine-grained velocity features. These features can be attributed to for example arm swinging while a person is walking or presence of something in context of strong background noise. These fine-grained features help to compute new types of information such as drone blade rotation~\cite{134} or allow analytical models to better predict information of interest. Raja et al.~\cite{37} denote that \gls{stft} is better suited for identification of movement direction, time of occurrence, and duration from a RSS signal.

The beamformer is an estimation technique that can be used to calculate a one dimensional set of complex numbers associated to a user-defined \gls{aoa}. It is used as an alternative to the \gls{aoa} \gls{fft}. The sensing application pipelines either use the \gls{mvdr}, i.e., the Capon beamformer~\cite{14,21,102,104,157} or do not specify the beamformer type~\cite{25,34,132}. The Capon beamformer computes the set with a steering vector and covariance matrix. The covariance matrix is a model describing spatial and frequency domain \gls{if} signal characteristics. The steering vector represents the phase rotation due to extra distance, which has to be traversed by a signal reflection at receiver channels in an array in phasor notation. Reflection path extraction is used to iteratively extract \gls{rss} signal components from the signal at the receiver associated to a path between reflector and receiver~\cite{20}. \gls{emd} decomposes a signal into several \glspl{imf}~\cite{75}. The output from both methods stays in the time domain. Matsuguma and Kajiwara~\cite{65} retrieve a range profile based on a convolution in time between an \gls{if} pulse, i.e., single chirp, and impulse echo response. Oka et al.~\cite{117} synchronize a Schottky diode output signal with an encoder distance signal to obtain an intensity image. Pawliczek et al.~\cite{119} perform surface normal calculation with a phase data spectrogram to improve defect visualization. Peak detection is used to count breath/heartrate from a filtered time-series \gls{rss} signal~\cite{26,66} or to determine if a Schottky diode output signal indicates presence of metallic or non-metallic objects. Reflection loss extraction isolates reflection loss from the total \gls{rss} loss with a set of equations~\cite{26,66}. Häfner et al.~\cite{137} measure azimuth angle and time of arrival by means of a maximum-likelihood based parameter estimator. Landmark extraction~\cite{158}, based on a profile measured at a given azimuth angle, returns a set of landmarks, i.e., set of range and azimuth value tuples. Landmark extraction performs a set of filtering operations, after which magnitude values are scaled according to the probability that the magnitude value indicates a landmark. Continuous peaks at certain ranges indicate a landmark. Pulse integration uses the integration operation with the purpose of improving signal-to-noise ratio~\cite{151,155}, help discover movement in a spectrogram~\cite{122}, and to deduce a spectrogram representing micro-doppler signatures~\cite{152,153}. Pulse integration types include incoherent integration~\cite{122}, coherent integration~\cite{151}, spectrogram integration across range~\cite{152,153}, and a wideband signal filter operation~\cite{155}. POSP is used in~\cite{3} to calculate an integral. Further details, including the long version of POSP, are not present. Frequency compression is not explained in~\cite{88}.

\begin{table*}[ht]
    \centering
    \caption{Summary of data reconstruction \& denoising pre-processing methods deployed in millimeter wave sensing pipelines. The denoising abbreviations are explained throughout Section~\ref{sec:data_reconstruction_denoising}.}
    \label{tab:data_reconstruction_denoising}
    \tiny\sffamily
    \setlength{\tabcolsep}{2pt}
    \begin{NiceTabular}{@{}lM{13mm}|ccM{7mm}cM{7mm}M{7mm}M{10mm}M{6mm}cM{7mm}M{6mm}M{6mm}M{8mm}M{7mm}M{10mm}M{7mm}M{7mm}M{7mm}M{7mm}M{7mm}@{}}[code-before = \rowcolors{3}{}{gray!12}]
        \noalign{\hrule height 0.8pt}
        & \Block{1-1}{\textbf{Reconstructon}} & \Block{1-20}{\textbf{Denoising}} & & & & & & & & & & & & & & & & & & & \TstrutTwo\\[1.2ex]
        & Phase unwrapping & CFAR & MTI & Thresh- olding & LLSE & Function fitting & Gaussian model  & Manual background subtraction & Otsu algo- rithm & SVD & Feed-forward NN & Band- pass filter & High- pass filter & Gaussian filter & Moving average filter & Offset version recombination & Doppler compen- sation & Deboun- cing & Mirror- ing & Distance correction & Window- ing \\[2.7ex] \hline

        \cite{4}   &   &   &   &   &   &   &   & X &   &   &   &   &   &   &   &   &   &   &   &   &   \\
        \cite{7}   &   & X &   &   &   &   &   &   &   &   &   &   &   &   &   &   &   &   &   &   &   \\
        \cite{14}  &   &   &   &   &   &   &   & X &   &   &   &   &   &   &   &   &   &   &   &   &   \\ 
        \cite{17}  &   & X & X &   &   &   &   &   &   &   &   &   &   &   &   &   &   &   &   &   &   \\ 
        \cite{21}  &   & X &   &   &   &   &   &   &   &   &   &   &   &   &   &   &   &   &   &   &   \\ 
        \cite{24}  & X &   &   &   & X &   &   &   &   &   &   &   &   &   &   &   &   &   &   &   &   \\ 
        \cite{25}  &   &   & X &   &   &   &   &   &   &   &   &   &   &   &   &   &   &   &   &   &   \\ 
        \cite{26}  &   &   &   &   &   &   &   &   &   &   &   & X &   &   &   &   &   &   &   &   &   \\ 
        \cite{34}  & X & X &   & X &   &   &   &   &   &   &   & X &   &   &   &   &   &   &   &   & X \\ 
        \cite{44}  &   &   &   & X &   &   &   &   &   &   &   &   &   &   &   &   &   &   &   &   &   \\ 
        \cite{57}  & X &   &   &   &   &   &   &   &   &   &   &   &   &   &   &   &   &   &   &   &   \\ 
        \cite{61}  &   &   &   &   &   &   &   &   & X &   &   &   &   &   &   &   &   &   &   &   &   \\ 
        \cite{62}  &   & X &   &   &   &   &   &   &   &   &   &   &   &   &   &   &   &   &   &   &   \\ 
        \cite{66}  &   &   &   &   &   &   &   &   &   &   &   & X &   &   &   &   &   &   &   &   &   \\ 
        \cite{74}  &   &   &   &   &   &   &   &   &   &   &   &   & X &   &   &   &   &   &   &   &   \\ 
        \cite{77}  &   &   &   &   &   &   &   & X &   &   &   & X &   &   &   &   &   & X &   &   &   \\ 
        \cite{81}  &   & X &   &   &   &   &   &   &   &   &   &   &   &   &   &   &   &   &   &   &   \\ 
        \cite{82}  &   & X &   &   &   &   &   &   &   &   &   &   &   &   &   &   &   &   &   &   &   \\ 
        \cite{85}  &   &   &   & X &   &   &   & X &   &   &   &   &   &   &   &   &   &   &   &   &   \\ 
        \cite{86}  &   &   &   &   &   &   &   & X &   &   &   &   &   &   &   &   &   &   &   &   &   \\ 
        \cite{94}  &   & X &   &   &   &   &   &   &   &   &   &   & X &   &   &   &   &   &   &   &   \\ 
        \cite{96}  &   & X &   &   &   &   &   &   &   &   &   &   &   &   &   &   &   &   &   &   &   \\ 
        \cite{100} &   &   &   &   &   &   &   &   &   &   & X &   &   &   &   &   &   &   &   &   &   \\ 
        \cite{102} &   & X &   &   &   &   &   & X &   &   &   &   &   &   &   &   &   &   &   &   &   \\ 
        \cite{104} &   &   &   &   &   &   &   & X &   &   &   &   &   &   &   &   &   &   &   &   &   \\ 
        \cite{112} &   &   &   &   &   &   &   & X &   &   &   &   & X &   &   &   &   &   &   & X &   \\ 
        \cite{118} &   &   &   & X &   &   &   &   &   & X &   &   &   &   &   &   &   &   &   &   &   \\ 
        \cite{119} & X &   &   &   &   & X &   &   &   &   &   &   &   &   &   &   &   &   &   &   &   \\ 
        \cite{122} & X &   &   &   &   &   &   &   &   &   &   & X &   &   &   &   &   &   &   &   &   \\ 
        \cite{123} &   &   &   &   &   &   &   &   &   &   &   & X &   &   &   &   &   &   &   &   &   \\ 
        \cite{124} & X &   &   &   &   &   &   &   &   &   &   & X &   &   &   &   &   &   &   &   &   \\ 
        \cite{129} &   &   &   &   &   &   &   &   &   &   &   &   &   &   &   & X &   &   &   &   &   \\ 
        \cite{132} &   & X &   &   &   &   &   &   &   &   &   &   &   &   &   &   & X &   &   &   & X \\ 
        \cite{133} &   &   &   &   &   &   &   & X &   &   &   &   &   &   &   &   &   &   &   &   &   \\ 
        \cite{134} &   &   &   &   &   &   &   & X &   &   &   &   &   &   &   &   &   &   &   &   &   \\ 
        \cite{140} &   & X &   &   &   &   &   &   &   &   &   &   &   &   &   &   &   &   &   &   &   \\ 
        \cite{142} &   &   &   &   &   &   &   &   &   &   &   &   &   &   &   &   &   &   & X &   &   \\ 
        \cite{146} &   & X &   &   &   &   &   & X &   &   &   &   & X & X &   &   &   &   &   &   &   \\ 
        \cite{147} &   & X &   &   &   &   &   &   &   &   &   &   &   &   &   &   &   &   &   &   &   \\ 
        \cite{148} &   &   &   &   &   &   & X &   &   &   &   &   &   &   &   &   &   &   &   &   &   \\ 
        \cite{149} &   & X &   &   &   &   &   &   &   &   &   &   &   &   &   &   &   &   &   &   &   \\ 
        \cite{150} &   &   &   & X &   &   &   &   &   &   &   &   &   &   & X &   &   &   &   &   &   \\ 
        \cite{151} &   &   &   & X &   &   &   &   &   &   &   &   &   &   & X &   &   &   &   &   &   \\ 
        \cite{153} &   & X &   &   &   &   &   &   &   &   &   &   &   &   &   &   &   &   &   &   &   \\ 
        \cite{157} &   & X &   &   &   &   &   &   &   &   &   &   &   &   &   &   &   &   &   &   &   \\ 
        \cite{164} &   &   &   &   &   &   &   & X &   &   &   &   &   &   &   &   &   &   &   &   &   \\ \noalign{\hrule height 0.8pt}
    \end{NiceTabular}
\end{table*}

After using pulse compression on a given set of positions, while scanning an area by means of free space scanning with a moving device, a two dimensional structure of complex values, representing combined \gls{if} \gls{iq} signals, also known as a \gls{sar} image is obtained. When using the magnitude of these complex values to create a grayscale image, the resulting raw \gls{sar} grayscale image is severely out of focus because it does not represent spatial information correctly yet. In most \gls{sar} pre-processing pipelines, range and azimuth reference functions are generated and convoluted with the \gls{sar} image in sequence to increase the focus. As a result, the \gls{sar} image correctly represents spatial information. These techniques are known as range~\cite{80} and azimuth compression~\cite{sar_intro,07110101}. More information can be found in~\cite{sar_intro,07110101}. \gls{mser} exploits the fact that reflections originating from the metal structures of vehicles, in contrast to reflections from the road surface, show up as bright stable area's in a \gls{sar} grayscale image. Candidate regions are selected that stay below a grayscale area variation rate~\cite{68}. Visual saliency detection, after a set of pre-processing steps performed on a \gls{sar} image, returns a binary image in which white area's indicate presence of objects~\cite{68} or parked vehicles~\cite{67}.

\subsubsection{Data Reconstruction and Denoising}
\label{sec:data_reconstruction_denoising}
Several sensing application pipelines deploy reconstruction and denoising techniques in the data domain. The techniques try to resolve data corruption, which refers to observing values in a data structure that are drastically different from the expected values in a certain context. In addition to filtering corruption causes already mentioned in Section~\ref{sec:signal_reconstruct_denoise}, data reconstruction and denoising also filter Doppler components caused by transmitter time multiplexing, remove redundant data parts, and filter data transients. Data transients are caused by persons that become stationary after walking into a room~\cite{77}.

Under a sampling and signal change assumption, one dimensional phase unwrapping is performed by~\cite{24}. The phases of complex numbers in the slow time direction are analyzed from a two dimensional data structure prior to radial velocity \gls{fft} operations. A wrapped phase change greater than \(\pi\) in a pair of consecutive complex numbers indicates that the phase of the second complex number should be corrected by adding or subtracting \(2\pi\) through means of phasor multiplication. One dimensional phase unwrapping can also be solved with a null range measurement obtained at the first chirp of the first frame. Subsequent chirp and frame processing is combined with multiplying the \gls{if} signal with the null range measurement~\cite{57}. Two dimensional phase unwrapping through a path following algorithm is considered in the sensing application pipeline explained in~\cite{119}.

\gls{cfar} is an adaptive thresholding technique that is used to extract, i.e., reduce a spectrogram containing magnitude values to, spectrogram parts, based on a sliding window, that indicate presence of targets against data corruption present in the spectrogram. Several \gls{cfar} types have been considered. One dimensional types include cell averaging~\cite{96,132,140,146,147}, cell averaging smallest of~\cite{21,62,102,132}, cell averaging greatest of~\cite{62}, clutter map~\cite{81,82}, and ordered statistics~\cite{94}. A two dimensional custom \gls{cfar} type is considered in~\cite{157}. The type indicates how the threshold is determined. More information on \gls{cfar} types can be found in~\cite{radar_basics,6042197,4104231}.

Background subtraction is performed by removing point cloud data in a frame with zero Doppler velocity~\cite{4,133} and Cartesian coordinates that fall outside certain boundaries~\cite{4}, removing measured profile and spectrogram (part) averages from profiles and spectrograms~\cite{14,77,112,164}, deleting a profile belonging to an empty \gls{fov} from target measurement profiles~\cite{86}, and statically removing the 0 Hz component from maximum Doppler frequency data across time~\cite{134}.

\gls{mti} is not explained in~\cite{17,25}. The papers suggest that it is a more general term used to indicate that denoising is performed. Static thresholding is used to extract spectrogram parts from which phase variation over time~\cite{34} or magnitude~\cite{150} exceeds a threshold or create binary radar images~\cite{44,118} (unknown purpose~\cite{44} or to make material faults visible~\cite{118}). \gls{llse} is used to estimate a shift caused by DC offset in complex numbers' imaginary and real parts of a 2D data structure, prior to radial velocity \gls{fft} (used to estimate chest vibration). The shift is used afterwards to adjust the complex number's imaginary and real parts~\cite{24}. 2D quadratic function fitting is used~\cite{119} to fit a function to phase data, which is later subtracted from the phase data to eliminate phase curvature. A per-pixel Gaussian model is used to subtract noise coming from unwanted background reflections from spectrograms~\cite{148}. The Otsu algorithm is an automatic threshold selection method supported by image segmentation. It is used for noise reduction in radar images~\cite{61}. Noise reduction with help from \gls{svd} can be applied to extract desired spectrogram parts indicating presence of target reflection~\cite{118}. A feed-forward neural network can be used to filter ghost targets (i.e., superimposed noise from unwanted background and target reflections) in automotive radar sensing~\cite{100}. The bandpass and highpass filters reported in this section are applied in the frequency domain. Profiles are multiplied with a transfer function that represents the bandpass or highpass filter's frequency response~\cite{freq_domain_filters}. A 2D Gaussian filter is convolved over a spectrogram for extra noise reduction in addition to background subtraction~\cite{146}. Moving average filtering is performed in time by first subtracting an empty background spectrogram from a measurement spectrogram. Afterwards, the background spectrogram is updated with the measurement spectrogram~\cite{150}. In offset version recombination, a noisy set of complex numbers created with \gls{stft} is offset with a simple addition of a frequency dependent anti-symmetric function~\cite{129}. Spectrograms originating from real and virtual receiver channels cannot be retrieved simultaneously at real receivers. Radar collection systems employ multiplexing strategies to retrieve the spectrograms~\cite{ti_mimo_radar}. When time multiplexing is used, spectrograms from real and virtual receivers will experience an unwanted Doppler induced phase shift. This shift is compensated for by means of phasor multiplication~\cite{mmwave_sdk_for_explanation}. To mitigate analytical model performance degradation caused by data transients, a debouncing logic can be implemented~\cite{77}. Radar images acquired through imaging do not properly represent the geometry of an environment due to multipath propagation. Image correction can be applied by means of mirroring techniques~\cite{142}. Distance correction~\cite{112} scales correlation profiles while taking into consideration that sample strength falls off according to a certain pattern. To remove motion corrupted segments from heartbeat data, the heartbeat data is segmented and segments are removed based on the outcome of a thresholding procedure~\cite{34}. Two dimensional windowing prior to performing a \gls{fft} operation to determine velocity~\cite{132} can be considered a brick-wall filter~\cite{37}.

\subsubsection{Data Transformation}
\label{sec:data_transform}
Data transformation refers to using mathematical operations on data structures that either cause them to change into new, higher-level data types or cause data structure aspects, e.g., value range or axis range considered, to change. There is a wide variety in goals that sensing application pipelines try to achieve through use of data transformation. For example, value range changes can be used to unify the value range of different variables. This will omit bias towards variables that have a bigger value range compared to other variables during analytical model training~\cite{scaling_normalization_raschka}. Data type changes allow higher-level information to be extracted from lower-level information. For example, position information in Cartesian coordinates can be extracted from lower-level range and \gls{aoa} information~\cite{164}, voxels created from position information encapsulate higher-level body shape information~\cite{8}, etc.

\begin{table*}[ht]
    \centering
    \caption{Summary of data transformation pre-processing methods deployed in millimeter wave sensing pipelines. The abbreviations are explained throughout Section~\ref{sec:data_transform}.}
    \label{tab:data_transform}
    \tiny\sffamily
    \setlength{\tabcolsep}{1pt}
    \begin{NiceTabular}{@{}lM{7mm}M{6mm}M{7mm}M{8mm}M{9mm}M{6mm}M{6mm}M{7mm}cM{8mm}M{6mm}M{6mm}cM{6mm}M{6mm}|M{7mm}M{8mm}M{10mm}M{12mm}cM{8mm}M{6mm}M{8mm}@{}}[code-before =\rowcolors{3}{}{gray!12}]
        \noalign{\hrule height 0.8pt}
        & \Block{1-15}{\textbf{Artifact (type) change}} & & & & & & & & & & & & & & & \Block{1-8}{\textbf{No Artifact (type) change}} & & & & & & & \TstrutTwo\\[1.2ex]
        & Peak detection & Auto- correl- ation & Thresh- olding & Spectral analysis & Dimension reduction & Voxeli- zation & Binari- zation & Math. morph- ology & LRMF & Edge detection & Coord- inate trans- form & Point trans- form & FWT & Grey- scaling & Ray casting & Normali- zation & ROI extraction & Edge region discovery & Windowing/ Segmentation & MRC & Manual smoothing & Conv- olution mask & Down- sampling \\[3.9ex] \hline

        \cite{1}   &   &   &   &   &   &   &   &   &   &   &   &   &   &   &   &   &   &   &   &   &   &   & X \\ 
        \cite{3}   &   &   &   &   &   &   &   &   &   &   & X &   &   &   &   &   &   &   &   &   &   &   &   \\ 
        \cite{4}   &   &   &   &   &   &   &   &   &   &   & X &   &   &   &   &   &   &   &   &   & X &   &   \\ 
        \cite{5}   &   &   &   &   &   &   &   &   &   &   & X &   &   &   &   &   &   &   &   &   &   &   &   \\ 
        \cite{7}   &   &   &   &   &   &   &   &   &   &   &   &   &   &   &   & X & X &   &   &   &   &   &   \\ 
        \cite{8}   &   &   &   &   &   & X &   &   &   &   & X &   &   &   &   &   &   &   & X &   &   &   &   \\ 
        \cite{9}   &   &   &   &   &   & X &   &   &   &   &   &   &   &   &   &   &   &   & X &   &   &   &   \\ 
        \cite{17}  &   &   &   &   &   &   &   &   &   &   & X &   &   &   &   &   &   &   &   &   &   &   &   \\ 
        \cite{21}  & X &   &   &   &   &   &   &   &   &   & X &   &   &   &   &   &   &   &   &   &   &   &   \\ 
        \cite{24}  &   &   &   &   &   &   &   &   &   &   &   &   &   &   &   &   & X &   &   &   &   &   &   \\ 
        \cite{25}  &   &   &   &   &   &   &   &   &   &   &   &   &   &   &   &   &   &   &   &   &   &   &   \\ 
        \cite{34}  & X &   &   &   &   &   &   &   &   &   &   &   &   &   &   &   &   &   &   &   &   &   &   \\ 
        \cite{41}  &   &   &   &   &   &   & X & X &   &   &   &   & X &   &   &   &   & X & X &   &   &   &   \\ 
        \cite{61}  &   &   &   &   &   &   &   & X & X &   &   &   &   &   &   &   &   &   &   &   &   &   &   \\ 
        \cite{62}  &   &   &   &   &   &   &   &   &   &   & X &   &   &   &   &   &   &   &   &   &   &   &   \\ 
        \cite{67}  &   &   &   &   &   &   &   & X &   &   &   &   &   &   &   &   &   &   &   &   &   &   &   \\ 
        \cite{72}  &   &   &   &   &   &   &   &   &   &   &   &   &   &   &   &   &   &   & X &   &   &   &   \\ 
        \cite{77}  &   &   &   &   &   &   &   &   &   &   &   &   &   &   &   &   &   &   &   &   &   &   &   \\ 
        \cite{80}  &   &   & X &   &   &   &   &   &   &   &   &   &   &   &   &   &   &   &   &   &   &   &   \\ 
        \cite{88}  &   &   & X &   &   &   &   &   &   &   &   &   &   &   &   &   &   &   &   &   &   &   &   \\ 
        \cite{94}  &   &   &   &   &   &   &   &   &   &   &   &   &   &   &   & X &   &   &   &   &   &   &   \\ 
        \cite{96}  &   &   &   &   &   &   &   &   &   &   & X &   &   &   &   &   &   &   &   &   &   &   &   \\ 
        \cite{102} &   &   &   &   &   &   &   &   &   &   & X &   &   &   &   &   &   &   &   &   &   &   &   \\ 
        \cite{106} &   &   &   &   &   &   &   &   &   &   &   &   &   &   &   &   &   &   &   &   &   &   &   \\ 
        \cite{113} &   &   &   &   &   &   &   &   &   &   & X & X &   &   &   & X &   &   &   &   &   &   &   \\ 
        \cite{118} &   &   &   &   &   &   &   &   &   &   &   &   &   &   &   &   &   &   &   &   &   & X &   \\ 
        \cite{122} &   &   &   & X &   &   &   &   &   &   &   &   &   &   &   &   &   &   &   &   &   &   &   \\ 
        \cite{123} & X &   &   &   &   &   &   &   &   &   &   &   &   &   &   &   & X &   &   &   &   &   &   \\ 
        \cite{124} & X & X &   &   &   &   &   &   &   &   &   &   &   &   &   &   &   &   &   &   &   &   &   \\ 
        \cite{130} &   &   &   &   & X &   &   &   &   &   &   &   &   &   &   &   &   &   &   &   &   &   &   \\ 
        \cite{132} &   &   &   &   &   &   &   &   &   &   & X &   &   &   &   &   &   &   &   &   &   &   &   \\ 
        \cite{134} &   &   &   &   &   &   &   &   &   & X &   &   &   & X &   &   &   &   &   &   & X &   &   \\ 
        \cite{139} &   &   &   &   &   &   &   &   &   &   &   & X &   &   &   &   &   &   &   &   &   &   &   \\ 
        \cite{140} &   &   &   &   &   &   &   &   &   &   & X &   &   &   &   &   &   &   &   &   &   &   &   \\ 
        \cite{142} &   &   &   &   &   &   &   &   &   &   &   &   &   &   & X &   &   &   &   &   &   &   &   \\ 
        \cite{146} &   &   &   &   &   &   &   &   &   &   &   &   &   &   &   & X &   &   & X &   &   &   &   \\ 
        \cite{148} &   &   &   &   &   &   &   &   &   &   &   &   &   &   &   & X &   &   &   &   &   &   &   \\ 
        \cite{149} &   &   &   &   &   &   &   &   &   &   &   &   &   &   &   &   &   &   &   & X &   &   &   \\ 
        \cite{151} &   &   &   &   &   &   &   &   &   &   &   &   &   &   &   &   & X &   &   &   &   &   &   \\
        \cite{164} &   &   &   &   &   & X &   &   &   &   & X &   &   &   &   & X &   &   &   &   &   &   &   \\ \noalign{\hrule height 0.8pt}
    \end{NiceTabular}
\end{table*}

Peak detection is used to retrieve velocity information from a range, azimuth, and velocity cube~\cite{21}, retrieve breath/heartbeat data by means of average inter-peak distance in a phase data sequence in time~\cite{34}, and determine magnitude peak position in a range profile~\cite{123}. Thresholding is either applied on a range-compressed down-conversion \gls{if} signal to determine vertical range between the collection system and the ground~\cite{80} or used to determine if space debris is detected or not~\cite{88}. Spectral analysis is a general term used to describe the fact that \glspl{imf} are converted
to a frequency domain representation to determine if they can be attributed to heartbeat or breathing data based on a given frequency range~\cite{122}. Dimension reduction is performed by transforming a two dimensional structure of complex values into a structure containing magnitude values and determining the position that corresponds to the dominant reflection of a lap joint. Data evaluation is performed at the position that yields a one dimensional structure~\cite{130}. Voxelization can be thought of as creating a three dimensional grid. The grid consists of voxels, i.e., custom values (or sets of custom values). Values found in the sensing application pipelines include number of points present~\cite{9} and accumulated velocity information of every point~\cite{164} in bounded coordinate regions of a point cloud frame. Voxelization is used to turn a variable number of point information rows~\cite{164}, associated to a point cloud frame, into a data structure with fixed dimensions~\cite{9,164}. Binarization refers to the creation of a binary image from a radar image. Binary images contain black and white values assigned through use of a threshold~\cite{41}. Mathematical morphology is a term used to describe a set of matrix operations that are used to transform binary images~\cite{41,67}. These matrix operations are called erosion, dilation, opening and closing, and involve an input image and a structuring element. The structuring element is similar to a kernel in the context of two dimensional convolution. Opening has been used to transform a binary saliency map into connected regions~\cite{67}. Dilation has been used to improve image visualization~\cite{61}. \gls{lrmf} is used to factorize a radar image into two: a foreground, i.e., concealed object, and background to detect concealed objects. Factorization is performed with a patch-based, i.e., segment-based, Gaussian mixture model~\cite{61}. Laplacian of Gaussian and Canny edge detection are used on a gray scale converted spectrogram containing magnitude values to discover Doppler frequency information~\cite{134}. Once range and \gls{aoa} values retrieved from spectrograms or cubes are known, the values can be transformed from range and \gls{aoa} information, through polar or spherical to two or three dimensional Cartesian coordinate conversion equations, to position information in Cartesian coordinates~\cite{4,5,8,17,21,62,96,102,113,132,140,164}. The new data structure is referred to as either a point cloud frame or point scan. The data structures differ in context in which measurements were conducted. Stationary radar collection systems are used to retrieve point cloud frames while moving radar collection systems on a robot or vehicle are used to retrieve point scans. The active imaging application pipeline presented in~\cite{3} transforms coordinates by means of integration. Point transformation is concerned with converting a point cloud frame based on fixed coordinate ranges into a radar image. To every pixel, a RGB value is assigned. The R value corresponds to a x-coordinate, the G value to a y/z-coordinate (separate xy/xz images), and B value to a magnitude value. Coordinates that contain no points are assigned a black RGB value~\cite{113}. The application pipeline in~\cite{139} uses a similar approach in which points converted to camera coordinates are used to compute RGB values. \gls{fwt} is used to transform a radar image into the wavelet domain with the intent to denoise it in the wavelet domain~\cite{41}. Greyscaling refers to converting a spectrogram consisting of magnitude values to a grayscale image~\cite{134}. Once uplink pilot signal \gls{aoa} and time of arrival, and a radar image consisting of reflective shapes are known, casting a ray, i.e., millimeter wave, at the estimated \gls{aoa} from a base station will eventually lead to a mobile user's range with respect to the base station under the assumption of reflection at a specular angle~\cite{142} with a computation on time of arrival.

Normalization refers to mapping real value ranges of several data structures. Spectrograms with magnitude values are normalized to unit scale~\cite{7}, between 0 and 1 and centered around the mean value~\cite{94}, by dividing every magnitude value with the sum of all magnitude values in the spectrogram~\cite{146} or by scaling magnitudes logarithmically and performing a max-min truncation~\cite{148}. RGB values assigned to a radar image with help from a point cloud frame with fixed coordinate ranges are based on normalized range, \gls{aoa}, and magnitude information~\cite{113}. Point cloud frame coordinates are max-min normalized based on fixed coordinate ranges~\cite{164}. \gls{roi} selection significantly reduces search dimensionality over a spectrogram
during feature extraction computations~\cite{7}. To isolate periodic chest movement, among all spectrogram columns corresponding to range in an acceptable \gls{fov}, the range column with the maximum average magnitude value is selected~\cite{24}. With background information on test person distance, a frequency domain \gls{roi} can be selected~\cite{123}. After thresholding, the closest detected range profile part in range is selected for each radar in case of multiple detected target peaks~\cite{151}. Through use of a pixel condition argument, person edges are discovered in image segments if a pixel value in the condition argument is greater than a certain threshold. The edge values are set to zero afterwards~\cite{41}. Windows containing a time dependent sequence of voxel grids are created and used to form a dataset~\cite{8,9}. An image segmentation procedure is presented in~\cite{41}. Dynamic sized temporal input data can be formatted by using the Markov Frame method~\cite{72}. Windowing can also be employed to track target position across spectrograms~\cite{146}. \gls{mrc} refers to the creation of mean range and velocity and range and azimuth spectrograms from a range, velocity, and azimuth cube containing magnitude values by computing a weighted mean~\cite{149}. Discontinuous range, azimuth sine, and velocity sequence values are smoothed. At sequence window ends, a discontinuous point is set to the value of its nearest neighbor. When eligible values are at both sides of the discontinuous point, the discontinuous point becomes the average of these values~\cite{4}. Frequency index smoothing is used in the pipeline explained in~\cite{134}. A convolution mask can be applied to an active imaging spectrogram to make a concealed tile image and the crack within it more visible~\cite{118}. Downsampling is used to convert a \gls{csi} frequency profile into a smaller feature vector~\cite{1}.

\subsubsection{Dataset Creation and Augmentation}
\label{sec:dataset_creation_augmentation}
After datasets, consisting of data samples, have been generated by both the signal and data transformation processes, several sensing application pipelines augment these datasets with artificially created data samples. In case of profiles, spectrograms, and cubes, it is assumed that real value structures consisting of magnitude values are generated. Reasons for data augmentation include improving analytical modeling performance on unseen data during inference~\cite{35,150,151,164} and to increase the dataset size to a size required during model training to achieve good analytical model performance~\cite{46,164} without the need for extra sampling. For each original spectrogram in the dataset or for an average spectrogram per class, artificial spectrograms can be generated by sampling pixel values according to a normal distribution~\cite{35,150}. Radar images can be fused with image segments at a random location~\cite{46}. Radar images and point cloud frames have also been rotated, scaled, and content inside the images and point cloud frames has been translated~\cite{151,164}. To prepare datasets for analytical model training in several sensing application pipelines, data samples are labeled with class labels and part of the dataset is designated for model training while other parts are designated for model validation and testing.

\begin{table}[!h]
    \centering
    \caption{Summary of dataset augmentation \& creation pre-processing methods deployed in millimeter wave sensing pipelines.}
    \label{tab:dataset_creation_augmentation}
    \tiny\sffamily
    \setlength{\tabcolsep}{4pt}
    \begin{NiceTabular}{@{}lM{15mm}M{15mm}M{20mm}|cc@{}}[code-before =\rowcolors{3}{}{gray!12}]
        \noalign{\hrule height 0.8pt}
        & \Block{1-3}{\textbf{Augmentation}} & & & \Block{1-2}{\textbf{Creation}} & \Tstrut\\[1.2ex]
        & Normal distribution & Object at random location fusion & Rotation, scaling, skew, and translation & Labeling & Splitting \\[1.7ex] \hline

        \cite{9}   &   &   &   &   & X \\ 
        \cite{14}  &   &   &   &   & X \\ 
        \cite{17}  &   &   &   & X &   \\ 
        \cite{35}  & X &   &   &   & X \\ 
        \cite{37}  &   &   &   & X & X \\ 
        \cite{39}  &   &   &   &   & X \\ 
        \cite{44}  &   &   &   & X &   \\ 
        \cite{46}  &   & X &   &   & X \\ 
        \cite{52}  &   &   &   & X & X \\ 
        \cite{74}  &   &   &   &   & X \\ 
        \cite{75}  &   &   &   &   & X \\ 
        \cite{82}  &   &   &   &   & X \\ 
        \cite{94}  &   &   &   &   & X \\ 
        \cite{95}  &   &   &   &   & X \\ 
        \cite{96}  &   &   &   & X & X \\ 
        \cite{113} &   &   &   &   & X \\ 
        \cite{146} &   &   &   &   & X \\ 
        \cite{148} &   &   &   & X & X \\ 
        \cite{149} &   &   &   &   & X \\ 
        \cite{150} & X &   &   &   & X \\ 
        \cite{151} &   &   & X & X & X \\ 
        \cite{152} &   &   &   &   & X \\ 
        \cite{153} &   &   &   &   & X \\ 
        \cite{163} &   &   &   &   & X \\ 
        \cite{164} &   &   & X &   &   \\ 
        \cite{165} &   &   &   & X & X \\ \noalign{\hrule height 0.8pt}
    \end{NiceTabular}
\end{table}

\subsubsection{Gaps and Challenges}
During the analysis of pre-processing methods, it was discovered that there is a lot of variety in the definition of what is considered to be pre-processing and way that pre-processing pipelines are constructed. We consider signal processing methods and typical data science pre-processing methods to be pre-processing methods. Other papers consider signal processing methods to be part of the signal collection system and only consider typical data science pre-processing methods to be pre-processing methods. Sensing application pipelines perform the processes mentioned in Figure~\ref{fig:pre-processing-pipeline-diagram} in different orders, skip processes, revert back to pre-processing after feature extraction or analytical modeling, etc.

When looking at Tables~\ref{tab:signal_reconstruct_denoise},~\ref{tab:signal_transform},~\ref{tab:data_reconstruction_denoising},~\ref{tab:data_transform}, and~\ref{tab:dataset_creation_augmentation}, there are a few observations that raise interest. Many of the table columns contain a few crosses while a large portion of the crosses is located in just a few columns. There are two explanations for this. Many sensing application pipelines rely on pre-processing methods that are well known. In signal transformation, many sensing application pipelines rely on \gls{fft} and \gls{stft} for signal conversion. However, estimation techniques such as Capon beamforming, Bartlett beamforming, \gls{music}, \gls{samv} beamforming, \gls{mle}, etc. exist that can be used in the pre-processing pipeline too. In~\cite{63,96}, test results have been reported for the \gls{music} algorithm. Many of the pre-processing methods are sensing application pipeline specific. For example, landmark extraction~\cite{158} works with a specific radar collection system that collects a signal at specific azimuth angles, Doppler compensation~\cite{132} is only used when transmit time multiplexing is used for measuring \gls{aoa} information, etc.

Another observation is that many table rows contain multiple crosses. This indicates use of multiple reconstruction, denoising, and transformation pre-processing methods. In~\cite{77}, multiple pre-processing pipeline paths are used in parallel to obtain different artifacts and artifact types that all use their own denoising and transformation methods. In~\cite{34}, there are two consecutive pre-processing stages. In the first stage, phase data is obtained in a pre-processing pipeline where \gls{cfar} is used to denoise spectrogram data. In the second stage, phase data is bandpass filtered before being used to extract breath/heartbeat data. Pre-processing methods are also used consecutively. In~\cite{6}, it is explained that signal reconstruction is specifically used to omit limitations encountered with denoising of the received signal phase through \gls{ddbr}.

\subsection{Feature Extraction}
\label{sec:feature_extraction}
The end result of the pre-processing phase in the application pipeline is a set of raw data samples in the form of, for example, sequence windows~\cite{1,163,165}, (voxelized) point cloud frames~\cite{8,9,164}, spectrograms~\cite{11,71,96}, radar images~\cite{19,113,139}, etc. This raw form of data, however, is not always used directly by analytical models. Degradation in modeling performance is sometimes caused by noise and redundancy in raw data samples. Extraction of relevant and informative data features from raw data samples is performed to compensate for this~\cite{StorcheusRK15}. Reducing the raw dataset to a limited number of features makes it easier to visualize for better understanding and gaining knowledge about the process that led to the generated raw data samples~\cite{feature_extraction_intro}.

The feature extraction methods used by millimeter wave sensing applications take in the raw dataset and map the entire dataset to a new feature space~\cite{StorcheusRK15}. The feature extraction phase is, however, not a required step. Recent deep learning techniques~\cite{9,46,95,113,115,146,147} and several modeling algorithms~\cite{6,22,23} perform well on raw data samples and, therefore, do not require the feature extraction phase. The border between feature extraction and analytical modeling has become blurry in recent papers on millimeter wave sensing applications. Several deep learning models contain layers designated for feature extraction~\cite{1,8,9,36,94,148,149,153,164}. Certain positioning and environment mapping algorithms resemble feature extraction methodologies~\cite{20,22}. Transfer learning~\cite{39,52,139} is not included in this section since the methodologies transfer analytical model parameters to another analytical model rather than extracted data features. Feature analysis~\cite{4,72,93,152} was only addressed by a small minority of papers and is therefore excluded from this paper as well.

Table~\ref{tab:feature_extraction_summary} presents our analysis of feature extraction approaches used in millimeter wave sensing applications. In what follows, we explain main feature extraction methods used in millimeter wave sensing applications. 

\begin{table}[!h]
    \centering
    \caption{Summary of feature extraction methods deployed in millimeter wave sensing pipelines. The automatic mapping abbreviations are explained throughout Section~\ref{sec:feature_extraction}.} \label{tab:feature_extraction_summary}
    \tiny\sffamily
    \setlength{\tabcolsep}{1pt}
    \begin{NiceTabular}{@{}lM{10mm}M{10mm}|M{10mm}M{11mm}M{8mm}cM{14mm}c@{}}[code-before =\rowcolors{3}{}{gray!12}]
        \noalign{\hrule height 0.8pt}
         & \Block{1-2}{\textbf{Manual mapping}} & & \Block{1-6}{\textbf{Automatic mapping}} & & & & & \Tstrut\\[1.2ex]
         & Time dependent & Frequency dependent & Clustering & A-priori estimation & Meta learning & t-SNE & Representation learning & PCA \\[1.7ex] \hline
         
          \cite{2}   & X & X &   &   &   &   &   &   \\ 
          \cite{4}   & X & X &   &   &   &   &   &   \\ 
          \cite{6}   & X &   &   &   &   &   &   &   \\ 
          \cite{7}   & X & X &   &   &   &   &   &   \\ 
          \cite{8}   &   &   & X &   &   &   &   &   \\ 
          \cite{9}   &   &   &   &   &   &   &   & X \\ 
          \cite{12}  & X &   &   &   &   &   &   &   \\ 
          \cite{13}  & X & X &   &   &   &   &   &   \\ 
          \cite{14}  &   &   &   &   &   &   &   & X \\ 
          \cite{19}  &   &   &   &   &   &   & X &   \\ 
          \cite{21}  &   &   &   & X &   &   &   &   \\ 
          \cite{24}  &   & X &   &   &   &   &   &   \\ 
          \cite{27}  & X &   &   &   &   &   &   &   \\ 
          \cite{29}  &   & X &   &   &   &   &   &   \\ 
          \cite{32}  & X & X &   &   &   &   &   &   \\ 
          \cite{34}  &   & X & X &   &   &   &   &   \\ 
          \cite{37}  & X & X &   &   &   &   &   & X \\ 
          \cite{61}  & X &   &   &   &   &   &   &   \\ 
          \cite{63}  &   & X &   &   &   &   &   &   \\ 
          \cite{65}  & X &   &   &   &   &   &   &   \\ 
          \cite{66}  & X & X &   &   &   &   &   &   \\ 
          \cite{67}  &   &   &   &   &   &   &   & X \\ 
          \cite{71}  &   &   &   &   &   &   &   & X \\ 
          \cite{74}  & X & X &   &   &   &   &   &   \\ 
          \cite{75}  & X &   &   &   &   &   &   &   \\ 
          \cite{77}  &   &   & X &   &   &   &   &   \\ 
          \cite{80}  & X &   &   &   &   &   &   &   \\ 
          \cite{81}  & X &   &   &   &   &   &   &   \\ 
          \cite{82}  &   & X &   &   &   &   &   &   \\ 
          \cite{93}  & X & X &   &   &   &   &   &   \\ 
          \cite{96}  &   &   & X &   &   &   &   &   \\ 
          \cite{99}  &   &   &   & X &   &   &   &   \\ 
          \cite{102} &   &   &   & X &   &   &   &   \\ 
          \cite{104} &   & X &   &   &   &   &   &   \\ 
          \cite{132} &   &   &   & X &   &   &   &   \\ 
          \cite{134} &   & X &   &   &   &   &   &   \\ 
          \cite{150} &   &   &   &   & X & X &   &   \\ 
          \cite{151} &   & X &   &   &   &   &   &   \\ 
          \cite{152} & X & X &   &   &   &   &   &   \\ 
          \cite{163} & X &   &   &   &   &   &   &   \\
          \cite{165} & X &   &   &   &   &   &   &   \\ \noalign{\hrule height 0.8pt}
    \end{NiceTabular}
\end{table}

\subsubsection{Manual Feature Mapping}
Manual feature mapping relies heavily on experience and knowledge of domain experts to extract features~\cite{feature_extraction_intro}. The main feature categories found in the pipelines include statistics, calculus, geometry, vision, and gesture recognition. Statistical features include (weighted) mean~\cite{7,12,13,37,63,66,93,104,163,165}, median~\cite{152}, total value~\cite{2,7,13,93}, standard deviation and variance~\cite{6,7,12,66}, root mean square~\cite{7,13,163,165}, moving average~\cite{65,104}, sequence window centroid~\cite{7,80,93}, range~\cite{4,59,66,152}, quartiles~\cite{66}, higher order cumulants~\cite{81}, coefficients~\cite{104,152}, value normalization~\cite{4,82,152}, value distribution~\cite{2,13,61,75,93}, probability~\cite{80}, and histogram features~\cite{61}. Calculus features include maximum~\cite{4,13,24,27,29,37,63,66,82,134,163,165}, minimum~\cite{13,37,151,163,165}, absolute value~\cite{13,65,163,165}, a signal pattern across time~\cite{6}, discrete differentiation~\cite{2,7,13,65,93}, discrete integration~\cite{7}, derivative sum~\cite{7}, integrated sum~\cite{7} and integrated delta~\cite{7}. Geometrical features include intensity area's~\cite{63}, roundness~\cite{63}, eccentricity~\cite{63}, perimeter~\cite{63} and shape slope~\cite{63}. A specific visual feature used is the so called local binary pattern~\cite{61}. The authors in~\cite{74} use a feature extraction algorithm comprised of steps that involve some of the features mentioned above.

In 2016, Lien et al.~\cite{7} conducted extensive research into the development of range-doppler spectrogram specific features in the context of gesture recognition. The goal was to reduce computational overhead on resource constrained devices. After \glspl{roi} have been selected, matrix calculations are applied to these regions to obtain multi-channel integration, multi-channel derivative and temporal derivative matrices. In 2018, Flintoff et al.~\cite{93} introduced a new gesture recognition feature in the form of a sonar value.

\subsubsection{Automatic Feature Mapping}
Automatic feature mapping methods used in the millimeter wave sensing applications mainly aim at dimensionality reduction. Several tracking applications that utilize point cloud frames either use clustering~\cite{8,96} or the Kalman filter's prediction process~\cite{21,99,102,132} to reduce point clouds into point cloud cluster centroid values. Since the number of point clouds per frame is unknown, a clustering algorithm that does not require a number of clusters to be defined a priori is required~\cite{8}. More information about the prediction process can be found in~\cite{ti_group_tracker}. Clustering is also used to reduce people counts corresponding to the same person~\cite{77} or to obtain a median heart rate measurement from a noisy spectrum~\cite{34}. \gls{pca} is used for linearly reducing an input data vector \(v\) into principle component vector \(p = W^Tv\) containing decorrelated principal components suitable for use with \glspl{svm} and decision trees. The symbol \(W \in \mathbb{R}^{(N,L)}\) denotes a projection matrix where \(N\) is the input vector size and \(L\) the number of desired principal features. The projection matrix elements are calculated such that the variance included in input data vectors used for training is maximized~\cite{6472238}. To reduce a dataset to a low feature dimension for visualization purposes, \gls{t-sne} can be used~\cite{150}.

Other automatic feature mapping methods include representation and meta learning~\cite{6472238,150}. Even though these methods also involve dimensionality reduction, it is considered to be a positive side effect rather than a goal. The main idea behind representation learning is to create a machine learning model that has the ability to extract features from a dataset. The features can be fed into a variety of downstream models used for solving a task that is loosely related to the task associated to the upstream model~\cite{6472238}. Meta learning differs from representation learning because it has the goal of providing features that can be used for a variety of different task types rather than loosely related tasks~\cite{150}.

\subsubsection{Gaps and Challenges}
\label{sec:feature_extraction_gaps_challenges}
Table~\ref{tab:feature_extraction_summary} shows that the majority of approaches that utilize feature extraction rely on manual feature mapping. In contrast to reducing computational overhead, manual feature mapping costs a lot of time and human resources. Automatic feature mapping in millimeter wave sensing applications is an unexplored field of research. Only one methodology found~\cite{150} uses a feature extraction technique for feature visualization to gain an understanding of extracted features. Lastly, meta and representation learning are severely underexploited in millimeter wave pipelines.

\subsection{Analytical Modeling}
\label{sec:analytical_modeling}
Most millimeter wave sensing applications presented in Figure~\ref{fig:application_taxonomy} cannot reach their goal by simply executing a set of pre-processing steps and extracting information relevant to the application. They require a model to translate information from pre-processed datasets or extracted feature sets to the application goal. Examples of these models include classification models~\cite{8,9,75}, filter models~\cite{8,91,131}, and measurement models~\cite{6,22,141,159}. Classification models assign class labels to a set of input data. The class can indicate performed activities, gestures or events/failures, presence of a specific person/object, etc. Filter models estimate trajectories of tracked entities, filter out false positive class predictions, etc. in an environment where noise causes trajectory measurements and class predictions to be stochastic in nature. Measurement models calculate the value of information such as position, yaw rate, or absolute velocity using mathematical models. In some cases, measurement models use a cost minimization technique for value calculation. The costs are computed with a function that uses input data as a function parameter. We classified the models used in millimeter wave sensing applications based on the model class (data driven vs. model driven) and the model type (black, grey, white models). Black, grey and white model types do not refer to the degree of model explainability. The model types refer to model determinism and use of physical knowledge and/or data for model construction. The model types were adopted from a physiological model identification study presented by Duun-Henriksen et al.~\cite{10.1177/193229681300700220}. Apart from occasionally explaining that a certain model fuses millimeter wave data with data originating from other sensing domains in Section~\ref{sec:hybrid_modeling} to avoid hybrid model explanation ambiguity, further discussion of models that fuse millimeter wave data with data originating from other sensing domains is outside the scope of this review. In what follows, we analyze the analytical model classes and model types used in millimeter wave sensing applications.

\subsubsection{Model driven modeling}
Model driven modeling relies on physical and/or mathematical knowledge about the technology or the environment to construct a model. The knowledge refers to information, a set of rules and/or a set of equations obtained from physical phenomena or mathematical proofs. The model driven approaches are classified either as white box or grey box models. White box models vary in determinism and solely rely on physical and/or mathematical knowledge for model construction. For example, phase tracking~\cite{6} deploys position initialization, resulting in an extra dependency on a random variable in addition to phase inputs for a certain position output. In contrary, instantaneous ego-motion estimation~\cite{159} relies on Monte-Carlo simulation to test a model. This indicates that the computations performed by the model are deterministic in principle. Grey box models are non-deterministic and also rely on physical and/or mathematical knowledge for model construction. They differ from white box models because parts of the model are continuously altered across time or completed with information extracted from input data and data features~\cite{10.1177/193229681300700220}. Examples include hidden state updates and use of cost and similarity parameters.

Table~\ref{tab:model_driven_analytical_models} presents our analysis of model driven approaches used in millimeter wave sensing application pipelines. Next, we explain different approaches used under white and grey box models.

\begin{table*}[ht]
    \centering
    \caption{Summary of model driven analytical models used in millimeter wave sensing pipelines. The abbreviations are explained throughout Section~\ref{sec:analytical_modeling}.} \label{tab:model_driven_analytical_models}
    \tiny\sffamily
    \setlength{\tabcolsep}{4pt}
    \begin{NiceTabular}{@{}lM{7mm}cM{12mm}M{10mm}c|M{6mm}M{7mm}M{10mm}cccM{10mm}ccc@{}}[code-before =\rowcolors{3}{}{gray!12}]
        \noalign{\hrule height 0.8pt}
         & \Block{1-5}{\textbf{White box}} & & & & & \Block{1-10}{\textbf{Grey box}} & & & & & & & & & \Tstrut\\[1.2ex]
         & Phase tracking & SVD & Velocity/Acker- man model & Angle/Time model & \acs{dtdoa} & Shadow RTI & Bayesian filter & Association/ Allocation & Scan matching & MLE & Triangulation & Angle/Time model & \acs{rsa} model & Rician model & Velocity model \\[1.7ex] \hline
         
        \cite{5}    &   &   &   &   &   &   & X &   &   &   &   &   &   &   &   \\ 
        \cite{6}    & X &   &   &   &   &   &   &   &   &   &   &   &   &   &   \\ 
        \cite{8}    &   &   &   &   &   &   & X & X &   &   &   &   &   &   &   \\ 
        \cite{15}   &   &   &   &   &   &   & X &   &   &   &   &   &   &   &   \\ 
        \cite{16}   &   &   &   &   &   &   & X &   &   &   &   &   &   &   &   \\ 
        \cite{21}   &   &   &   &   &   &   & X & X &   &   &   &   &   &   &   \\ 
        \cite{22}   &   &   &   & X &   &   &   &   &   &   &   &   &   &   &   \\ 
        \cite{25}   &   &   &   &   &   &   & X &   &   &   &   &   &   &   &   \\ 
        \cite{27}   &   &   &   &   &   &   &   &   &   &   & X & X &   &   &   \\ 
        \cite{31}   &   &   &   &   &   &   & X &   &   &   &   &   &   &   &   \\ 
        \cite{32}   &   &   &   &   &   &   & X &   &   &   &   &   &   &   &   \\ 
        \cite{49}   &   &   &   &   &   &   &   & X &   &   &   &   &   &   &   \\ 
        \cite{80}   &   &   &   &   &   &   &   &   &   &   &   &   &   &   & X \\ 
        \cite{87}   &   &   &   &   &   &   &   &   &   &   &   &   &   &   & X \\ 
        \cite{91}   &   &   &   &   &   &   & X & X & X &   &   &   &   &   &   \\ 
        \cite{92}   &   &   &   &   &   &   &   &   &   &   &   &   &   & X &   \\ 
        \cite{99}   &   &   &   &   &   &   & X & X &   &   &   &   &   &   &   \\ 
        \cite{101}  &   &   &   &   & X &   &   &   &   &   &   &   &   &   &   \\         
        \cite{102}  &   &   &   &   &   &   & X & X &   &   &   &   &   &   &   \\ 
        \cite{107}  &   &   &   &   &   & X &   &   &   &   &   &   &   &   &   \\ 
        \cite{130}  &   &   &   &   &   &   &   &   &   & X &   &   &   &   &   \\ 
        \cite{131}  &   &   &   &   &   &   & X & X &   &   &   &   &   &   &   \\ 
        \cite{132}  &   &   &   &   &   &   & X & X &   &   &   &   &   &   &   \\ 
        \cite{133}  &   &   &   &   &   &   & X &   &   &   &   &   &   &   &   \\ 
        \cite{135}  &   &   &   &   &   &   &   &   &   &   &   &   & X &   &   \\ 
        \cite{158}  &   & X &   &   &   &   &   & X &   &   &   &   &   &   &   \\ 
        \cite{159}  &   &   & X &   &   &   &   &   &   &   &   &   &   &   &   \\ 
        \cite{160}  &   &   &   &   &   &   &   &   & X & X &   &   &   &   &   \\
        \cite{161}  &   &   &   &   &   &   & X &   & X &   &   &   &   &   &   \\ \noalign{\hrule height 0.8pt}
    \end{NiceTabular}
\end{table*}

\paragraph{White Box Models}
The phase tracking model~\cite{6} calculates a two dimensional position in Cartesian coordinates based on successive phase shifts. The two dimensional position is calculated with a combination of distance equation sets and triangulation. The initial position is measured with a special acquisition module that is independent of the input data. The velocity and the Ackerman model~\cite{159} first calculates a radar velocity vector containing longitudinal and lateral velocities based on the sinusoidal progress of measured radial velocities over the azimuth angle using a least-square approach. Afterwards, object's absolute velocity and yaw rate are calculated using a model based on the Ackerman condition involving a velocity and yaw rate equation. The angle/time model~\cite{22} calculates mobile object positions based on a configuration involving multi-path component \gls{tof} and \gls{aoa} information or a configuration involving multi-path component \gls{tof} and \gls{aod} information. The information is retrieved from a ray tracing tool. This information, combined with meta-data such as a map layout, results in object locations. The eventual location is determined based on majority voting. Relative motion between two radar scans can be measured in a least-square sense by using \gls{svd} on a set of landmark matches~\cite{158}. \gls{dtdoa} is a positioning model consisting of a set of equations that use pseudo time of arrival to compute a position estimate~\cite{101}.

\paragraph{Grey Box Models}
Bayesian filters originate from stochastic filtering theory and Bayesian statistics. As indicated in Section~\ref{sec:data_preprocessing}, millimeter wave sensing applications obtain measurements as discrete samples. Therefore, the explanation assumes a discrete Bayesian filter modeled as a \gls{hmm}. The model contains one hidden state that varies with time. The state describes information that is interesting for the millimeter wave sensing application. The exact information is unobservable and thus hidden. Furthermore, action variables are omitted. The hidden state \(x_t\) is dependent on one previous state \(x_{t-1}\). Measurements \(s_t\) in the form of raw data samples or data features are observed at, i.e., sampled from, the pre-processing or feature extraction phase. These measurements are dependent on the current hidden state. The state transition function includes noise to account for a distribution of different outputs the function can yield at a given time instant. The measurement observation function accounts for noise to take care of measurement inaccuracies~\cite{bayesian_filter_survey,5713818}.


The state transition and the state observation functions \(x_t=g(x_{t-1}, pn_t)\) and \(s_t=h(x_t, mn_t)\) are modeled as \glspl{pdf} \(p(x_t|x_{t-1})\) and \(p(s_t|x_t)\)~\cite{5713818}. \glspl{pdf} explicitly represent the uncertainty in variables taking on a particular value in a range of values at a specific time instant. The Bayesian filter continuously executes two functions called the predict and update functions with a recursive algorithm. The role of the predict function is to estimate a new hidden state \gls{pdf} \(p(x_t|s_{1:t-1}) = \sum_{x_{t-1}} p(x_t|x_{t-1}) p(x_{t-1}|s_{1:t-1})\) based on the previous hidden state \gls{pdf} and state transition \gls{pdf}. The new hidden state \gls{pdf} can be used to derive hidden state values~\cite{mit_bayesian_filters}. The role of the update function is to adjust the hidden state \gls{pdf} when measurements are observed. The update function is based on Bayes' theorem. The update function ensures that the hidden state \gls{pdf} can always be used to derive hidden state values that closely represent, and do not drift away from, the exact hidden state values. Calculation and/or modeling of \(p(x_t|s_{1:t-1})\), \(p(s_t|x_t)\), and \(p(x_t|x_{t-1})\) is a core task of Bayesian filtering. More information can be found in a comprehensive review authored by Chen~\cite{bayesian_filter_survey}.

Several Bayesian filters exist, such as the custom Bayesian~\cite{7,131}, particle~\cite{5,162}, \(\alpha-\beta\)~\cite{151}, Kalman~\cite{8,17,25}, extended Kalman~\cite{21,99,102,132,133,141,161}, fusion extended Kalman~\cite{32,100}, unscented Kalman~\cite{91}, fusion adaptive Kalman~\cite{15} and adaptive Sage-Husa Kalman~\cite{16,31} filter. The Kalman filter~\cite{bayesian_filter_survey}, under linear, quadratic and Gaussian assumptions, can represent the state transition and observation functions \(x_t=g(x_{t-1}, pn_t)\) and \(s_t=h(x_t, mn_t)\) as a set of linear equations. All other Kalman filters loosen these assumptions. The extended Kalman filter, for example, allows the functions to be approximated as a Jacobian matrix~\cite{bayesian_filter_survey}. Particle filters~\cite{5713818,bayesian_filter_survey} approximate \(p(x_t|s_{1:t-1})\) as a set of particles \(\{<x_t^{(i)}, w_t^{(i)}>\}_{i=1}^l\) where \(w_t^{(i)} = p(s_t|x_t^{(i)})\). The \(\alpha-\beta\) filter~\cite{151} is a simplified Bayesian filter that limits the number of hidden states to two. Predictions and updates are executed with a simple set of equations. Symbols \(\alpha\) and \(\beta\) refer to manually set correction gains used during the update process.

Scan matching models are concerned with finding the best rotation and translation operations that have the ability to align two point scans or a point scan to an existing area map. The \gls{icp} model involves two iterative steps until convergence. In the first step, points from two are matched based on closeness to one another in a given space. Closeness is measured by means of a distance metric. In the second step, to find the optimal rotational angle and translation, the sum of squared distances between the points is used~\cite{161}. \gls{ndt} scan matching~\cite{91,160} requires a grid of probability functions created from a map for matching.

Association/allocation models either use combinatorial optimization~\cite{8,131}, a gating function~\cite{21,99,102,132}, template model matching~\cite{49} or landmark association~\cite{91, 158}. These models associate input data to an existing entity or create new entities based on an optimization criterium. This criterium is dependent on a cost metric involving the input data. For example in tracking, the models determine if incoming input data belong to a certain track that is already known or indicate a new track.

Other grey box models include Shadow RTI~\cite{107}, triangulation, angle/time model~\cite{27}, \gls{rsa} model~\cite{135}, velocity model~\cite{80}, Rician model~\cite{92}, and \gls{mle}~\cite{130,160}. Shadow RTI~\cite{107} measures link \gls{rss} attenuation changes across a network of millimeter wave transceivers. The knowledge utilized for model construction is that when people walk by, the \gls{rss} attenuation changes. Every link is made of a set of `pixels'. Link \gls{rss} attenuation contribution is computed for every pixel by solving a least square problem based on a measured \gls{rss} change vector. The triangulation and angle/time models in~\cite{27} use \gls{aoa} spectrums coming from multiple access points to determine a mobile client's position from a set of measured anchor positions based on an associated cost metric. Additional input data include meta-data that are not measured such as room boundaries and a permanent obstacle set. The \gls{rsa} model~\cite{135} measures object characteristics based on separate models. The object characteristics include surface curvature, surface boundary, and material. The surface boundary is determined with a surface reflection model that predicts \gls{rss} series of a reflection surface with fixed-size surface boundaries. The surface boundary of a measured \gls{rss} series is determined by matching the \gls{rss} series to a set of predicted \gls{rss} series produced with the surface reflection model based on a similarity metric. The Rician model in~\cite{92} was derived from the Rician model in~\cite{10.1117/12.488039}. The model compares a measured radar cross section with pre-simulated radar cross sections to identify a class based on log-likelihood computations. The measured cross section receives the class of the pre-simulated radar cross section that results in the largest computed log-likelihood value. \gls{mle} is used to estimate data of interest from noisy measurement signals. The model in~\cite{130} uses reference data and a likelihood function to return a data estimate that maximizes the likelihood function value. Joint spatial and Doppler-based ego-motion estimation in~\cite{160} computes state \glspl{pdf} for a vehicle's yaw rate, longitudinal velocity and lateral velocity by using a joint optimization problem. Mean velocity~\cite{80,87} is measured with a set of equations and pulse-pair phase estimation, which relies on a correlation product that depends on the input data.

\begin{table*}[ht]
    \centering
    \caption{Summary of data driven analytical models used in millimeter wave sensing pipelines. The abbreviations are explained throughout Section~\ref{sec:analytical_modeling}.} \label{tab:data_driven_analytical_models}
    \tiny\sffamily
    \setlength{\tabcolsep}{4pt}
    \begin{NiceTabular}{@{}lccccccc|cccccccc@{}}[code-before = \rowcolors{3}{}{gray!12}]
        \noalign{\hrule height 0.8pt}
         & \Block{1-7}{\textbf{Shallow black box}} & & & & & & & \Block{1-8}{\textbf{Deep black box}} & & & & & & & \Tstrut\\[1.2ex]
         & SVM & SVDD & MPM & SOM & Decision tree & Clustering & k-NN & LVQ & One-versus-one SVM & AdaBoost SVM & Random forest & Feed-forward NN & CNN & LSTM & CNN+LSTM \\[0.4ex] \hline
         
        \cite{1}   &   &   &   &   &    &   &   &   &   &   &   &   & X &   &   \\ 
        \cite{2}   &   &   &   &   &    &   &   &   &   &   & X &   &   &   &   \\ 
        \cite{3}   &   &   &   &   &    &   &   &   &   &   &   &   & X &   &   \\ 
        \cite{4}   &   &   &   &   & X  &   &   &   &   &   &   &   &   &   &   \\ 
        \cite{6}   &   &   &   &   & X  &   &   &   &   &   &   &   &   &   &   \\ 
        \cite{8}   &   &   &   &   &    &   &   &   &   &   &   &   &   & X &   \\ 
        \cite{9}   &   &   &   &   &    &   &   &   & X &   &   & X &   & X & X \\ 
        \cite{12}  &   &   &   &   &    &   &   &   &   &   & X &   &   & X &   \\ 
        \cite{13}  &   &   &   &   &    &   &   &   &   &   & X &   &   &   &   \\ 
        \cite{14}  &   &   &   &   &    &   &   &   &   &   & X &   &   &   &   \\ 
        \cite{19}  &   &   &   &   &    &   &   &   &   &   &   &   & X &   &   \\ 
        \cite{23}  &   &   &   & X &    &   &   & X &   &   &   &   &   &   &   \\ 
        \cite{35}  &   &   &   &   &    &   &   &   &   &   &   &   &   &   & X \\ 
        \cite{36}  &   &   &   &   &    &   &   &   &   &   &   &   &   &   & X \\ 
        \cite{37}  &   &   &   &   &    &   &   &   & X &   &   &   &   &   &   \\ 
        \cite{39}  &   &   &   &   &    &   &   &   &   &   &   &   & X &   &   \\ 
        \cite{46}  &   &   &   &   &    &   &   &   &   &   &   &   & X &   &   \\ 
        \cite{52}  &   &   &   &   &    &   &   &   &   &   &   &   & X &   &   \\ 
        \cite{61}  &   &   &   &   &    &   &   &   &   & X &   &   &   &   &   \\ 
        \cite{62}  &   &   &   &   &    & X &   &   &   &   &   &   &   &   &   \\ 
        \cite{63}  &   &   &   &   &    &   & X &   &   &   &   &   &   & X &   \\ 
        \cite{65}  &   &   &   &   &    & X &   &   &   &   &   &   &   &   &   \\ 
        \cite{66}  &   &   &   &   & X  &   &   &   &   &   &   &   &   &   &   \\ 
        \cite{67}  & X &   &   &   &    &   &   &   &   &   &   &   &   &   &   \\ 
        \cite{71}  &   &   &   &   &    &   &   &   &   &   &   &   &   & X &   \\ 
        \cite{72}  &   &   &   &   &    &   &   &   &   &   &   &   &   &   & X \\ 
        \cite{74}  &   &   &   &   &    &   &   &   & X &   &   &   &   &   &   \\ 
        \cite{75}  &   &   &   &   &    &   &   &   & X &   &   &   &   &   &   \\ 
        \cite{81}  &   & X &   &   &    &   &   &   &   &   &   &   &   &   &   \\ 
        \cite{82}  &   &   & X &   &    &   &   &   &   &   &   &   &   &   &   \\ 
        \cite{93}  &   &   &   &   &    &   &   &   &   &   & X &   &   &   &   \\ 
        \cite{95}  &   &   &   &   &    &   &   &   &   &   &   & X &   &   &   \\ 
        \cite{96}  &   &   &   &   &    &   &   &   &   &   &   &   & X &   &   \\ 
        \cite{113} &   &   &   &   &    &   &   &   &   &   &   &   & X &   &   \\ 
        \cite{115} &   &   &   &   &    &   &   &   &   &   &   & X &   &   &   \\ 
        \cite{139} &   &   &   &   &    &   &   &   &   &   &   &   & X &   &   \\ 
        \cite{146} &   &   &   &   &    &   &   &   &   &   &   &   & X &   &   \\ 
        \cite{147} &   &   &   &   &    &   &   &   &   &   &   &   & X &   &   \\ 
        \cite{148} &   &   &   &   &    &   &   &   &   &   &   &   &   &   & X \\ 
        \cite{149} &   &   &   &   &    &   &   &   &   &   &   &   &   &   & X \\ 
        \cite{150} &   &   &   &   &    & X & X &   &   &   &   &   &   &   &   \\ 
        \cite{152} &   &   &   &   &    &   & X &   &   &   &   &   &   &   &   \\ 
        \cite{153} &   &   &   &   &    &   &   &   &   &   &   &   & X &   &   \\ 
        \cite{163} &   &   &   &   &    &   &   &   &   &   & X &   &   &   &   \\ 
        \cite{164} &   &   &   &   &    & X &   &   &   &   &   &   &   &   & X \\
        \cite{165} &   &   &   &   &    &   &   &   & X &   &   &   &   &   &   \\ \noalign{\hrule height 0.8pt}
    \end{NiceTabular}
\end{table*}

\subsubsection{Data driven modeling}
\label{sec:data_driven_modeling}
This type of modeling solely relies on information contained in the input data itself as well as the feature sets to construct a model~\cite{10.1177/193229681300700220}. Model construction typically involves choosing a model type and associated hyperparameters, fitting the model parameters to a dataset designated for training and testing the model performance on a dataset designated for validation. Fitting and validation are performed continuously in an iterative manner based on a cost or reward function by changing the model type or hyperparameters until the model exhibits the desired performance on the validation dataset~\cite{Goodfellow-et-al-2016,10.1145/1961189.1961199,Murthy1998}. Data driven models are classified either as a shallow black box or deep black box model. Both types are non-deterministic and solely rely on data for model construction. The difference between the two is that in shallow black box models, the input only passes through a single model structure before obtaining a model output while in a deep black box model, the input goes through multiple submodels such as layers, decision trees, or \glspl{svm}. 

Table~\ref{tab:data_driven_analytical_models} presents our analysis of data driven approaches used in millimeter wave sensing applications. Next, we explain different approaches used under shallow and deep black box models.

\paragraph{Shallow Black Box Models}
Supervised shallow black box models, such as \glspl{svm} and decision trees, used for classification can perform well with limited datasets of small size~\cite{cristianini_shawe-taylor_2000}. However, this comes at the cost of requiring carefully crafted features. The features are either crafted manually~\cite{4,6,66,74,75} or automatically in a linear fashion~\cite{9,14,37,67}. This process is labor intensive and/or can limit the eventual model performance for more complicated tasks during model validation.

The pipelines deploying \glspl{svm} use it to learn and execute a \gls{csvc} task. \glspl{svm} are applicable to binary classification tasks in which an optimal hyperplane is learned from training data. The hyperplane divides the data sample space into two sections, each corresponding to a given class. The classification is based on a decision function, which uses a set of support vector samples and a kernel function. The kernel function introduces non-linearity to the optimal hyperplane. Additional information on how \glspl{svm} learn can be found in~\cite{10.1145/1961189.1961199}.

Decision trees learn a tree structure containing nodes, branches, and leaves from a training dataset. Nodes represent binary input attribute tests, branches represent test outputs, and leaves represent a class that can be assigned. An unseen input dataset goes throughout the entire tree and is ultimately assigned to a single class label~\cite{4}. More information regarding decision tree construction can be found in~\cite{Murthy1998}.

Most approaches using a \gls{svm} or decision tree have been classified as an ensemble of several classifiers. These ensembles are formed with special multi-class and ensemble learning strategies such as one-versus-one~\cite{9,37,74,75,165}, random forest~\cite{2,7,12,13,14,93,163} and AdaBoost~\cite{61}. Commonly, by letting several weak classifiers assign class labels to a given input, the final classification output is determined via majority voting.

Other models include clustering~\cite{17,20,65,162}, \gls{svdd}~\cite{81}, \gls{mpm}~\cite{82}, \gls{som}~\cite{23}, \gls{lvq}~\cite{23}, and \gls{knn}~\cite{63,150,152}. Clustering is an unsupervised learning algorithm to divide data into several clusters of data points that are similar to one another. Different techniques such as K-means~\cite{20,65}, DBSCAN~\cite{17,164}, Mean shift~\cite{62}, or DenStream~\cite{162} clustering have been used. \gls{svdd} and \gls{mpm} are similar to a \gls{svm}. They differ from a \gls{svm} because both learn a hypersphere rather than a hyperplane that separates all data and feature samples into two classes~\cite{81,82}. A \gls{som} is an unsupervised one layer neural network that reduces n-dimensional input vectors into a 2D feature map. A class is assigned to an unseen input sample based on how close the input sample corresponds to a feature map weight vector based on a distance metric. The \gls{lvq} is a supervised two layer neural network containing a competitive and fully connected layer. The competitive layer is similar to a \gls{som} apart from having a limited and predefined number of outputs instead of having an output for every feature map weight vector and containing a transfer function. The output of the transfer function is fed into a fully connected layer to return a classification result for a set of user defined classes~\cite{23}. A \gls{knn} model assigns a class to an input based on a majority vote of labels from k-nearest samples in a given data space.

\paragraph{Deep Black Box Models}
Most deep black box models belong to the machine learning paradigm called deep learning. The basic idea of deep learning is that by concatenating a number of submodels (layers), increasing the number of computation nodes in a submodel (neurons), and combining every submodel with an activation function, a non-linear vector mapping function that varies in complexity is learned from training data. The varying complexity comes at the cost of requiring large training datasets and a long training time. Additional information on deep learning can be found in~\cite{Goodfellow-et-al-2016}.

\glspl{cnn} found in the pipelines are mainly combined with 2D spectrograms~\cite{11,17,94,96,146,150,153} and 2D radar images~\cite{19,39,46,52,113,139}. \glspl{cnn} have been used extensively for images originating from vision and object detection domains in the past. Spectrograms and radar images share similar characteristics with these images. The data types are 2-dimensional and features relevant to the mapping function are made up of values local to one another in the value matrix. Another observation is that many approaches base their model on existing vision and object detection models such as VGG~\cite{94,96}, ResNet~\cite{94,146}, ZFnet~\cite{52}, Faster R-CNN~\cite{39,153}, YOLO~\cite{46} and FCOS~\cite{139}. The pipelines therefore rely on experience gathered with images in the vision and object detection domains. In addition, a few approaches have tried using a temporal \gls{cnn} with 1D profile~\cite{1} data and a feature distributed \gls{cnn} with a single point cloud frame~\cite{147} as input.

\begin{table*}[ht]
    \centering
    \caption{Summary of hybrid analytical models used in millimeter wave sensing pipelines. The abbreviations are explained throughout Section~\ref{sec:analytical_modeling}.} 
    \label{tab:hybrid_analytical_models}
    \tiny\sffamily
    \setlength{\tabcolsep}{4pt}
    \aboverulesep=0ex
    \belowrulesep=0ex
    \begin{NiceTabular}{@{}lcccccM{10mm}M{10mm}c|ccccc@{}}[code-before =\rowcolors{3}{}{gray!12}]
        \noalign{\hrule height 0.8pt}
         & \Block{1-8}{\textbf{Model driven}} & & & & & & & & \Block{1-5}{\textbf{Data driven}} & & & & \Tstrut\\[1.2ex]
         & \Block{1-3}{\textbf{White box}} & & & \Block{1-5}{\textbf{Grey box}} & & & & & \Block{1-1}{\textbf{Shallow black box}} & \Block{1-4}{\textbf{Deep black box}} & & & \\
         \cmidrule(lr){2-4}\cmidrule(lr){5-9}\cmidrule(lr){10-10}\cmidrule(lr){11-14}
         & Path geometry & Triangulateration & Trilateration & Bayesian filter & Scan matching & Association/ Allocation & Angle/Time model & \gls{rss} model & Clustering & Random forest & CNN & LSTM & CNN+LSTM \TstrutTwo\\[1.7ex] \midrule
         
        \cite{7}   &   &   &   & X &   &   &   &   &   & X &   &   &   \\ 
        \cite{11}  &   &   &   &   &   & X &   &   & X &   & X &   &   \\ 
        \cite{17}  &   &   &   & X &   &   &   &   & X &   & X &   &   \\ 
        \cite{20}  & X &   &   &   &   &   &   &   & X &   &   &   &   \\ 
        \cite{94}  &   &   &   & X &   &   &   &   &   &   & X &   & X \\ 
        \cite{100} &   &   &   & X &   &   &   &   & X &   &   &   &   \\ 
        \cite{141} &   & X &   & X &   &   & X & X & X &   &   &   &   \\ 
        \cite{151} &   &   & X & X &   &   &   &   &   &   & X & X &   \\
        \cite{162} &   &   &   & X & X &   &   &   & X &   &   &   &   \\ \noalign{\hrule height 0.8pt}
    \end{NiceTabular}
\end{table*}

An approach that can be combined with time sequence data is the \gls{rnn}. \glspl{rnn} are very good at discovering time dependencies among these sequences. Most approaches adopt the \gls{lstm} network which is a special type of \gls{rnn} that deals with the vanishing gradient problem noted when training traditional \glspl{rnn}~\cite{151}. The authors in~\cite{72} use a \gls{gru} layer. This layer is comparable to a \gls{lstm} layer apart from not having separate memory cells. The \gls{lstm} network is mainly combined with a flattened (voxelized) point cloud frame sequence~\cite{8,9}, Manually extracted 2D spectrogram sequence features~\cite{63}, \gls{rss} fingerprint matrix~\cite{12}, \gls{pca} extracted feature sequence~\cite{71} and alpha-beta filtered trajectory~\cite{151}. An important observation here is that \gls{lstm} requires some kind of feature extraction prior to using the time sequences for training.

\gls{cnn}/\gls{lstm} combinations harness the strengths from both models. Comparable to local spectrogram and radar image features, point cloud frames contain clouds that are located very sparsely. In contrast, the positioning of points inside a given cloud is very compact. Therefore, time distributed \glspl{cnn} can be used to obtain features from a single spectrogram, radar image and point cloud frame in a sequence while \gls{lstm} models the time dependency in the sequence afterwards in an end-to-end fashion~\cite{72,148,164}. The authors in~\cite{8} denote that bi-directional \glspl{lstm} converge faster than \gls{cnn}/\gls{lstm} combinations for a person identification task. However, this does not mean that they are better since the authors in~\cite{9} note that a \gls{cnn}/\gls{lstm} combination outperforms a bi-directional \glspl{lstm} network in a human activity detection task. Therefore, model testing remains the norm when creating new models.

\subsubsection{Hybrid Modeling}
\label{sec:hybrid_modeling}
In Tables~\ref{tab:model_driven_analytical_models} and~\ref{tab:data_driven_analytical_models} several rows include multiple crosses, indicating use of multiple models. There are reasons for combining models. The authors in~\cite{9,23,27} compare several self-created models to each other. Authors in~\cite{8,12,21,63,91,99,102,131,132,150,158,160,161} incorporate different processes in their sensing application pipeline that use a distinct model. Several papers use pipelines that include a tree or sequential set of multiple models coming from both the data and model driven paradigms. These sets are considered to be hybrid models. Table~\ref{tab:hybrid_analytical_models} presents our analysis of hybrid models used in millimeter wave sensing application pipelines.

In principle, hybrid models are constructed by putting the mathematical and data driven models together in a sequential or tree-based manner. Hybrid models constructed in a sequential manner were found in~\cite{7,11,17,20,94}. In~\cite{7}, a Bayesian filter was combined with a random forest output to reduce sporadic false-positive errors. A patient behavior detection task first tracks the patient with a combination of clustering and a Bayesian filter. Tracking information is used to construct two dimensional Doppler spectrograms that can be used with a \gls{cnn} for behavior detection. This limits computational complexity since additional pre-processing with STFT or FWT is omitted~\cite{17}. Environmental mapping~\cite{20} is performed with a specific set of steps. First, spatial channel profiles consisting of \gls{aoa}, \gls{aod} and \gls{rss} information are associated to a potential reflector through clustering. Afterwards, reflector points are retrieved through elementary geometry. An issue experienced in moving target classification is having micro-Doppler signatures spread across many different range bins in profiles across time. This issue is solved by tracking the position of these bins with a Bayesian filter and association/allocation combination. Afterwards, the bins are passed through STFT and fed into different neural networks for classification~\cite{94}. Motion behavior detection in~\cite{11} is performed by first applying a clustering algorithm on point cloud data to form micro-Doppler signature data. The signature data is afterwards fed into a \gls{cnn} to predict motion behavior.

\begin{table*}[hb]
    \centering
    \caption{Summary of evaluation metrics used for measuring performance of the analytical models mentioned in Section~\ref{sec:analytical_modeling}. The data driven evaluation metric abbreviations are explained throughout Section~\ref{sec:performance_eval_metrics}.} 
    \label{tab:eval_metrics_summary}
    \tiny\sffamily
    \setlength{\tabcolsep}{4pt}
    \begin{NiceTabular}{@{}lcccccccccc|ccc@{}}[code-before = \rowcolors{3}{}{gray!12}]
        \noalign{\hrule height 0.8pt}
         & \Block{1-10}{\textbf{Data driven}} & & & & & & & & & & \Block{1-3}{\textbf{Model driven}} & & \Tstrut\\[1.2ex]
         & Accuracy & Invalid gesture rate & Precision & Recall & Specificity & F1 score & Confusion matrix & ROC & Error & Visual inspection & Accuracy & Error & Visual inspection \\[0.4ex] \hline
         
          \cite{1}   & X &   &   &   &   &   &   &   &   & X &   &   &   \\ 
          \cite{2}   & X &   &   &   &   &   &   &   &   &   &   &   &   \\ 
          \cite{4}   & X & X &   &   &   &   &   &   &   &   &   &   &   \\ 
          \cite{5}   &   &   &   &   &   &   &   &   &   &   &   & X & X \\ 
          \cite{6}   & X &   &   &   &   &   &   &   &   &   &   & X &   \\ 
          \cite{7}   & X &   &   &   &   &   &   &   &   &   & X &   &   \\ 
          \cite{8}   & X &   &   &   &   &   & X &   &   &   &   & X &   \\ 
          \cite{9}   & X &   &   &   &   &   & X &   &   &   &   &   &   \\ 
          \cite{11}  & X &   &   &   &   &   &   &   &   &   &   &   &   \\ 
          \cite{12}  & X &   &   &   &   &   & X &   &   &   &   &   &   \\ 
          \cite{13}  & X &   &   &   &   &   & X &   &   &   &   &   &   \\ 
          \cite{14}  &   &   &   &   &   &   & X &   &   &   &   &   &   \\ 
          \cite{15}  &   &   &   &   &   &   &   &   &   &   & X & X &   \\ 
          \cite{16}  &   &   &   &   &   &   &   &   &   &   &   & X &   \\ 
          \cite{17}  & X &   &   &   &   &   &   &   &   &   &   &   &   \\ 
          \cite{19}  & X &   &   &   &   &   &   &   &   &   &   &   &   \\ 
          \cite{20}  & X &   &   &   &   &   &   &   &   &   & X & X &   \\ 
          \cite{21}  &   &   &   &   &   &   &   &   &   &   & X & X & X \\ 
          \cite{22}  &   &   &   &   &   &   &   &   &   &   &   & X &   \\ 
          \cite{23}  & X &   &   &   &   &   & X &   &   &   &   &   &   \\ 
          \cite{25}  &   &   &   &   &   &   &   &   &   &   &   & X &   \\ 
          \cite{26}  & X &   &   &   &   &   & X &   &   &   &   &   &   \\ 
          \cite{27}  &   &   &   &   &   &   &   &   &   &   &   & X &   \\ 
          \cite{31}  &   &   &   &   &   &   &   &   &   &   &   & X &   \\ 
          \cite{32}  &   &   &   &   &   &   &   &   &   &   &   & X &   \\ 
          \cite{35}  & X &   &   &   &   &   & X &   &   &   &   &   &   \\ 
          \cite{36}  & X &   &   &   &   &   &   &   &   &   &   &   &   \\ 
          \cite{37}  & X &   &   &   &   & X & X &   &   &   &   &   &   \\ 
          \cite{39}  &   &   & X & X &   & X &   & X &   &   &   &   &   \\ 
          \cite{41}  &   &   & X & X &   &   &   &   &   &   &   &   &   \\ 
          \cite{44}  &   &   &   &   &   &   &   &   &   &   &   &   & X \\ 
          \cite{46}  &   &   & X &   &   &   &   &   &   &   &   &   &   \\ 
          \cite{49}  &   &   &   &   &   &   &   &   &   &   &   & X &   \\ 
          \cite{52}  &   &   & X &   &   &   &   &   &   &   &   &   &   \\ 
          \cite{61}  &   &   & X & X &   &   &   &   &   &   &   &   &   \\ 
          \cite{63}  & X &   & X & X &   & X & X &   &   &   &   &   &   \\ 
          \cite{65}  & X &   &   &   &   &   &   &   &   &   &   &   &   \\ 
          \cite{66}  & X &   &   &   &   &   & X &   &   &   &   &   &   \\ 
          \cite{67}  &   &   &   & X &   &   &   &   &   &   &   &   &   \\ 
          \cite{71}  &   &   & X & X &   & X & X &   &   &   &   &   &   \\
          \cite{72}  & X &   &   &   &   &   &   &   &   &   &   &   &   \\ \noalign{\hrule height 0.8pt}
    \end{NiceTabular}
\end{table*}

\begin{table*}[ht]
    \ContinuedFloat
    \centering
    \caption{Continued}
    \tiny\sffamily
    \setlength{\tabcolsep}{4pt}
    \begin{NiceTabular}{@{}lcccccccccc|ccc@{}}[code-before = \rowcolors{3}{}{gray!12}]
        \noalign{\hrule height 0.8pt}
         & \Block{1-10}{\textbf{Data driven}} & & & & & & & & & & \Block{1-3}{\textbf{Model driven}} & & \Tstrut\\[1.2ex]
         & Accuracy & Invalid gesture rate & Precision & Recall & Specificity & F1 score & Confusion matrix & ROC & Error & Visual inspection & Accuracy & Error & Visual inspection \\[0.4ex] \hline
         
          \cite{74}  & X &   &   &   &   &   & X &   &   &   &   &   &   \\ 
          \cite{75}  & X &   &   &   &   &   & X &   &   &   &   &   &   \\ 
          \cite{80}  &   &   &   &   &   &   &   &   &   &   &   &   & X \\ 
          \cite{81}  &   &   &   &   &   &   &   &   &   & X &   &   &   \\ 
          \cite{82}  &   &   &   &   &   &   &   &   &   & X &   &   &   \\ 
          \cite{91}  &   &   &   &   &   &   &   &   &   &   &   & X &   \\ 
          \cite{92}  &   &   &   &   &   &   &   &   &   &   & X &   &   \\ 
          \cite{93}  & X &   &   &   &   &   &   &   &   &   &   &   &   \\ 
          \cite{94}  & X &   &   &   &   &   &   &   &   &   &   &   &   \\ 
          \cite{95}  &   &   &   & X &   &   &   &   &   &   &   &   &   \\ 
          \cite{96}  &   &   & X & X &   &   &   &   &   &   &   &   &   \\ 
          \cite{99}  &   &   &   &   &   &   &   &   &   &   & X & X &   \\ 
          \cite{100} &   &   &   &   &   &   &   &   &   &   &   & X &   \\ 
          \cite{102} &   &   &   &   &   &   &   &   &   &   &   &   & X \\ 
          \cite{107} &   &   &   &   &   &   &   &   &   &   &   & X & X \\ 
          \cite{113} &   &   &   &   &   &   &   &   & X &   &   &   &   \\ 
          \cite{115} & X &   &   & X & X &   &   &   &   &   &   &   &   \\ 
          \cite{130} &   &   &   &   &   &   &   &   &   &   &   & X &   \\ 
          \cite{131} &   &   &   &   &   &   &   &   &   &   & X & X &   \\ 
          \cite{132} &   &   &   &   &   &   &   &   &   &   & X & X & X \\ 
          \cite{133} &   &   &   &   &   &   &   &   &   &   &   & X &   \\ 
          \cite{135} &   &   &   &   &   &   &   &   &   &   &   & X &   \\ 
          \cite{139} &   &   & X & X &   &   &   &   &   &   &   &   &   \\ 
          \cite{141} &   &   &   &   &   &   &   &   &   &   &   & X &   \\ 
          \cite{146} & X &   & X & X &   & X & X &   &   &   &   &   &   \\ 
          \cite{147} & X &   &   &   &   &   & X &   &   &   &   &   &   \\ 
          \cite{148} & X &   &   &   &   &   & X &   &   &   &   &   &   \\ 
          \cite{149} &   &   &   &   &   &   & X &   &   &   &   &   &   \\ 
          \cite{150} & X &   &   &   &   & X & X &   &   & X &   &   &   \\ 
          \cite{151} & X &   &   &   &   &   & X &   &   &   &   &   &   \\ 
          \cite{152} & X &   &   &   &   &   & X &   &   &   &   &   &   \\ 
          \cite{153} &   &   & X & X &   &   & X &   &   &   &   &   &   \\ 
          \cite{158} &   &   &   &   &   &   &   &   &   &   &   & X &   \\ 
          \cite{159} &   &   &   &   &   &   &   &   &   &   &   & X &   \\ 
          \cite{160} &   &   &   &   &   &   &   &   &   &   &   & X &   \\ 
          \cite{161} &   &   &   &   &   &   &   &   &   &   &   & X &   \\ 
          \cite{162} &   &   &   &   &   &   &   &   &   &   &   & X &   \\ 
          \cite{163} & X &   &   &   &   &   & X &   &   &   &   &   &   \\ 
          \cite{164} &   &   &   &   &   &   &   &   & X &   &   &   &   \\
          \cite{165} & X &   &   &   &   &   & X &   &   &   &   &   &   \\ \noalign{\hrule height 0.8pt}
    \end{NiceTabular}
\end{table*}

There are two versions of tree-based hybrid models as addressed in~\cite{100,141,151,162}. In the first version, the tree-based hybrid model either branches off to perform multiple processes simultaneously or fuses the output of several simultaneously performed processes to perform a single consecutive process. In addition to vision data fusion with a Bayesian filter, millimeter wave data is first clustered~\cite{100}. \gls{slam} is solved by first clustering a raw radar scan to make it more sparse. Secondly, the scan is fused with odometry sensor data in a particle filter to estimate an odometry pose. Meanwhile, scan matching with a reference scan is implemented on the sparse radar scan for retrieving new parameters used for updating the cluster model~\cite{162}. In the second tree-based hybrid model version, a model decision is taken based on a particular context at particular points in the tree. An air writing pipeline~\cite{151} tracks consecutive three dimensional locations with an \(\alpha-\beta\) filter. The locations are determined via trilateration with range estimates coming from several radars. Afterwards, both a \gls{cnn} and \gls{lstm} are tested on a character recognition task. Another model pipeline used for solving the \gls{slam} problem~\cite{141} first chooses a triangulateration, angle/time, or \gls{rss} model for both client and anchor node localization based on whether the environment map is known or not. Secondly, a decision between a \gls{rss} or angle/time model is made for estimating obstacle surfaces. Thirdly, the surface limits are determined either by analyzing the difference in \gls{rss} at \glspl{aoa} situated at the limits or interaction of obstacle sides if the shape is two dimensional. Fourthly, clustering is used to reduce measurement bias due to noise. Lastly, an extended Kalman filter is deployed to improve the results.

\subsubsection{Gaps and Challenges}
An interesting observation derived from Tables~\ref{tab:model_driven_analytical_models} and~\ref{tab:data_driven_analytical_models} is that use of white box and shallow black box models in the context of millimeter wave sensing applications is restricted. White box models solely rely on physical and/or mathematical knowledge for model construction. Several shallow black box models, such as \glspl{svm} and decision trees, rely on manually crafted data features for model construction~\cite{4,6,66,74,75}. Crafting features manually relies heavily on experience and knowledge of domain experts. The observation and information regarding white box and shallow black box model types suggest that there is ample room for research geared towards better understanding the millimeter wave sensing environment, i.e., ample room for analyzing millimeter wave sensing environment dynamics and uncertainties. Secondly, Table~\ref{tab:data_driven_analytical_models} indicates that most research with deep black box models is restricted to \glspl{cnn} and \glspl{lstm}. This leaves room for exploring the use of \glspl{tcn} for modeling time dependency in raw data sample or data feature sequences. Lastly, Tables~\ref{tab:model_driven_analytical_models} and~\ref{tab:data_driven_analytical_models} indicate that the variety of models used in the millimeter wave sensing application pipelines is limited. Many application pipelines use \glspl{cnn}, \glspl{lstm}, Bayesian filters, and/or association/allocation. It is a challenge to increase the model variety and thus the knowledge on how analytical modeling in the millimeter wave sensing environment can be approached.

\subsection{Modeling Evaluation}
In this section, we review evaluation metrics used to measure performances of models employed in millimeter wave sensing applications as mentioned in Section~\ref{sec:analytical_modeling}. We also describe techniques used to improve data driven and grey box model training performance and evaluation.

\subsubsection{Performance Evaluation Metrics}
\label{sec:performance_eval_metrics}
Table~\ref{tab:eval_metrics_summary} summarizes performance metrics that are used in the reviewed papers. It can be seen that confusion matrix and accuracy are the most often used metrics in data driven model performance assessment. In training and validation of data driven models, the validation dataset is used to make choices about the model, including its hyperparameters~\cite{Goodfellow-et-al-2016}. After iteratively training and validating the model, the final model performance in relation to the application goal is evaluated on a held out test dataset consisting of samples that the model has not seen before. The outcome of the testing is compared to data labels and results are presented in the form of a confusion matrix and/or metrics that summarize the content of a confusion matrix. The confusion matrix~\cite{74,146} shows model classification performance for every class in the test dataset. The rows and columns represent ground truth and predicted classes respectively. Diagonal elements in the matrix denote the number, or fraction, of true positives for every class. When the diagonal elements are removed, the remaining row elements denote the false positives and remaining column elements the false negatives for every class. Positive samples are those that have been predicted to belong to a certain class. Negative samples are those that have been predicted to not belong to a certain class. Accuracy is defined as the overall proportion of predicted test dataset labels that match with the ground truth test dataset labels. Precision is a ratio of the number of true positive predictions to the number of all positive predictions made for a certain class. Recall is a ratio of the the number of true positive predictions to the number of all samples with a positive label in the ground truth dataset for a certain class~\cite{Goodfellow-et-al-2016}. Sometimes the inverse of accuracy and precision, the so called misclassification rate~\cite{4,15,99} and false discovery rate~\cite{41,61}, are also used. F1 score is the harmonic mean of precision and recall~\cite{146}. Specificity is a ratio of the number of true negative predictions to the number of all samples with a negative label in the ground truth dataset for a certain class~\cite{115}. The area under the \gls{roc} curve has also been used, which shows the degree of output separability~\cite{39}. A paper on gesture recognition used invalid gesture rate in conjunction with accuracy and misclassification rate to test data driven modeling performance for a set of gesture classes. Invalid gesture rate is defined as a ratio of samples classified as invalid gesture compared to the total number of samples in the test dataset~\cite{4}. Several model driven model results are also reported with the accuracy metric.

Model driven model performance is measured based on output comparison with a baseline, i.e., via an error metric with a well-established measurement technique. Error is defined as output deviation compared to the baseline and can be measured through various parameters, such as \gls{rms}, \gls{crlb}, mean, and standard deviation. \gls{crlb} is defined as the minimum achievable variance~\cite{130} of a certain parameter. A few data driven models were also tested with an error metric. Both data and model driven models sometimes rely on visual inspection to determine model performance. Visual inspection can be used to assess feature separability~\cite{1,81,82,150} or model behavior~\cite{5,21,44,80,102,107,132}.

\begin{table*}[ht]
    \centering
    \caption{Summary of model improvement, optimization, and evaluation techniques used to improve and evaluate data driven and grey box model performance.} \label{tab:data_driven_model_optimization_methods}
    \tiny\sffamily
    \setlength{\tabcolsep}{4pt}
    \begin{NiceTabular}{@{}lM{10mm}cM{8mm}M{8mm}M{8mm}|ccc|ccc|M{12mm}M{12mm}M{10mm}c@{}}[code-before =\rowcolors{3}{}{gray!12}]
        \noalign{\hrule height 0.8pt}
         & \Block{1-5}{\textbf{Regularization}} & & & & & \Block{1-3}{\textbf{Hyperparameter tuning}} & & & \Block{1-3}{\textbf{Cross validation}} & & & 
         \Block{1-4}{\textbf{Training stabilization}} & & & \Tstrut\\[1.2ex]
         & Loss constraint & Dropout & Noise training & Weight decay & Early stopping & Manual & Grid search & Adaptive & K fold & Monte carlo & Leave one out & Batch normalization & Weight initialization & Architecture addition & Momentum \\[1.7ex] \hline
         
         \cite{1}   & X &   &   &   &   & X &   & X &   &   &   &   &   &   &   \\ 
         \cite{2}   &   &   &   &   &   & X &   &   &   &   &   &   &   &   &   \\ 
         \cite{4}   &   &   &   &   &   & X &   &   &   &   &   &   &   &   &   \\ 
         \cite{6}   &   &   &   &   &   & X &   &   &   &   &   &   &   &   &   \\ 
         \cite{7}   &   &   &   &   &   & X &   &   &   &   &   &   &   &   &   \\ 
         \cite{8}   &   & X &   &   &   & X &   & X &   &   &   &   &   &   &   \\ 
         \cite{9}   & X & X &   &   &   & X & X & X & X &   &   &   &   &   &   \\ 
         \cite{11}  &   & X &   &   &   & X &   &   &   &   &   &   &   & X &   \\ 
         \cite{14}  &   &   &   &   &   &   & X &   & X &   &   &   &   &   &   \\ 
         \cite{17}  &   & X &   &   &   & X &   & X &   &   &   &   &   & X &   \\ 
         \cite{19}  &   &   &   &   &   & X &   & X &   &   &   &   &   &   &   \\ 
         \cite{23}  &   &   &   &   &   & X &   &   &   &   &   &   &   &   &   \\ 
         \cite{35}  &   & X &   &   &   & X &   & X &   &   &   & X & X &   &   \\ 
         \cite{36}  &   &   &   &   &   &   &   &   &   &   &   & X &   &   &   \\ 
         \cite{39}  &   &   &   & X &   & X &   & X &   &   &   &   &   &   & X \\ 
         \cite{46}  &   &   &   &   &   & X &   & X &   &   &   & X & X &   &   \\ 
         \cite{52}  & X &   &   &   &   & X &   & X &   &   &   &   &   &   & X \\ 
         \cite{63}  &   &   &   &   &   & X &   & X &   &   &   &   &   &   &   \\ 
         \cite{71}  &   &   &   &   &   & X &   & X &   &   &   &   &   &   &   \\ 
         \cite{72}  &   & X & X &   &   & X &   &   &   &   &   &   &   & X &   \\ 
         \cite{74}  & X &   &   &   &   & X & X &   & X &   &   &   &   &   &   \\ 
         \cite{75}  & X &   &   &   &   & X & X &   & X &   &   &   &   &   &   \\ 
         \cite{93}  &   &   &   &   &   & X &   &   & X &   &   &   &   &   &   \\ 
         \cite{94}  &   & X &   & X &   & X &   & X & X &   &   & X &   & X &   \\ 
         \cite{96}  &   &   &   &   &   & X &   & X &   &   &   &   &   &   &   \\ 
         \cite{107} & X &   &   &   &   &   &   &   &   &   &   &   &   &   &   \\ 
         \cite{113} &   & X &   &   & X & X &   & X &   &   &   &   &   &   &   \\ 
         \cite{115} &   &   &   &   &   & X &   &   &   &   &   &   &   &   &   \\ 
         \cite{139} &   &   &   & X &   & X &   & X &   &   &   &   & X & X & X \\ 
         \cite{146} &   &   &   &   &   & X &   & X &   & X &   & X &   & X &   \\ 
         \cite{147} &   &   &   &   &   & X &   & X &   &   &   & X &   & X &   \\ 
         \cite{148} &   & X &   &   &   & X &   & X & X &   &   &   &   &   & X \\ 
         \cite{149} &   & X &   &   &   & X &   & X &   &   &   &   & X &   &   \\ 
         \cite{150} &   & X &   &   &   & X &   & X &   &   &   &   & X &   &   \\ 
         \cite{151} &   & X &   & X & X & X &   & X &   &   &   & X & X &   &   \\ 
         \cite{152} &   &   &   &   &   & X &   &   & X &   &   &   &   &   &   \\ 
         \cite{153} &   & X &   &   &   & X &   & X &   &   &   &   &   &   &   \\ 
         \cite{163} &   &   &   &   &   &   &   &   & X &   & X &   &   &   &   \\ 
         \cite{164} &   & X &   &   &   &   &   &   &   &   &   & X &   & X &   \\
         \cite{165} &   &   &   &   &   &   &   &   & X &   & X &   &   &   &   \\ \noalign{\hrule height 0.8pt}
    \end{NiceTabular}
\end{table*}

\subsubsection{Model Improvement and Evaluation Techniques}
\label{sec:data_driven_model_optimization_methods}
After developing mathematical proofs or conducting numerous quantitative experiments, physical and/or mathematical information required for solving a particular problem becomes known. Once this information is known, the white box model construction process for solving the problem is straightforward and the resulting model will exhibit good performance~\cite{math_model_intro}. This is not the case for grey and black box models. Grey box models involve hidden state updates and/or use of cost and similarity functions that depend on input data or features. This requires grey box model evaluation through numerous experiments every time the sensing application context or input data or feature distribution changes. As explained in Section~\ref{sec:data_driven_modeling}, black box models are constructed through an iterative process in which finally a model is created that performs well on a test dataset or feature set. This section elaborates on several problems that occur during grey box and black box model evaluation and black box model creation, and how these problems can be solved with special model improvement, evaluation, and/or optimization techniques. These techniques are summarized in Table~\ref{tab:data_driven_model_optimization_methods}. Because new model creation is explained in Section~\ref{sec:analytical_modeling}, we do not focus on it in this section. Elements causing training instability are not addressed. Prediction denoising through \gls{nms} or fusion performed by~\cite{39} is out of context in this section and is not addressed either. Next we explain main model improvement, optimization, and evaluation techniques used in data driven and grey box models of millimeter wave sensing applications.

\paragraph{Regularization}
An issue frequently encountered during model training is model overfitting, in which model parameters and mapping function completely adapt to the training data or feature set. This will cause the model to exhibit sub-optimal performance on held-out (unseen) data. Various regularization techniques try to avoid overfitting and help the model to perform well on a wider variety of data it might encounter when deployed in specific applications~\cite{Goodfellow-et-al-2016}. For example, the ability of a cost function to minimize or a reward function to maximize itself can be limited with special constraint~\cite{1,9,52,74,75,107} and weight decay~\cite{39,94,139,151} parameters added to the function. It can be decided to occasionally stop a parameter update for a limited number of model parameters in a particular training cycle through means of dropout~\cite{8,9,11,17,35,72,94,113,148,149,150,151,153,164}. In early stopping, the validation data or feature set can be used to inspect the model performance after every training cycle. In case the performance starts to diminish compared to previous training cycles, model training will be stopped~\cite{113,151}. A certain percentage of the training set can include noise samples to make the eventual model robust to input noise~\cite{72}.

\paragraph{Hyperparameter Tuning}
Most data driven models involve some kind of manual hyperparameter tuning as shown in Table~\ref{tab:data_driven_model_optimization_methods}. Hyperparameters include number of layers, neurons, forest size, forest depth, number of training cycles, number of data or feature inputs before an update is applied to model parameters, etc. Sometimes it is difficult to set the right initial hyperparameters due to a lack of a priori knowledge and experience. In this case, an exhaustive hyperparameter search using a grid search~\cite{9,14,74,75} can be implemented, through which a complete grid of hyperparameter values is constructed. Afterwards, for every possible hyperparameter combination, a model is created and its performance is evaluated. The set of final hyperparameters is selected as the one that gives the best model performance. Not adapting the learning rate, i.e., a hyperparameter that controls how much parameter update is used to change the model parameters, during training cycles can result in long execution time for model training or training instability. To combat these problems, adaptive learning rate techniques such as \gls{adam} estimation~\cite{1,8,9,17,35,71,94,113,146,147,149,150,151,153} or manual learning rate decay~\cite{19,148} can be employed.

\paragraph{Cross Validation}
Cross validation refers to training a model and evaluating model performance multiple times on a variety of different data or feature set splits. Splitting refers to partitioning a data or feature set into a subset designated for model training, a subset designated for model validation, and a subset designated for model testing. If model performance results are retrieved from a single training, validation, and test split, the model performance results will have a split specific bias. This means that the performance result will strongly deviate from the mean performance result that could be expected based on a given data or feature set. To combat this, model training, validation and testing can be executed on a variety of splits for more robust performance result evaluation. This can be implemented very exhaustively with leave-one-out cross validation in which the train-validate-test split is repeated for every possible combination~\cite{163,165}. Instead, methods such as Monte Carlo~\cite{146} and k-fold cross validation~\cite{9,14,74,75,93,94,148,152,163,165} can be used in which the data or feature set splits are limited to a certain number.

\paragraph{Training Stabilization}
Elements like gradient exploding, gradient vanishing, internal covariate shift, noisy parameter updates, and bad parameter initialization have a negative impact on training stability. Training stability refers to an analytical model whose cost or reward function gradually minimizes or maximizes over time when new training samples are encountered during model training. Training instability examples include the cost or reward function prematurely becoming constant, a cost function that suddenly starts to increase with continuous acceleration, etc. Batch normalization~\cite{35,36,46,94,146,147,151,164}, initializing the model parameters prior to training with Xavier initialization~\cite{46,149,150,151}, making changes to existing model types with for example the ResNet layer~\cite{94,146} or LeakyRELU activation function~\cite{11,17,72,164} and parameter update averaging over multiple training cycles with momentum~\cite{39,52,139,148} have been proposed to solve the problem of training instability.

\subsubsection{Gaps and Challenges}
A challenge often encountered in model evaluation is determining a set of metrics that can completely and objectively assess model performance. From Table~\ref{tab:eval_metrics_summary}, it can be concluded that there is no such universal metric and none of the existing works covered all or even a majority of the given performance metrics. This, in combination with model variation, makes benchmarking between models difficult. A widely used metric that does not portray a complete performance picture is the accuracy metric. Accuracy cannot discriminate between a good or bad performing model in case data or feature sets are heavily skewed in terms of class distribution. More information regarding the selection of a suitable set of performance metrics can be found in~\cite{eval_metric_review}. Designing models that perform well in both lab and real-life contexts is still a major challenge. Model simulation~\cite{22,49,141,159}, controlled experiments~\cite{9,81,82,115}, and system challenges related to generalization~\cite{118} and impact~\cite{123} are frequently observed in scientific studies. Examples of controlled experiments and restricted experimental environment include test subjects performing activities directly in front of a radar~\cite{9} and object detection with very limited and pre-defined objects placed at specific positions in a radar's \gls{fov}~\cite{81,82}. Table~\ref{tab:data_driven_model_optimization_methods} shows that many papers do not harness the strength of multiple regularization methods. It is reasonable to think that combining loss constraint, dropout, and/or weight decay is not required or it is undesired because a single one of these methods may already result in good model performance on unseen data or they might negatively influence the training process~\cite{Goodfellow-et-al-2016}. Therefore, combining loss constraint, dropout, and/or weight decay to improve model performance on unseen data is always subject to extensive evaluation. However, using early stopping in combination with other regularization methods can omit the need for exhaustively tuning the required number of training cycles to achieve optimal model performance while not negatively influencing the training process~\cite{Goodfellow-et-al-2016}.

\section{Challenges, Trends, and Future Perspective}
\label{sec:challenges_future_trends}
This section presents and explains identified scientific and technological challenges and trends for applications of millimeter wave as a sensing technology. The challenges and trends have been categorized into several challenge and trend categories: hardware, unsupervised representation learning, support from other sensing domains, application integration, and crowd analysis. In addition, this section also provides a future perspective for applications of millimeter wave as a sensing technology.

\subsection{Hardware}
Several challenges related to the hardware used to collect data have an effect on how well a millimeter wave sensing methodology performs. Human vital sign monitoring pipelines are currently not reliable enough in the context of multiple humans that are all located at varying distances from each other and at various positions in reference to the data collection system. Issues include different relative error results compared to a baseline in the context of changing measurement positions (measuring in front or at a side of a test subject)~\cite{66,123}, impact of random body movement~\cite{26,34,121}, and measurement occlusion issues when humans are standing too close to each other~\cite{26,66}. One study focused on finding the most optimal vital sign sensing position~\cite{53}. Several pipelines experience performance deterioration at large distances from the data collection system~\cite{75,86,103,112,129,132,133,147}, in the presence of solids such as occluding objects~\cite{96,106,112,132} and brick/concrete wall barriers~\cite{75,128,129}, not using 3D printed placement constraints~\cite{165}, and in the context of hard to distinguish entities~\cite{74,93,95,113,134,135}. A challenge in concealed object detection and pedestrian detection is the variable reflection intensity caused by different types of clothing people wear~\cite{61,108}. One gesture recognition methodology in a car analyzed the impact of measurement position~\cite{2}. The optimal sensor placement was found to be on the center console (for use by front seat passenger and the driver) or in between the backs of the front seats (for use by the back-seat passengers). Performance deterioration was encountered when the sensor was placed too close to detectable objects such as the gear shift. Robustness against environmental effect analysis is limited to ego-motion estimation in~\cite{158}. Other challenges for manufacturers are to bring the cost of millimeter wave systems down to allow large deployments on a budget (e.g., for crowd analytics) and to improve the design and the form factor such that these systems do not cause architectural and aesthetic concerns for environments where they are deployed.

\subsection{Unsupervised Representation Learning}
Unsupervised representation learning has been used to pre-train a denoising autoencoder~\cite{19} in millimeter wave application pipelines. The network can be used afterwards to extract features from a dataset that generalize to a wide variety of end tasks~\cite{6472238}. Features are transferred from a pre-trained neural network to a supervised end learning task network. The networks used during pre-training and the end task share no learning task relation~\cite{10.1145/1273496.1273592}. More information can be found in a review written by Bengio et al.~\cite{6472238}. In 2014, a new breakthrough in unsupervised pre-training was realized by Dosovitskiy et al.~\cite{10.5555/2968826.2968912}. A concept called self-supervised learning was envisioned in which a supervised neural network tries to classify data transformations applied to unlabeled data in the pre-training stage. Since millimeter wave data is unlabeled too when it is sampled from the underlying hardware, it remains to be determined if unsupervised representation learning can reduce the amount of labeled data required for training end task deep learning models by using either unsupervised or self-supervised learning in a given pre-training stage. Millimeter wave datasets that can be used for experimentation include the UWCR radar mini, NuScenes, RadHAR, and Solinteraction datasets~\cite{9,96,nuscenes2019,165}. However, the ability to explore is currently hindered by a lack of more datasets as indicated in~\cite{113} for pose estimation.

\subsection{Support From Other Sensing Domains}
Some pipelines apply sensor fusion of millimeter wave data with data originating from other sensing domains. Sensing domains include vision~\cite{15,31,32,43,44,62,100,121,139}, depth~\cite{5,121}, lidar~\cite{62}, inertial measurements~\cite{91} and exerted forces~\cite{93}. Data originating from these domains complement millimeter wave data and therefore cause more accurate analytical modeling performance in several situations. Analytical modeling performance in on-road detection and tracking suffers from limited spatial resolution of, and noise in, millimeter wave data~\cite{15,31,44,100,139}. Odometry information model performance also suffers from noise in millimeter wave data~\cite{91}. Data originating from other sensing domains have been used to guide a radar collection system to the most accurate vital sign sensing position~\cite{121} and have lead to increased analytical model performance~\cite{5,43} and model state accuracy in object detection for visually impaired people~\cite{5}. Millimeter wave data also provide benefits to the other sensing domains in return. The data are used for example to differentiate objects based on material, density or volume where exerted forces only measure similar object geometries~\cite{93}. In the future, fusion with other sensing domains can be extended. For example, when measuring crowd density in a limited space inside a building, fusion with heat energy obtained through a temperature sensor can be explored.

\subsection{Application Integration}
A variety of different sensing applications integrate different application types. Examples include combining communication and sensing~\cite{76,166}, combining vital sign, activity, and gesture sensing for more robust occupancy detection~\cite{77}, \gls{slam}~\cite{131,141,142,162}, detection/tracking and identification~\cite{8,12,146}, and detection and activity recognition~\cite{72}. Detection and tracking make identification and activity recognition more robust in unknown environments. Research combining communication and sensing is limited to vision~\cite{76} and range simulations~\cite{166}. Due to performance deficiencies with communication-only access point sensing~\cite{76}, sensing capabilities built into special access points are envisioned to enable better elderly monitoring and building analytics without extra costs related to installation of dedicated sensor hardware. The combination of communication and sensing in \gls{v2v} scenario's is envisioned to enable robust driver assistance systems~\cite{6127923}. Yassin et al.~\cite{141} denote that \gls{slam} research with millimeter waves is still at its infancy.

\subsection{Crowd Analysis}
Throughout the review process no papers have been found that explore crowd analysis with millimeter wave sensing. The non-image based crowd counting review by Kouyoumdjieva et al.~\cite{8660451} is recommended as an introductory read since millimeter wave crowd analysis applications in the future will belong to the non-image based crowd analysis application category. Future millimeter wave crowd analysis research will include density/size, flow/trajectory/movement, and activity/behavior analysis. Density/size analysis refers to counting the number of people and their distribution in a given area. Kouyoumdjieva et al.~\cite{8660451} denote that for disaster management and city-wide public transportation crowd counting tier models should be developed that have the ability to provide a macro crowd count based on local estimates calculated over different micro or meso area's. Flow/trajectory/movement analysis refers to determining major movement flows and directions in a given area, as well as identifying minimum and maximum bounds in terms of target numbers and dynamicity. Activity/behavior analysis refers to determining meso and macro level crowd activities, as well as identifying the correlation between location and activities and minimum and maximum bounds in terms of target numbers and granularity of activities. Major challenges for millimeter wave crowd analysis include methodology scalability to major events including thousands of people and operation security. The cost to deploy a grid of millimeter wave data collection systems is currently too high and deployed grids, in case the grid is not concealed in the environment, will interfere with decor design. Kouyoumdjieva et al.~\cite{8660451} denote that no non-image based crowd counting methodologies explore security in the form of robustness to output manipulation and malicious users trying to disable methodology functionality.

\subsection{Future Perspective}
We notice that most millimeter wave sensing research is currently limited to a personal lab scope. This means that research is most often performed with one person, a few people, one object or a few objects in simulated or controlled experiment environments. We believe that there is a lot of potential for research geared towards exploring millimeter wave sensing in large scale active and dynamic industrial and urban area's involving numerous people, other mammals, and or objects in the future.

Millimeter wave sensing infrastructures are unobtrusive in nature and will be deployed ubiquitously. Future millimeter wave infrastructures will also co-exist and collaborate with other sensing and communication infrastructures, and serve multiple sensing applications in parallel. For example, sensing capabilities will be integrated into special communication access points in the future to enable better elderly monitoring and building analytics without costs related to installation of dedicated sensor hardware~\cite{76}. Integration of communication and sensing in \gls{v2v} scenario's will enable robust driver assistance systems~\cite{6127923}.

Analytical modeling will become more prominent in millimeter wave sensing application pipelines. Jiang et al.~\cite{1} have presented an analytical model that has the ability to extract input data features that are environment and user specific information (i.e. domain) independent in the context of human activity detection. We believe that the ability to learn extraction of domain independent input data features is an important goal for future research in the context of analytical models that work with millimeter wave data. This ability will allow analytical models to be robust against influences from a variety of domains and therefore to perform well in real-life scenario's in the future. Analytical modeling research should be conducted to determine which domain influences can or cannot be mitigated, which domains can or cannot be integrated in an analytical model, how an analytical model can learn latent domains from the input data, and which analytical model types work best in a wide variety of different application types. For example, several hardware challenges can be regarded as a domain, rather than trying to mitigate hardware challenge influence and application robustness deterioration with cancellation methods~\cite{s20051454} per challenge, using sensor fusion, or integrating application types. A variety of different models that extract domain independent input data features can be found in papers which do not consider millimeter wave sensing. For example, papers that consider WiFi \gls{csi}~\cite{8808929,10.1145/3307334.3326081,10.1145/3161183}. An important step in analytical modeling, due to the labor intensive nature of labelling millimeter wave data, is the development of unsupervised models that achieve performance that is on par with supervised counterparts. We believe that it is also important to investigate how learning domain independent input data feature extraction can be integrated with creation of analytical models through means of unsupervised or self-supervised representation learning.

Stimuli such as sound, light, stress, anxiety, overtraining, temperature, humidity, etc. have an effect on our health and vitality~\cite{ross_ans_health_vit}. The COVID-19 pandemic and confinement measures taken during the pandemic result in stress and anxiety in a wide variety of different people~\cite{covid_mental_health}. We believe that this will spark debate in the general public about the effect of stress and anxiety and how we can reduce stress and anxiety in our daily lives in the near future. We also believe that this will stimulate research regarding the effect of stimuli on health and well-being with millimeter wave sensing systems. Vital sign sensing with millimeter wave systems can be used in niches where so called contact sensors, i.e., sensors involving electrodes, air analysis with nasal cannula or mask, a strap-on system, smart watch, smart phone, etc.~\cite{s20051454,pmid28290720}, cannot be used. Examples include, but are not limited to, skin irritation and allergic reactions, damaged skin (rashes, burns, hives, etc.), in the presence of clothes and obstacles~\cite{8732355}, multiple beings, and humans that have certain behavioral conditions (e.g. severe autism, dementia, etc.).

\section{Conclusion}
\label{sec:conclusion}
This is the first review that completely covers millimeter wave sensing application pipelines and pipeline building blocks in the form of a systematic literature review to the best of our knowledge. The millimeter wave technology covers a wide bandwidth and its short wavelength gives limited range, giving low signal interference. This means transceivers can be packed very densely in an area without disrupting each others’ communication signals. These properties of the technology not only yield high communication rates, but also provide a great opportunity for sub-millimeter accuracy level sensing of the surroundings, easily penetrating through simple obstacles like plastic and fabric.

Our analysis of the literature showed that there are indeed a variety of application types for millimeter sensing that we group into three domains; namely, human, object, and environment sensing. The application pipelines in the literature are made up of (a subset of) five common building blocks: data collection, pre-processing, feature extraction, analytical modeling, and modeling evaluation. There is naming confusion in the literature in terms of which models and techniques take part in which building blocks. In this paper, we provided sharp descriptions of millimeter wave sensing building blocks; i.e., the hardware, algorithms, analytical models, and or model evaluation techniques that are covered by each block.

For different applications in the literature, each building block may select from a variety of models and techniques. A close look into many application instances reveals that a large majority of them stick to a combination of a few common solutions in their sensing pipelines. The rest of the models and techniques remain application-specific and their usage is not explored widely in other applications. This is not surprising as this field of research and development is still young and it is safer to rely on widely accepted solutions. For example, it is very common to utilize deep black box models employing \glspl{cnn} and \gls{lstm} networks in the more frequently used data driven modeling, whereas Bayesian filters and association/allocation seem to be the first choices among model driven approaches. In terms of feature extraction, manual feature mapping is the predominant choice of researchers. We therefore encourage the researchers entering, or planning to conduct new research in, the millimeter wave sensing environment to try out new models. This will increase the model variety and thus the knowledge on how analytical modeling in the millimeter wave sensing environment can be approached.

\end{document}